\ifdefined\DoubleSided
  \documentclass[twoside,openright,a4paper,12pt]{nusthesis}
\else
  \documentclass[a4paper,12pt]{nusthesis}
\fi

\dsp 

\usepackage[utf8]{inputenc}

\usepackage{indentfirst} 

\usepackage{silence} 
\usepackage{bookmark}
\usepackage[
  backend=biber,
  style=ieee,
  citestyle=numeric,
  giveninits=true,
  sorting=nyt,
  maxbibnames=99,
  dashed=false,
  doi=false]{biblatex}
\DeclareFieldFormat*{title}{``#1''\newunitpunct} 
\usepackage{microtype} 
\usepackage{tabulary} 
\usepackage{hhline}

\usepackage{enumitem}

\usepackage{hyperref}

\usepackage{listings}

\usepackage{amsthm}
\newtheorem{theorem}{Theorem}
\theoremstyle{definition}
\newtheorem{defn}[theorem]{Definition} 
\theoremstyle{plain}
 
\newtheorem{thm}[theorem]{Theorem} 
\newtheorem{lemma}[theorem]{Lemma} 
 
\newtheorem{conjecture}{Conjecture} 

\usepackage{mathtools}

\usepackage{interval}
\intervalconfig{soft open fences}

\usepackage[linesnumbered,ruled,vlined]{algorithm2e}
\SetAlgorithmName{Algorithm}{Algorithm}{Algorithm}
\IncMargin{0.5em}
\SetCommentSty{textnormal}
\SetNlSty{}{}{:}
\SetAlgoNlRelativeSize{0}
\SetKwInput{KwGlobal}{Global}
\SetKwInput{KwPrecondition}{Precondition}
\SetKwProg{Proc}{Procedure}{:}{}
\SetKwProg{Func}{Function}{:}{}
\SetKw{And}{and}
\SetKw{Or}{or}
\SetKw{To}{to}
\SetKw{DownTo}{downto}
\SetKw{Break}{break}
\SetKw{Continue}{continue}
\SetKw{SuchThat}{\textit{s.t.}}
\SetKw{WithRespectTo}{\textit{wrt}}
\SetKw{Iff}{\textit{iff.}}
\SetKw{MaxOf}{\textit{max of}}
\SetKw{MinOf}{\textit{min of}}
\SetKwBlock{Match}{match}{}{}

\usepackage{bbm}

\usepackage{array}
\newcolumntype{L}[1]{>{\raggedright\let\newline\\\arraybackslash\hspace{0pt}}m{#1}}
\newcolumntype{C}[1]{>{\centering\let\newline\\\arraybackslash\hspace{0pt}}m{#1}}
\newcolumntype{R}[1]{>{\raggedleft\let\newline\\\arraybackslash\hspace{0pt}}m{#1}}

\usepackage{color}
\usepackage[usenames,dvipsnames]{xcolor}
\usepackage{soul}
\soulregister\cite7
\soulregister\ref7
\soulregister\pageref7
\soulregister\autoref7
\soulregister\eqref7

\renewcommand{\eqref}{Equation~\ref}

\newcommand*{\RootPicDir}{pic}
\newcommand*{\PicDir}{\RootPicDir}

\newcommand*{\SetPicSubDir}[1]{\renewcommand*{\PicDir}{\RootPicDir /#1}}

\usepackage{lipsum} 

\usepackage{amsmath,amssymb}

\newcommand{\block}[2]{{#1}_{1}, {#1}_{2}, \ldots, {#1}_{#2}}
\newcommand{\Share}{\texttt{Sh}}

\newcommand{\eps}{\varepsilon}
\newcommand{\bit}{\{0, 1\}}

\newcommand{\cS}{\mathcal{S}}
\newcommand{\cX}{\mathcal{X}}
\newcommand{\cT}{\mathcal{T}}
\newcommand{\E}{\mathbb{E}}

\newcommand{\minH}{H_\infty}

\newcommand{\Rec}{\texttt{Rec}}
\newcommand{\Sh}{\texttt{Sh}}

\newcommand{\cL}{\mathcal{L}}
\newcommand{\cA}{\mathcal{A}}
\newcommand{\cQ}{\mathcal{Q}}
\newcommand{\cR}{\mathcal{R}}
\newcommand{\bE}{\ensuremath{\mathbb{E}}}
\newcommand{\bN}{\ensuremath{\mathbb{N}}}

\newcommand{\cF}{\ensuremath{\mathcal{F}}}
\newcommand{\cH}{\ensuremath{\mathcal{H}}}
\newcommand{\cP}{\ensuremath{\mathcal{P}}}
\newcommand{\cY}{\ensuremath{\mathcal{Y}}}
\newcommand{\tY}{\ensuremath{\widetilde{Y}}}

\newcommand{\from}{\ensuremath{ \leftarrow }}
\newcommand{\dist}[2]{\ensuremath{ \Delta \left(#1; #2 \right) }}
\newcommand{\distCond}[3]{\ensuremath{ \Delta \left(#1; #2 \left| #3 \right. \right) }}
\newcommand{\supp}[1]{\ensuremath{\mathbf{supp}({#1})}}
\newcommand{\unif}[1]{\ensuremath{ U_{#1} }}
\newcommand{\minEnt}[1]{\ensuremath{H_\infty(#1)}}
\newcommand{\condMinEnt}[2]{\ensuremath{H_\infty(#1|#2)}}
\newcommand{\avgCondMinEnt}[2]{\ensuremath{\tilde{H}_\infty(#1|#2)}}
\newcommand{\blocks}[2]{\ensuremath{ #1_{1}, \ldots, #1_{#2} }}
\newcommand{\functionBlocks}[3]{\ensuremath{ #3(#1_{1}), \ldots, #3(#1_{#2}) }}
\newcommand{\givenBy}{\ensuremath{\sim}}
\newcommand{\closeTo}[1]{\ensuremath{\approx_{#1}}}
\newcommand{\concat}{\ensuremath{\circ}}
\newcommand{\mutualInfoCond}[3]{\ensuremath{I(#1 ; #2 | #3)}}
\newcommand{\indicate}[1]{\ensuremath{\mathbbm{1}\{{#1}\}}}

\newcommand{\ecc}{\ensuremath{\mathtt{EC}}}

\newcommand{\extractorParam}[5]{\ensuremath{ #1 : [ #2 , #3 \mapsto #4 \sim #5 ] }}

\newcommand{\sext}{\ensuremath{\mathbf{SExt}}}
\newcommand{\mac}{\ensuremath{\mathtt{MAC}}}
\newcommand{\samp}{\ensuremath{\mathtt{samp}}}

\newcommand{\ext}{\ensuremath{\mathbf{ext}}}
\newcommand{\fnmext}{\ensuremath{\mathbf{FNMExt}}}
\newcommand{\twonmext}{\ensuremath{\mathbf{2NMExt}}}
\newcommand{\nmExt}{\ensuremath{\mathbf{2NMExt}}}
\newcommand{\betterTwoNMExt}{\ensuremath{\mathbf{nmRaz}}}

\newcommand{\liExt}{\ensuremath{\mathbf{Li}}}
\newcommand{\razExt}{\ensuremath{\mathbf{Raz}}}
\newcommand{\treExt}{\ensuremath{\mathbf{Tre}}}
\newcommand{\crTreExt}{\ensuremath{\mathbf{crTre}}}
\newcommand{\outerExt}{\ensuremath{\mathbf{NMExt}}}
\newcommand{\innerExt}{\ensuremath{\mathbf{CRExt}}}

\newcommand{\outerOutputLength}{\ensuremath{m}}
\newcommand{\innerOutputLength}{\ensuremath{d}}

\newcommand{\Enc}{\mathbf{Enc}}
\newcommand{\Dec}{\mathbf{Dec}}

\newcommand{\treError}{\ensuremath{\eps_T}}
\newcommand{\razError}[1]{\ensuremath{2^{-(1.5)#1}}}
\newcommand{\liError}{\ensuremath{\eps_L}}
\newcommand{\collisionError}{\ensuremath{\eps_{Collision}}}
\newcommand{\fnmError}{\ensuremath{\eps_{fnm}}}
\newcommand{\twonmError}{\ensuremath{\eps_{tnm}}}
\newcommand{\outerError}{\ensuremath{\delta_{\outerExt}}}
\newcommand{\innerError}{\ensuremath{\delta_{\innerExt}}}

\newcommand{\modifiedTampering}{\ensuremath{T_{f, g}}}

\newcommand{\YLeft}{\ensuremath{Y_\ell}}
\newcommand{\YRight}{\ensuremath{Y_r}}
\newcommand{\YLeftTampered}{\ensuremath{g(Y)_\ell}}

\newcommand{\YLeftLength}{\ensuremath{n_\ell}}
\newcommand{\YRightLength}{\ensuremath{n_r}}
\newcommand{\poly}{\text{poly}}

\newcommand{\Tamp}{\mathsf{Tamp}}
\newcommand{\Sim}{\mathsf{Sim}}

\newcommand{\same}{\mathsf{same}^*}

\newtheorem{definition}[theorem]{Definition}
\newtheorem{corollary}[theorem]{Corollary}
\newtheorem{claim}[theorem]{Claim}

\newtheorem{question}[theorem]{Question}
\newtheorem{remark}[theorem]{Remark}

\newcommand{\cD}{\mathcal{D}}
\newcommand{\cZ}{\mathcal{Z}}

\newcommand{\cG}{\mathcal{G}}

\newcommand{\sh}{\mathsf{sh}}

\newcommand{\cV}{\mathcal{V}}

\newcommand{\cB}{\mathcal{B}}

\newcommand{\messageLength}{b}

\newcommand{\zPair}{(Z_1, Z_2)}
\newcommand{\fgZ}{(f(Z_1), g(Z_2))}
\newcommand{\fgZCondOnSi}{(f(Z_1), g(Z_2) | Z_1 \in \mathcal{S}_i)}
\newcommand{\ZCondOnSi}{(Z_1, Z_2 | Z_1 \in S_i)}

\newcommand{\bitsB}{\bits^b}

\newcommand{\bitsL}{\bits^\ell}

\newtheorem{lem}[theorem]{Lemma}

\newtheorem{coro}[theorem]{Corollary}

\newcommand{\cI}{\mathcal{I}}

\newcommand{\cM}{\mathcal{M}}
\newcommand{\cC}{\mathcal{C}}

\newcommand{\bits}{\{0, 1\}}

\newcommand{\Leak}{\mathsf{Leak}}

\let\originalleft\left
\let\originalright\right
\renewcommand{\left}{\mathopen{}\mathclose\bgroup\originalleft}
\renewcommand{\right}{\aftergroup\egroup\originalright}

\newcommand{\dg}{\mathsf{deg}}

\newcommand{\leftEntropy}{\left(\beta - (\frac{i-1}{r})\right)\messageLength}
\newcommand{\rightEntropy}{\left(\frac{i-1}{r} \right)\messageLength}

\newcommand{\closeToLeft}{D_{1, i}}
\newcommand{\closeToRight}{D_{2, i}}
\newcommand{\closeToPair}{(\closeToLeft, \closeToRight)}
\newcommand{\fgcloseToPair}{(f(\closeToLeft), g(\closeToRight))}

\newcommand{\negl}{\mathsf{negl}}

\newcommand{\numberRep}[1]{v_{#1}}
\newcommand{\butterflyEdge}[3]{e_{#1}(#2, #3)}
\newcommand{\groupSize}{\lvert \cG \rvert}

\newcommand{\patrasku}{P\v{a}tra\c{s}cu}

\usepackage{amssymb,amsmath,amsthm}

\begin{document}

\title{THINKING INSIDE THE BOX: PRIVACY AGAINST STRONGER ADVERSARIES}

\author{Eldon Chung}
\prevdegrees{%
  B.Comp. (Hons.), National University of Singapore}
\degree{Doctor of Philosophy}
\field{Computer Science}
\degreeyear{2024}
\supervisor{Associate Professor Divesh Aggarwal}
\cosupervisor{Dr. Maciej Obremski}

\examiners{%
Professor Marco Tomamichel\\
Assistant Professor Prashant Nalini Vasudevan\\
Associate Professor Stefano Tessaro,
University of Washington 
}

\maketitle

\declaredate{02-Feb-2024}
\declaresign{\RootPicDir /signature.jpg} 
\declarationpage

\begin{frontmatter}
  \begin{acknowledgments}

I am very fortunate to have been advised by Prof. 
Divesh Aggarwal. He has been one of the most supportive professors I have ever met and has always had the best interest of his students in mind. He has always let me explore any questions I express an interest in and encouraged me to work on it to the best of my ability. 
I am also very grateful to have been co-advised by Prof. Maciej Obremski, for the amount of care, time, and effort, he has put into mentoring me, and for the endless list of interesting questions he has given me to work on. Both of them have been amazing guiding figures and I would not have chosen anyone else to be in their place if given a choice. I hope to carry their lessons with me and eventually make changes in the world that they can be proud of.

I would also like to thank the many amazing people I have had an opportunity to collaborate and discuss ideas with.
I would like to thank Zeyong Li and Grace Tan for all the times we have collaborated and worked together. The two of them have always spared time to work on my ideas with me, no matter how silly they turn out to be. While some of the ideas may not have panned out, it was always a fun learning experience. I would also like to thank Prof. Siyao Guo and Prof. Kasper Green Larsen for hosting me at NYU Shanghai and Aarhus University respectively. I would like to thank Prof. Noah Stephens-Davidowitz, and Prof. Alexander Golovnev, for opportunity they gave me to collaborate with them, and for all the fun and unforgettable moments while doing so. I would like to thank the people I've shared an office with in the last few years. Teodora Baluta, Kareem Shehata, Bo Wang, and Rui Shi Chew, for all the random chats and discussions we have had. 

Finally and very importantly, I want to thank Duong Pham, and Grace Tan for being my emotional support all these years. I'm not sure I would have been able to make it this far without the two of you.

\end{acknowledgments}

  \tableofcontents 
  \chapter{Summary}

In this thesis, we study extensions of statistical cryptographic primitives. In particular we study leakage-resilient secret sharing, non-malleable extractors, and immunized ideal one-way functions. The thesis is divided into three main chapters.

In the first chapter, we show that 2-out-of-2 leakage resilient (and also non-malleable) secret sharing requires randomness sources that are also extractable. This rules out the possibility of using min-entropic sources. 

In the second, we introduce collision-resistant seeded extractors and show that any seeded extractor can be made collision resistant at a small overhead in seed length. We then use it to give a two-source non-malleable extractor with entropy rate 0.81 in one source and polylogarithmic in the other. The non-malleable extractor lead to the first statistical privacy amplification protocol against memory tampering adversaries.

In the final chapter, we study the hardness of the data structure
variant of the $3$SUM problem which is motivated by a recent construction to immunise random oracles against pre-processing adversaries. We give worst-case data structure hardness for the $3$SUM problem matching known barriers in data structures for adaptive adversaries. We also give a slightly stronger lower bound 
in the case of non-adaptivity. Lastly, we give a novel result in the bit-probe setting.

\section{Papers}

Chapter 3 is based on \textbf{On Secret Sharing, Randomness, and Random-less Reductions for Secret Sharing} which was accepted at TCC 2022 \cite{ACOR21}.

Chapter 4 is based on \textbf{Extractors: Low Entropy Requirements Colliding With Non-Malleability}, which
was accepted at CRYPTO 2023 \cite{ACO23}.

Chapter 5 is based on \textbf{Stronger 3SUM-Indexing Lower Bounds}, which was accepted at SODA 2023 \cite{CL23}.
\end{frontmatter}

\SetPicSubDir{ch-Intro}

\chapter{Introduction}
\vspace{2em}

Cryptography is the study of protocols between multiple 
parties that provide some notion of privacy against adversaries. 
Notable examples include:
\begin{enumerate}
    \item (\textbf{Multi Party Computation, MPC}) Multiple parties coming together to compute a function $f$ on their joint inputs without revealing their inputs to each other.
    \item (\textbf{Privacy Amplification, PA}) Communicating over an eavesdropped channel, to ``strengthen'' a weakly random secret key into a fully uniform one.
    \item (\textbf{Signature Schemes}) Appending ``signatures'' to messages that can later be verified using some previously distributed information. 
\end{enumerate}

Typically, such protocols are built using vital 
components. From our examples listed above:
\begin{enumerate}
    \item MPC can be built using a primitive called \emph{secret sharing} \cite{CDN15,E22}.
    \item PA can be built using a primitive called a \emph{randomness extractor} \cite{DLWZ14}.
    \item Signature Schemes can be built using a primitive called \emph{one-way functions} \cite{R90}.
\end{enumerate}

However, the privacy guarantee of the above protocols 
are against adversaries whose access is only limited to the public 
communication between the parties or information disclosed by the protocols 
themselves. Such adversaries are sometimes 
referred to as ``black-box'' or ``out-of-the-box'' 
adversaries\footnote{Here, the box 
refers to the device that is performing the computation.}. 
Real life attacks on a cryptographic protocol or primitive need not 
be restricted in such ways. Adversaries might have: (1) Access to the 
``box'' to gain additional information about the inputs or computations 
(which we refer to as \emph{leakage}), or (2) the ability to tamper with the 
cryptographic primitives, or similarly, (3) the ability to inject 
``back-doors''. We can think of all three of such adversaries as having some additional 
access to or influence on certain parts \emph{inside the box}, 
which potentially circumvent the security guarantees of the protocols that are run. 
As they say, a chain is only as strong as its weakest link, similarly, a cryptographic protocol
is only as strong as its weakest primitive.
We will study strengthening these primitives.

We first give background on some of these types of attacks (referred to as side-channel attacks),
and then on certain cryptographic primitives that we will be studying.

\section{Side-Channel Attacks}
A \emph{side-channel attack} refers to any attack that 
is based on either obtaining additional 
information or influencing the execution of a 
protocol rather than being based on the design of the 
protocol itself \cite{WG21}. In the real world,
these attacks range from monitoring CPU cache accesses to leak
information about the secret key \cite{AGM16,IIES14,YF14+}, flipping targeted bits
by running other programs on the same device \cite{KDK+14}, or even introducing backdoors
into specific implementations (see 
the infamous Dual\_EC\_DRBG PRG algorithm standardised in NIST that was suspected to have a
backdoor for the NSA \cite{M13}).
We will study three types of cryptography that address these types of attacks. 

\paragraph{Leakage Resilience Cryptography.} Leakage Resilience 
Cryptography is the study of cryptography 
against adversaries that obtain additional information, such as 
computational states or secret keys or inputs. Such attacks have been 
implemented in real life \cite{HSHCPCFAF09,LSGP20,KHFGGHHLM20} to obtain 
portions of secret keys or other information about other running processes 
on the same machine.

Generally speaking, adversaries are modelled as having additional 
information via some arbitrary \emph{leakage function} $f$ (we will focus on the case that $f$ is computationally unbounded), and we refer to the output of $f$ as the \emph{leakage}. 
Relevant works in this area include designing symmetric key protocols in the face of bounded-length
leakage of secret keys \cite{D06,DLW06} and public key protocols in the face of arbitrary length
leakage of secret keys for \cite{AGV09}.

For a general survey on leakage-resilient cryptography, one can refer to \cite{DP08}.


\paragraph{Non-malleable Cryptography.} Non-malleable Cryptography 
on the other hand, is the study of cryptography where the adversary takes on 
a more active role by tampering with the execution of the protocol. For example, by tweaking inputs, or changing 
states of computation, or overwriting certain information.
There are several works introducing non-malleability for various 
cryptographic primitives \cite{DCM03,DPW18,FNMV14}. 

One particular application of interest is again the 
task of having two parties Alice and Bob who have to transform an agreed-upon weak secret (entropic but not necessarily uniform) into a fully uniform one -- 
except this time the adversary is not only allowed to eavesdrop, but 
to actively tamper with the communications between the two parties, or 
corrupt some of the internal state of one of the two parties. Prior works on this include \cite{DW09,Li12,CKOS19}.

\paragraph{Backdoors, and Immunisation.} The last model of adversaries that 
we will study is the type of adversary that 
can covertly influence protocols. For example,
they could influence protocols into using backdoored 
one-way functions for which they have an advantage in. 
This is part of a larger study called \emph{Kleptography}.
One way of creating primitives or protocols that 
resist such attacks is via \emph{immunization} which take 
established constructions and strengthening them against this type of adversary.

Along these lines, the works in \cite{DGGJR15,BFM18,RTYZ18,FJM18,GGHPV20} study immunizing 
one-way functions and pseudorandom generators in various settings. 
The setting introduced by \cite{GGHPV20} (which is the one we will focus on) is where the 
adversary who introduces the backdoor also has unbounded computational power create an advice string (of bounded length)
which they subsequently have bounded query access to. 

\section{Cryptographic Primitives}

\subsection{Randomness in Cryptography, and Extractors}
Algorithms for cryptography (and otherwise) are commonly 
designed while assuming access to uniformly random and 
independent bits as a random source.
However, it is not clear 
that random sources used by computers in real life (such as 
voltage measurements or timings between hardware events) can
provide such a guarantee. This is an issue because 
without this assumption, we might lose the security guarantees 
that the algorithms provide. 

One way to mitigate this is to use a \emph{randomness extractor} (denoted by $Ext$), which is an algorithm that is
applied on random sources $X$ to output (statistically close to)
$m$ uniformly random bits which can then be used by any algorithm. In the 
context of cryptography, the statistical distance $\varepsilon$ (or total variation 
distance) needs to be negligible\footnote{A function is called \emph{negligible} if it is smaller than any inverse polynomial.} in the security parameter of the protocol. We also note that extractors have interesting relationships to other constructs, such as Ramsey graphs \cite{C17}, expander graphs \cite{GUV09}, hash functions \cite{Vad12}, and more.

Ideally, we should consider constructing extractors for one of the most 
general class of random sources --- the class of random 
sources with \emph{min-entropy}. A random source $X$ has min-entropy $k$ ( denoted by $H_\infty(X) \geq k$)
if $\min_x \left\{-\log( \Pr[X = x] )\right\} \geq k$. We refer to the quantity $\frac{k}{n}$ as the \emph{min-entropy rate}.
Unfortunately, this class of randomness sources
is not extractable, even if we only consider $n$-bit sources $X$ with min-entropy $n - 1$
--- consider any proposed extractor $Ext'$ that outputs even a nearly uniform single bit for this class,
we can give a source $X'$ with $\minEnt{X'} \geq n - 1$ such that $Ext'(X')$ 
is the nearly constant distribution. 

To mitigate this, one natural relaxation is to instead 
provide the extractor with additional random sources. 
We first consider the case where extractors 
additionally have access to an independent and uniform 
source. 
For ease of exposition, when a random variable $X$ is supported on the set of $n$ bits $\bit^n$
and has min-entropy $k$ is called an \emph{$(n, k)$-source}.

\paragraph{(Seeded Extractors)}
Seeded extractors $Ext : \bit^n \times \bit^d \to \bit^m$ are algorithms that take in a $(n, k)$-sources $X$, 
and additionally a uniform $d$-bit source called the \emph{seed} $S$. 

Via the probabilistic argument, we know that
seeded extractors exist for arbitrary values of min-entropy $k$, 
with output length $m = k + d - 2\log(1/\eps) - O(1)$, and seed 
length $d = \log(n - k) + 2\log(1/\eps) + O(1)$, with total variation
distance $\eps$. Thus, the goal is to find explicit (or equivalently, efficient in $n$) constructions that meet such parameters.
The best known explicit constructions are due to \cite{RRV02} and 
\cite{GUV09}, which work for all values of $k$, and are either: (1) optimal 
in entropy loss, but have a slightly larger seed seed requirement or (2) 
optimal in seed requirement.

\paragraph{(Two-source Extractors)}
A generalisation of seeded extractors are \emph{two-source} extractors 
$Ext : \bit^{n_1} \times \bit^{n_2} \to \bit^m$, 
where the second source need not be uniform, but also has min-entropy.
Again, the probabilistic argument shows that two-source extractors 
exist for where the min-entropy requirement in both sources are 
$log(n) + 2\log(1/\eps) + O(1)$. However, the best known 
constructions fall short of meeting this requirement \emph{in general}. 

For example, Raz's extractor
in \cite{R05} requires one of the sources to have min-entropy at least 
$\frac{n}{2}$, and $\Omega(\log(n))$ min-entropy in the other source. 
The extractors by Chattopadhyay and Zuckerman \cite{CZ19}, and by 
Li in \cite{Li16} match the entropy requirement but only achieve inverse polynomial distance to uniform, which is insufficent for applications in cryptography.

\paragraph{(Non-malleable Extractors)}
An extension of extractors that will be of particular interest are 
extractors that are \emph{non-malleable}, (denoted by $nmext$). These are extractors that 
take in two sources $X$, and $Y$, and provide the extraction guarantee $nmext(X, Y) \approx_\eps U_m$, even when $nmext(f(X), g(Y))$ is revealed, where
$f$ and $g$ can be thought of as functions that tamper with the random sources $X$ and $Y$ respectively.
\emph{Seeded non-malleable extractors} fulfil this guarantee when $X$ has min-entropy, $Y$ is fully uniform, $f$
is the identity function, and $g$ is any fixed-point-free function. \emph{Seedless non-malleable extractors} (which are two-source extractors)
are when at least one of $f$ or $g$ are fixed-point-free, and if suffices that $X$ and $Y$ have min-entropy.

Seeded non-malleable extractors 
were first introduced and shown to exist 
(via the probabilistic argument) in \cite{DW09} for $(n, k)$ sources, with seed length $\log(n - k + 1) + 2\log(1/\eps) + O(1)$, and output length $\frac{k}{2} - \frac{3}{2}\log(1/\eps) - \log(d) - O(1)$.
Since then a line of works \cite{Li17,L20,CGL20}
resulted in seeded non-malleable extractors min-entropy $k = \Omega( \log^2(n / \eps))$,
output length $\Omega(k)$, and seed length $O(\log^2(n/\eps))$.

As for seedless non-malleable extractors, Li in \cite{L20} gives one such that 
the entropy requirement is $c \cdot n$ in both $n$-bit sources with exponentially small statistical distance to uniform. 
The exact value for $c$ for the entropy requirement is not known, though it is very close to $1$.

\subsection{Secret Sharing, and its extensions}
Informally, secret sharing \cite{B79,S79} is a cryptographic 
primitive that enables parties (using some sharing algorithm $\texttt{Share}$) to
``split'' a secret $m$ into a few ``shares'' $\block{\Share}{n}$, where the 
recovery of the secret is possible only when specific subsets of shares are 
used (as inputs for the recovery algorithm $\texttt{Rec}$). A special case 
of this is \emph{threshold} secret sharing, where recovery is 
possible when the amount of shares provided is above some threshold $t$, 
and impossible otherwise.
We also refer to this as $t$-out-of-$n$ secret sharing.

Secret sharing has also been used as a primitive for secure multi-party
computation \cite{CDN15,CCD88,BGW88}, generalised oblivious transfer \cite{SSR08},
threshold cryptography \cite{DF91}, and zero knowledge PCPs \cite{HVW22}. 

The ``standard'' construction for secret sharing is given by Shamir in \cite{S79},
and provides the guarantee that any adversary that holds less than $t$ shares is 
unable to distinguish between the original secret $m_0$ and some other secret $m_1 \neq m_0$.

\paragraph{(Leakage Resilient Secret Sharing)}
Introduced in \cite{BDIR21}, \emph{local leakage-resilient secret sharing} (LRSS) studies secret sharing 
for \emph{local leakage}. In addition to the usual guarantees, LRSS is furthermore secure against 
an adversary that is able to use leakage functions $f_1, f_2, \ldots, f_n$
to obtain leakage information $f_1(Sh_1), \ldots, f_n(Sh_n)$ from each share, where the total 
output length of each function $f_i$ is bounded.


\subsection{Privacy Amplification}
Introduced by Bennett, Brassard, and Robert in \cite{BBR88},
Privacy Amplification is the cryptographic protocol that 
allows Alice and Bob to create a fully uniform bit string, 
using a common shared weakly random string. The challenge here is that 
Alice and Bob have to communicate 
via a compromised channel on which 
the adversary is allowed to eavesdrop on
or tamper with messages (that are sent on the channel).
This protocol itself has been used as a building block for other applications in cryptography, such as quantum bit commitment \cite{BCJL93}, and quantum oblivious transfer \cite{BBCS92}.

The parameters of interest are: (1) The number 
of rounds of communication needed, (2) the 
entropy-rate required of the initially shared string, (3) the length of the resulting output string, (4) the total variation distance of the output string from the truly uniform distribution. Letting $k$ be the initial amount of min-entropy in the shared string, and letting $m$ be the length of the output, we can define the loss to be $k - m$. The goal is to minimize loss,
minimize the entropy rate required, whilst maximising output length, and keeping the total variation distance of the output string exponentially close to uniform.

Recently, Li in \cite{Li17} gave a two-round protocol
for arbitrary min-entropy, where the loss is linear in the amount of 
initial min-entropy, and the total variation distance is 
exponentially small in the amount of min-entropy. However, their 
protocol still relies on access to additional uniform randomness.

\subsection{One-Way Functions, the Random Oracle Model}
\emph{One-way functions} are efficient functions 
$f : \bit^n \to \bit^m$ such that for any efficient adversary $Adv$,
where $Adv$ has access to $f$, the probability that $Adv^f(f(x))$ (where
$x$ is chosen uniformly at random)
finds an $x'$ such that $f(x') = f(x)$ is negligible in $n$. 

They have been used as primitives to create other objects, such as 
pseudo-random generators \cite{Hol06}, or bit-commitment schemes \cite{HR07}.

Impagliazzo and Rudich in \cite{IR89} studied using a black-box 
fully random permutation $R$ to model the ideal one-way function. 
Indeed, a line of works \cite{IR89,Z98} show the hardness of 
inverting random oracles for efficient adversaries.

\paragraph{(One-Way Functions against Preprocessing Adversaries)} 
However, one-way functions in practice (such as the SHA family of algorithms
are not keyed), and instead are fixed in advance. This allows adversaries to pre-process 
information about the functions which can subsequently be used during the
inversion process. For example, consider an adversary that builds a massive inversion table for SHA256
which can then by used to arbitrarily invert it on any point.

In light of this, \cite{DTT10} consider a model of adversary $Adv$
where it additionally has access to an advice string of length $S$
which may depend on the random oracle $R$, and is allowed 
to make $T$ oracle accesses into the advice string, 
and random oracle, and they show an upper bound on the probability
(as a function of $S$ and $T$)
that an adversary is able to invert a random query. In particular
they show the probability $Adv$ inverts $f(x)$ for a uniformly
random $x$ is at most $O(\frac{T(S + n)}{N})$, with 
$x \in \bit^n$, and $N = 2^n$.

Guo et al. in \cite{GGHPV20} refer to the aforementioned type of adversary
as a \emph{preprocessing adversary} and observe that 
any privacy guarantees are lost when the length of the advice $S$ exceeds $N$
(basically having a large inversion table).
In light of this, they introduce \emph{one-way functions in the random
oracle model against $S$ preprocessing}. Which are functions 
$f = A^R$ ($A$ is an efficient algorithm 
with oracle access to $R$) 
which are efficiently computable given oracle access to
some random oracle $R$.

Interestingly, they further link the existence of such one-way functions (against adversaries that 
use $S$ space, and $T$ oracle accesses)
to the average case hardness of data structure problems, and show that both notions are \emph{equivalent}.

\begin{remark}[Informal, \cite{GGHPV20}]
    There exists a data structure problem that 
    is hard on average where the query distribution
    can be efficiently generated if and only if 
    there exists a one-way function against adversaries that 
use $S$ space, and $T$ oracle accesses in the random oracle model.
\end{remark}

\section{The General Questions and Our Results}

In this thesis, the overarching focus is on studying cryptographic primitives against adversaries that, 
in way or another, are ``in-the-box''. We view these results as steps in bridging the gap between 
theoretical cryptography, and modelling real-world attacks. Some of our results are tight, in the sense that any improvement
upon them will necessitate longstanding breakthroughs in either circuit complexity or two-source extractors, or 
data structure lower bounds.

\subsection{Randomness for Leakage-Resilient and Non-Malleable {Cryptography}}
As mentioned, it is impossible to construct
good extractors from min-entropic sources.
However, this does not mean that it is impossible to use such sources for specific applications.
Indeed, in the case of algorithms, Zuckerman in \cite{Z96} showed that min-entropic sources suffice for BPP
algorithms. On the other hand, for information-theoretic secret key encryption, Bosley and Dodis in \cite{BD07}
showed that encrypting an $m$-bit long message using an $n$-bit random source (with $m > \log(n)$) implies
a deterministic extractor for that source that outputs nearly $m$ bits, thus concluding that encryption requires randomness
sources which are extractable (i.e. min-entropic sources do not suffice).

The question remains open on other cryptographic primitives in terms of their randomness requirements.
Given how one can build 2-out-of-2 secret sharing from secret key encryption, a natural extension to the Bosley and Dodis result
is to also ask whether the same can be said out of 2-out-of-2 secret sharing. Indeed, this is an open question 
posed by Bosley and Dodis in \cite{BD07}.

\paragraph{(Our Contribution)}
We consider two extensions of secret sharing.

We show that 2-out-of-2 \emph{leakage resilient secret sharing} (LRSS) requires access to a randomness sources that are extractable. 
Moreover, if the 2-out-of-2 LRSS is efficient, then efficient extractors exist. Let $(\Share, \Rec)$ be a 
secret sharing scheme, and let $\cY$ be a class of 
random sources used by $\Share$. And informally, let it be that:
\begin{enumerate}
    \item Any adversary that has only $1$ share 
    has at distinguishing advantage at most $\eps_1$.
    \item Any adversary that obtains leakages $
    f(\Sh_1)$, and $g(\Sh_2)$ has distinguishing advantage at most $\eps_2$.
\end{enumerate}

Then we have the following:
\begin{thm}[Formally stated in \ref{thm:final}]\label{thm:lrss-ext}
    Let $(\Share,\Rec,\cY)$ be an $(\eps_1,\eps_2)$-leakage-resilient secret sharing scheme for $b$-bit messages (with $1$ bit of leakage).
    Then, either:
    \begin{enumerate}
        \item The scheme uses exponentially (in $b$) random bits;
        
        \item There exists an extractor that extracts $m$ bits with statistical distance $\delta$ from uniform. Here,
        $\delta$ on the order of $2^{-\Omega(b)}$, and $\poly(\eps_2)$. And $m$ is on the order of $b$, and 
        $\log(1/\eps_2)$.
    \end{enumerate}
    Furthermore, if $\Share$ is an efficient algorithm,
    the extractor can be given as a family of $\poly(b)$-sized
    circuits.
\end{thm}

The above result can also be extended to 2-out-of-2 non-malleable secret sharing with essentially the same parameters.

Informally, this shows that either one must use either: (1) 
use superpolynomially long random sources, or (2) use 
extractable sources to share secrets that are 
leakage-resilient or non-malleable.

Furthermore, we note that we can combine our results
here with a certain combinatorial object called 
\emph{distribution designs}. 
Distribution designs are combinatorial objects that ``distribute'' a set of $n$ items 
\{1, 2, \ldots, n\} into a collection of $m$ sets $\cS_1, \cS_2, \ldots, \cS_m$.
Then, consider any set of indices of sets $\cI \subseteq [m]$, the union 
of the sets indexed by $\cI$ yields $[n]$ if and only if $\lvert \cI \rvert$
is at least some threshold $t$.

To see how this would be useful, consider for example an $n$-out-of-$n$ 
secret sharing scheme (i.e. all $n$ shares need to be present for the secret
to be recovered). Then we could use a destribution design to distribute the $n$
shares that were created (without randomness) into $m$ sets. Note here that each set 
will be our new share, and the secret can only be recovered if and only if at least $t$ of
these new shares are present (since they recover all $n$ of the original shares).

Hence, we obtain the following corollaries:
\begin{corollary}[Informal]
    If every $(t=2,n,\eps)$-secret sharing scheme for $b$-bit messages using $d$ bits of randomness requires a $(\delta,m)$-extractable class of randomness sources, then so does every $(t',n',\eps)$-secret sharing scheme for $b$-bit messages using $d$ bits of randomness whenever $n\leq \binom{n'}{t'-1}$.
    Moreover, this is the best reduction possible with $t=2$.
\end{corollary}

\begin{corollary}[Informal]
    If every $(t,n,\eps)$-secret sharing scheme for $b$-bit messages using $d$ bits of randomness requires a $(\delta,m)$-extractable class of randomness sources, then so does every $(t'=n',n',\eps)$-secret sharing scheme for $b$-bit messages using $d$ bits of randomness whenever $n'\geq \binom{n}{t-1}$.
    Moreover, this is the best reduction possible with $t'=n'$.
\end{corollary}

We note that while this falls short of showing it for 2-out-of-2 secret sharing, we cover a wide class of secret sharing schemes (for both leakage resilient and non-malleable). Furthermore, we can view our above two corollaries as philosophically showing that the ``true randomness'' requirement for 2-out-of-2 secret sharing requires showing it for $t$-out-of-$n$ schemes with a broad parameter regime (in $n$ and $t$). This also means that to show a negative result (for 2-out-of-2), one can also show use a non-extractable source for one of these
choices of parameters.

\subsection{Explicit (Collision-Resistant/Non-malleable) Extractors}

Collision resistance is a common concept for cryptographic hash functions, which 
can be used to create other primitives 
like one-time signature schemes. Formally, a collision resistant 
hash function $h$ is defined as follows:
\begin{definition}[Collision Resistance Hash Function (CRHF)]
    A hash function family $\cH$ of hash functions $h : \bit^n \to \bit^m$ (with $m < n$) is collision-resistant if
    for all efficient (i.e. randomised, polynomial time in $n$) adversaries $A$:

    \begin{equation*}
        \Pr_{A, h \from \cH}[(x_0, x_1) \from A(h) : 
        x_0 \neq x_1, h(x_0) = h(x_1)] \leq \eps(n),
    \end{equation*}
    where $\eps(n)$ is negligible in $n$. 
\end{definition}

We further note an interesting connection between collision resistance 
and randomness extraction.

A universal hash family is a family of hash functions $\cH$,
each of which of the form $h : \bit^n \to \bit^m$ such that
for all $x \neq y$:
\begin{equation*}
	\Pr_{h \from \cH}[ h(x) = h(y) ] \leq 2^{-m}.
\end{equation*}

I.e. the probability of two inputs colliding is small.
The leftover hash lemma states that $\cH$ is actually
a good seeded extractor. 

\begin{theorem}
	Let $X$ be a random variable such that $H_\infty(X) \geq k$.
	Then for any $\eps > 0$, letting $\cH$ be a universal hash family 
	that outputs $m = k - 2\log(1/\eps)$ bits, then the extractor $ext(X, S)$
	that uses $S$ as a uniform seed source to sample $h \from \cH$ is such that
	$ext(X, S) \approx_\eps \unif{m}$.
\end{theorem} 

\paragraph{(Our Contribution)}
We introduce a novel and analogous extension for seeded extractors called \emph{collision resistant} extractors.

\begin{definition}[Collision Resistant Extractors]
    An extractor $crExt : \bit^n \times \bit^d \to \bit^m$ is $\delta$-collision resistant if for all fixed-point-free
    functions $f$, we have that $\Pr[crExt(X, S) = crExt(f(X), S)] \leq \delta$, where the probability 
    is taken over the randomness of $S$ and $X$.
\end{definition}

Such an object can be seen as having parallels to 
collision resistant hash functions in cryptography.
We also note
that such extractors are somewhat weaker than non-
malleable extractors since we only require that the outputs
differ, but potentially still arbitrarily correlated.

We show that any efficient seeded extractor
$ext : \bit^n \times \bit^d \to \bit^m$ can be made collision
resistant with the cost of a slightly longer seed $d' = d + O(\log(1/\delta))$.

\begin{theorem}
    Let $ext : \bit^n \times \bit^d \to \bit^m$ be a strong seeded extractor. Then there exists a strong seeded extractor $crExt : \bit^n \times \bit^{d + z} \to \bit^m$ with collision probability $\delta$ and $z = O(\log(1/\delta)\log^2(\log(1/\delta))\log(1/\eps))$,
    where $\delta$ is the collision probability, and $\eps$ 
    is the statistical distance of the output of the extractor
    from the distribution that is uniform over $m$ bits.
\end{theorem}

Turning our attention to existing extractors with this new notion, 
we also show that Raz's (two-source) extractor is collision resistant.
\begin{theorem}[Informal]
    There exists a two-source extractor $rExt : \bit^{n_1} \times \bit^{n_2} \to \bit^m$, 
    $rExt(X_1, X_2) \approx_\eps U_m$, with:
    (1) The first source $X_1$ on $n_1$ bits that has min-entropy $\Omega(\log(n_1))$,
    (2) second second source $X_2$ on $n_2$ has min-entropy $(\frac{1}{2} + \delta) n_2$,
    and (3) the statistical distance $\eps$ is at most $ 2^{-\frac{3m}{2}}$ with (4) collision probability $2^{-m + 1}$,
    and (5) $m = \Omega(k_1, n_2)$.
\end{theorem}

Collision resistant extractors are not just interesting in isolation, we further
use both of our results in collision resistant extractors to give better two-source non-malleable extractors with 
lower entropy requirements.

As mentioned, the prior best 2-source non-malleable extractor 
is by Li in \cite{L20}. Namely, their extractor requires 
the min-entropy rate in both sources nearly $1$ (i.e. both $n$-bit sources have min-entropy nearly $n$). This is less than ideal
for applications like privacy amplification, where 
the entropy requirement of the extractor translates to 
the initial min-entropy of the shared weakly random string between Alice and Bob. 

Combining our extractors with the standard technique of alternating extraction, we obtain breakthrough the following breakthrough result:
\begin{theorem}
    There exists a two-source non-malleable extractor $2nmext : \bit^{n_1} \times \bit^{n_2} \to \bit^m$, 
    $2nmext(X_1, X_2) \approx_\eps U_m$, with:
    (1) The first source $X_1$ on $n_1$ bits that has min-entropy $\Omega(\log(n_1))$,
    (2) second second source $X_2$ on $n_2$ has min-entropy $(\frac{4}{5} + \delta) n_2$,
    and (3) the statistical distance $\eps$ is at most $ 2^{-\frac{3m}{2}}$ with (4) collision probability $2^{-m + 1}$,
    and (5) has an output length $m = \Omega(k_1, n_2)$.
\end{theorem}

We note that this nearly matches the current best known two-source extractor (without non-malleability) with negligible error. (Raz's 
extractor uses $>0.5n$ entropy in one source, and $poly(\log(n))$ entropy in the other). Thus, there cannot be asymptotic improvements
to our construction without further improvements in two-source extractors as well (since non-malleable two-source extractors are stronger objects than two-source extractors). We further note, that doing so constitutes a significant breakthrough on a nearly two decade old problem.

One other aspect we will point out, is that, as noted by Eshan Chattopadhyay in his survey \cite{C20}, prior to our work, 
the other best known method for constructing $t$-source seed non-malleable extractors is via two objects known as \emph{advice correlation breakers}, and \emph{advice generators}.\footnote{Note the subtle difference here that the extractors being considered only have their seed tampered}. We view our work as taking a completely different approach and breaking away from the norm. As such, we hope this novel construction and approach will also be of independent interest. 

\paragraph{(Applications)}
Let us turn our attention to privacy amplification. 
Known constructions of privacy amplification either additionally require access to a fully random source,
or to a nearly full min-entropy source. Furthermore, prior protocols limit the adversary's ability to only tamper messages 
that were sent. 
We use our two-source non-malleable extractor to give a 6-round protocol with lower entropy requirements, and where the adversary is additionally 
allowed to arbitrarily tamper the internal memory of one of the two parties (we refer to such adversaries as
\emph{memory-tampering} adversaries, introduced by Aggarwal et al in \cite{AORSS20}). 
In fact, we would like to point out that this construction directly addresses a remark made at the end of 
Section 3 in \cite{AORSS20}: \textit{``We currently do not know explicit constructions of non-malleable two-source 
extractors with
parameters matching those required for Corollary 2, although it is known that there exist such (in-
efficient) extractors with significantly better parameters [CGGL20]. Thus, we leave this connection
between non-malleable two-source extractors and privacy amplification with a very strong adversary
as an interesting motivation for further study of such extractors with lower min-entropy require-
ment in the information-theoretic setting.''}.

\begin{theorem}[Informal]
    There is a 6-round protocol protocol where the shared random string $W$ has min-entropy $0.81n$, Alice and Bob additionally have independent sources of randomness
    that have min-entropy $0.001n$, and the protocol outputs $0.8n$ bits.
    Furthermore, the privacy guarantee holds even against memory-tampering adversaries.
\end{theorem}

\subsection{Stronger Bounds for Immunizing Random Backdoored Oracles}
Finally, we end our overview on immunizing backdoors, and mention some interesting links to the study of static 
data structure lower bounds.

As mentioned, prior work models one-way functions in the ideal way --- as a random oracle $R$, and pre-processing
adversaries are allowed $T$ query accesses into $R$, and into an $S$-length advice string $s(R)$. And in this 
setting, lower bounds are known. Namely, \cite{DTT10} show the probability of inverting a uniformly and randomly chosen
$x$ is at most $O \left(\frac{T(S + N)}{N}\right)$, with $N = 2^n$.

However, this begs the following questions: 
\begin{enumerate}
    \item If $R$ is compromised in some way, can we still build one-way functions?
    \item Can we use $R$ in some smarter way to bootstrap ourselves to build a better one-way function? 
\end{enumerate}

Guo et al in \cite{GGHPV20} proposed a way 
to immunize the random oracle $R$ against 
pre-processing adversaries by relying on the 
hardness of $k$SUM. In other words, they created one-way functions that were immune against adversaries
that were allowed to pre-process information about $R$.
They also proved their construction in two
steps: (1) By showing a direct equivalence between
immunizing $R$ against pre-processing adversaries 
and the existence of corresponding 
hard-on-average data 
structure problems, and (2) giving data 
structure lower bounds for 3SUM.

\begin{definition}[Data Structure Solution]
    A data structure solution is a pair of algorithms $A_1, A_2$ where:
    \begin{enumerate}
        \item $A_1$ on some input $x \in \mathcal{X}$ outputs $S$ memory cells, each of $w$ bits (as a data structure).
        \item $A_2$ on some input $q \in \mathcal{Q}$, obtains at most $T$ memory cells (each of at most $w$ bits) from the data structure, outputs an answer.
    \end{enumerate}

    We say the solution is adaptive, if $A_2$'s 
    memory accesses may depend on the answers
    to its previous accesses. We call it non-adaptive otherwise.

    Typically $\mathcal{X}$ and $\mathcal{Q}$ are 
    clear from context.
\end{definition}

We begin by sketching their construction.
Let $(G, +)$ be an abelian group, and let 
$R:[N] \to G$ be a random oracle. Then,
the proposed one-way function $f = A^R$
is given as: $f(x) = R(x_1) + R(x_2)$,
where the input distribution is uniform over
all strings $x = x_1, x_2$, where $x_1 \neq x_2$.

They then show that inverting a random output
of this function reduces to solving the following 
data structure problem:

\begin{definition}[3SUM-Indexing]
On input $\Vec{Z} = (z_1, z_2, \ldots, z_n) \in G^n$, 
$A_1(Z)$ outputs at most $S$ memory cells of $w$ bits,
such that for any $q \in G$, $A_2(q)$ accesses 
at most $T$ memory cells (by probing them) to output either:
$(i, j)$ such that $z_i + z_j = q$ or $\bot$ if 
no such pair exists.
\end{definition}

However, their lower bound has a few shortcomings:
(1) They only rule out \emph{non-adaptive} algorithms. That is to say, the algorithm's access pattern into the 
    memory cells can only depend on its input
    and not subsequent contents of the memory 
    cells it accesses. This rules out common 
    data structures such as hash tables, or 
    search trees and so on. Which, in turn, rules out security against adversaries that are can make adaptive queries.
(2) They require the group size $\lvert G \rvert$ to be at least 
quadratic in the number of input elements $n$.

In this thesis, we give various stronger data structure lower bounds for variants of 3SUM,
thereby giving stronger guarantees to their 
construction. Of the three lower bounds, two of them match the barrier for existing data structure lower bounds. 
And any further advancement necessitates a breakthrough in circuit lower bound results. The last lower bound, is 
a novel result in the special case of bit-probes.

\paragraph{(Our Contribution)}
Our first contribution improves upon their result
in a both ways: (1) Our lower bound now holds 
for adaptive data structures, and (2) the lower 
bound holds for group sizes nearly linear in the 
number of input elements.

\begin{theorem}
    Any data structure answering $3$SUM-Indexing queries for input sets of size $n$ for abelian groups $([m], + \mod m)$, with $m = O(n^{1+\delta})$ and \\
    $(\bit^{(1+\delta)\log(n)+O(1)}, \oplus)$ for a constant $\delta>0$, using $S$ cells of $w = \Omega(\lg n)$ bits must have query time $T = \Omega( \log n / \log(Sw/n) )$.
\end{theorem}

Next, we also give a slightly stronger result for non-adaptive data structures (when $|G|$ is superpolynomial in $n$).
\begin{theorem}
    Any non-adaptive cell probe data structure answering $3$SUM-Indexing queries for input sets of size $n$ for an abelian group $G$ of size $\omega(n^2)$, using $S$ cells of $w = \Omega(\lg n)$ bits must have query time $T = \Omega(\min\{ \log |G| / \log(Sw/n), n/w\})$.
\end{theorem}

Both the above results are tight: A stronger lower bound for $T$ would imply circuit lower bounds.

Lastly, we given a novel result for the special case of $w = 1$ (i.e. bit-probe data structures),
and show tight results in this regime.
\begin{theorem}
    Any non-adaptive data structure for $3$SUM-Indexing such that $T = 2$ and $w = 1$ requires $S = \Omega(|G|)$ for an abelian group $(G,+)$.
\end{theorem}

This result is the first of its kind, and we view this as a first step in potentially improving lower bounds for the regime of 
constant $T$ and/or constant $w$. We also note that the core techniques used for this result may be general enough to be used 
for other well known data structure problems.

\section{Preliminaries}
In this section we will cover commonly used theorems and lemmas. We will defer more chapter-specific background to a different preliminary section in each chapter.
\subsection{Notation}

We usually denote sets by uppercase calligraphic letters like $\cS$ and $\cT$, and write $[n]$ for the set $\{1,2,\dots,n\}$.
Random variables are denoted by uppercase letters such as $X$, $Y$, and $Z$, and we write $U_m$ for the uniform distribution over $\bit^m$.
For any set $\cS$, we denote by $\unif{\cS}$  the uniform distribution over the set $\cS$.

For any random variable $X$, we denote the support of $X$ by $\supp{X}$. Also, for any random variable $X$ and event $E$, we denote by $X|_E$ the random variable $X'$ such that for all $x \in \supp{X}$, $\Pr[X'=x] = \Pr[X=x|E]$. 

Given a vector $x\in\cS^n$ and set $\cT\subseteq [n]$, we define $x_\cT=(x_i)_{i\in\cT}$.
We denote the $\mathbb{F}_2$-inner product between vectors $x,y\in\bit^n$ by $\langle x,y\rangle$.
All logarithms in are taken with respect to base $2$.

\subsection{Probability Theory}
In this section, we introduce basic notions from probability theory that will be useful throughout this work.

\begin{defn}\label{def:SD}
    The \emph{statistical distance between random variables $X$ and $Y$} over a set $\cX$, denoted by $\Delta(X,Y)$, is defined as
    \begin{equation*}
        \Delta(X,Y)=\max_{\cS\subseteq \cX}|\Pr[X\in\cS]-\Pr[Y\in\cS]|=\frac{1}{2} \sum_{x\in\cX} \lvert \Pr[X = x] - \Pr[Y = x] \rvert.
    \end{equation*}
    Moreover, we say that $X$ and $Y$ are \emph{$\eps$-close}, denoted by $X\approx_\eps Y$, if $\Delta(X,Y)\leq \eps$, and \emph{$\eps$-far} if this does not hold.
\end{defn}
The following lemma is a version of the well-known XOR lemma (see~\cite{Gol11} for a detailed exposition of these types of results).
\begin{lemma}[XOR Lemma]\label{lem:XOR}
    If $X$ and $Y$ are distributions supported on $\bit^t$ such that
    \begin{equation*}
        \langle a, X \rangle \approx_\eps \langle a, Y \rangle
    \end{equation*}
    for all non-zero vectors $a \in \bit^t$, then
    \begin{equation*}
        X \approx_{\eps'} Y
    \end{equation*}
    for $\eps'=2^{t/2}\eps$.
\end{lemma}

For any random variables $A, B, C$, and event $E$, we shorthand $\dist{A,C}{B,C}$ by $\distCond{A}{B}{C}$, and $\dist{A|_E}{B|_E}$ by $\distCond{A}{B}{E}$ i.e.,
    \begin{equation*}
        \distCond{A}{B}{C} = \dist{A, C}{B, C}\;,
    \end{equation*}
    and 
    \[
      \distCond{A}{B}{E} = \dist{A|_E}{B|_E}\;.
    \]

The following lemma is immediate from the definitions and triangle inequality.
\begin{lemma}
    Let $A, B, C$ be random variables such that $A, B \in S$ and $\supp{C} = \mathcal{T}$ with $\mathcal{T} = \mathcal{T}_1 \cup \mathcal{T}_2, \mathcal{T}_1 \cap \mathcal{T}_2 = \emptyset$. Then:

    \begin{enumerate}
        \item $\distCond{A}{B}{C} \le \sum_{c \in \mathcal{T}} \Pr[C = c] \distCond{A}{B}{C = c}$ 
        \item $\distCond{A}{B}{C} \le \Pr[C \in \mathcal{T}_1] \distCond{A}{B}{C \in \mathcal{T}_1} + \Pr[C \in \mathcal{T}_2] \distCond{A}{B}{C \in \mathcal{T}_2}$
    \end{enumerate}
\end{lemma}

\begin{lemma}[Corollary of Lemma \ref{lem:XOR}, where $Y = U_m$ ]\label{lma:XORLemma}
    Let $\blocks{X}{m}$ be binary random variables and for any non-empty $\tau \subseteq [m]$, $\lvert \Pr[\bigoplus_{i \in \tau} X_i = 0] - \frac{1}{2} \rvert \leq \eps$, then $\dist{\blocks{X}{m}}{\unif{m}} \leq \eps \cdot 2^{\frac{m}{2}}$.
\end{lemma}

\subsection{Min-entropy}

\begin{definition}[Min-entropy]
	Given a distribution $X$ over $\cX$, the \emph{min-entropy of $X$}, denoted by $\minEnt{X}$, is defined as
	\begin{equation*}
		\minEnt{X}=-\log\left(\max_{x\in\cX} \Pr[X=x]\right).
	\end{equation*}
\end{definition}

\begin{definition}[Average min-entropy]
	Given distributions $X$ and $Z$, the \emph{average min-entropy of $X$ given $Z$}, denoted by $\avgCondMinEnt{X}{Z}$, is defined as
	\begin{equation*}
		\avgCondMinEnt{X}{Z}=-\log\left(\mathbb{E}_{z\leftarrow Z}\left[\max_{x\in\cX} \Pr[X=x|Z=z]\right]\right).
	\end{equation*}
\end{definition}

\begin{lemma}[\cite{DORS08}]\label{lem:avgminH}
    Given arbitrary distributions $X$ and $Z$ such that $|\supp{Z}|\leq 2^\lambda$, we have
    \begin{equation*}
        \avgCondMinEnt{X}{Z}\geq \minEnt{X,Z}-\lambda\geq \minEnt{X}-\lambda\;.
    \end{equation*}
\end{lemma}

\begin{lemma}[\cite{MW97}]\label{lem:fixavgminH}
    For arbitrary distributions $X$ and $Z$, it holds that
    \begin{equation*}
        \Pr_{z\leftarrow Z}[\minEnt{X|Z=z}\geq \avgCondMinEnt{X}{Z}-s]\geq 1-2^{-s}.
    \end{equation*}
\end{lemma}

\begin{definition}[$(n, k)$-sources]
    We say that a random variable $X$ is an \emph{$(n, k)$-source} if $\supp{X} \subseteq \bit^n$ and $\minEnt{X} \ge k$. Additionally, we say that $X$ is a \emph{flat $(n, k)$-source} if for any $a \in \supp{X}$, $\Pr[X = a] = 2^{-k}$, i.e., $X$ is uniform over its support. 
\end{definition}
$X \givenBy (n, k)$ denotes the fact that $X$ is an $(n, k)$-source. Further, we call $X$ \emph{$(n, k)$}-flat if $X \givenBy (n, k)$ and is flat. We say that $X$ is $\eps$-close to a flat distribution if there exists a set $\cS$ such that $X \closeTo{\eps} \unif{\cS}$.

\begin{definition}[$\eps$-smooth min-entropy]
    A random variable $X$ is said to have $\eps$-smooth min-entropy at least $k$ if there exists $Y$ such that $\dist{X}{Y} \leq \eps$, and

    \begin{equation*}
     \minEnt{Y} \ge k\;.
    \end{equation*}
\end{definition}

\subsection{Extractors}
\begin{definition}[(Strong) Two-Source Extractor, Collision Resistance]\label{defn:twosourceExt}
    Call $\ext:\bit^{n_1} \times \bit^{n_2} \to \bit^m$ a \emph{two-source extractor} for input lengths $n_1, n_2$, min-entropy $k_1, k_2$, output length $m$, and error $\eps$ if for any two independent sources $X, Y$ with $X \givenBy (n_1, k_1)$, $Y \givenBy (n_2, k_2)$, the following holds:
    \begin{equation*}
        \dist{\ext(X, Y)}{\unif{m}} \leq \eps
    \end{equation*}
    If $n_2= k_2$, we call such an extractor \emph{seeded}. We use $\extractorParam{\ext}{(n_1, k_1)}{(n_2, k_2)}{m}{\eps}$ to denote the fact that $\ext$ is such an extractor.

    Additionally, we call the extractor $\ext$ \emph{right strong}, if:
    \begin{equation*}
        \distCond{\ext(X, Y)}{\unif{m}}{Y} \leq \eps \;,
    \end{equation*}
    and we call the extractor $\ext$ \emph{left strong}, if:
    \begin{equation*}
        \distCond{\ext(X, Y)}{\unif{m}}{X} \leq \eps \;.
    \end{equation*}
    
\textbf{(Strong Extractors)} We call an extractor $\ext$ \emph{strong} if it is both left strong and right strong.
    
\textbf{(Collision Resistance)} The extractor is said to be $\collisionError$-collision resistant if $\Pr_{X, Y}[\ext(X, Y) = \ext(f(X), Y)] \leq \collisionError$ for all fixed-point-free functions $f$.
\end{definition}

\begin{lemma}\label{lma:avgCaseMinExtraction}
    If $\extractorParam{\ext}{(n, k)}{(d, d)}{m}{\eps}$ is a strong seeded extractor, then for any $X, W$ such that $\supp{X} \subseteq \bit^n$ and $\avgCondMinEnt{X}{W} \geq k + \log(1/\eta)$ with $\eta > 0$, it holds that:

    \begin{equation*}
        \distCond{\ext(X, \unif{d})}{\unif{m}}{\unif{d}, W} \leq \eps + \eta 
    \end{equation*}
\end{lemma}

\begin{proof}
    Let $\ext$, $X$ and $W$ be defined as above. Then, given that $\avgCondMinEnt{X}{W} \geq k + \log(1/\eta)$, 
    it follows from Markov's inequality that there exists a ``bad'' set $\mathcal{B}$ such that $\Pr[W \in \mathcal{B}] \leq \eta$, and 
    for all $w \notin \mathcal{B}$, $\condMinEnt{X}{W = w} \geq k$.
    Then,
    \begin{align*}
        \distCond{\ext(X, \unif{d})}{\unif{m}}{\unif{d}, W}
        &\leq \distCond{\ext(X, \unif{d})}{\unif{m}}{\unif{d}, W \in \mathcal{B}} \Pr[W \in \mathcal{B}]\\
        &\hspace{2em} + \distCond{\ext(X, \unif{d})}{\unif{m}}{\unif{d}, W \notin \mathcal{B}} \Pr[W \notin \mathcal{B}]\\
        &\leq 1 \cdot \Pr[W \in \mathcal{B}] + \distCond{\ext(X, \unif{d})}{\unif{m}}{\unif{d}, W \notin \mathcal{B}}\\
        &= \Pr[W \in \mathcal{B}] + \sum_{w \notin \mathcal{B}} \distCond{\ext(X, \unif{d})}{\unif{m}}{\unif{d}, W = w}\\
        &\leq \eta + \eps\;.
    \end{align*}
\end{proof}

\chapter{Randomess Requirements for Leakage-Resilient and Non-malleable Secret Sharing}

In this chapter (based on the paper \cite{ACOR21}), we first study leakage-resilience.
Modern cryptographic primitives are typically designed in such a way where adversaries are only allowed access to information that the primitive needs to publicise (perhaps by sending over to other parties). Metaphorically, there is a line drawn between what information must be made ``publicly available'', and what can be kept private from the rest of the world. This does not account for adversaries that are capable of crossing this boundary. Indeed, adversaries might have physical access to the machines running these primitives and thus measure voltages \cite{KJJ99}, 
or EM radiation emitted from the hardware \cite{AARR02}, etc \cite{LSGP20,CDHKS00}. In doing so, they potentially obtain even more information about the execution of the primitive than assumed for the proof of security. To address this, leakage-resilient cryptography broadly studies cryptography where the adversary is able to obtain such additional information (via a side-channel) in the form of a leakage
function $g$, or several leakage functions $g_1, g_2, \ldots, g_t$ (depending on the model).

For example, the adversary may obtain information about intermediate values or steps that were performed during the computation (referred to as computational leaks\footnote{For examples, see \cite{MR03,ISW03}}), or the adversary may obtain information from bits stored in memory {(referred to as memory leakage)}. For a general survey on leakage-resilient cryptography, one may refer to \cite{KR19,DP08}. Leakage-resilient storage was first introduced by Davi, Dziembowski, and Venturi in \cite{DDV10} as a 
way to store information in an encoded manner such that any adversary 
that is able to obtain a leakages about the encoded bits is unable to distinguish between two possible original messages. Dziembowski and Faust in \cite{DF11} also scheme for a similar case but where the adversary is allowed to make small but multiple leakages over a period of time. 

We will focus on a different leakage-resilient cryptographic primitive --- secret sharing. 
Secret sharing, introduced by Blakley~\cite{B79} and Shamir~\cite{S79}, strikes a meaningful balance between availability and 
confidentiality of secret information.
This fundamental cryptographic primitive has found a host of applications, most notably to threshold cryptography and multi-party computation (see~\cite{CDN} for an extensive discussion).
In a secret sharing scheme for $n$ parties, a dealer who holds a secret $s$ chosen from a domain $\cM$ 
can compute a set of $n$ \emph{shares} by evaluating a randomized function on $s$ which we write as
$\Share(s) = (\Sh_1, \ldots,\Sh_n)$.
The notion of \emph{threshold} secret sharing is particularly important: A $t$-out-of-$n$ secret sharing scheme ensures that any $t$ shares are sufficient to recover the secret $s$, but any $t-1$ shares reveal no information about the secret $s$.

Motivated by practice, several variants of secret sharing have been suggested which guarantee security under stronger adversarial models.
The notion of \emph{leakage-resilient} secret sharing was put forth in order to model and handle side-channel attacks to secret shared data.
In more detail, the adversary, who holds an unauthorized subset of shares, is furthermore allowed to specify a leakage function $\Leak$ from a restricted family of functions and learn $\Leak(\Sh_1,\dots,\Sh_n)$.
The goal is that this additional side information reveals almost no information about the secret.
Typically one considers \emph{local leakage}, where $\Leak(\Sh_1,\dots,\Sh_n)=(\Leak_1(\Sh_1),\dots,\Leak_n(\Sh_n))$ for local leakage functions $\Leak_i$ with bounded output length.
This makes sense in a scenario where shares are stored in physically separated locations.
The alternative setting where adversaries are allowed to \emph{corrupt} all shares (e.g., by infecting storage devices with viruses) led to the introduction of \emph{non-malleable} secret sharing.
In this case, the adversary specifies tampering functions $f_1, f_2,\ldots, f_n$ which act on the shares, and then the reconstruction algorithm is
applied to the tampered shares $f_1(\Sh_1),\dots,f_n(\Sh_n)$.
The requirement, roughly speaking, is that
either the original secret is reconstructed or it is destroyed, i.e., the reconstruction result is unrelated to the original secret.
Both leakage-resilient and non-malleable secret sharing have received significant attention in the past few years.

\paragraph{Cryptography with weak randomness.}
It is well-known that randomness plays a fundamental role in cryptography and other areas of computer science.
In fact, most cryptographic goals cannot be achieved without access to a source of randomness.
Almost all settings considered in the literature assume that this source of randomness is perfectly random: It outputs uniformly random and independent bits.
However, in practice it is extremely hard to generate perfect randomness.
The randomness needed for the task at hand is generated from some physical process, such as electromagnetic noise or user dependent behavior.
While these sources have some inherent randomness, in the sense that they contain entropy, samples from such sources are not necessarily uniformly distributed.
Additionally, the randomness generation procedure may be partially accessible to the adversary, in which case the quality of the randomness provided degrades even further.
The difficulty in working with such imperfect randomness sources not only arises from the fact that they are not uniformly random, but also because the exact distribution of these sources is unknown. 
One can at best assume that they satisfy some minimal property, for example that none of the outcomes is highly likely as first considered by Chor and Goldreich~\cite{CG88}.

The best one can hope for is to deterministically extract a nearly perfect random string for direct usage in the desired application.
While there are source models which allow for determinisitc randomness extraction, such as von Neumann sources~\cite{vN51}, bit-fixing sources~\cite{CGH85}, affine sources~\cite{Bou07}, and other efficiently generated or recognizable sources~\cite{Blu86,SV86,LLS89,TV00,DGW09,KRVZ11,Dvi12,BGLZ15,CL16}, all these models make strong assumptions about the structure of the source.
On the other hand, the most natural, flexible, and well-studied source model where we only assume a lower bound on the min-entropy of the source\footnote{A source is said to have \emph{min-entropy $k$} if the probability that it takes any fixed value is upper bounded by $2^{-k}$.} does not allow deterministic extraction of even $1$ almost uniformly random bit~\cite{CG88}.
This holds even in the highly optimistic case where the source is supported on $\bits^d$ and has min-entropy $d-1$.
Nevertheless, it has been long known, for example, that min-entropy sources are sufficient for simulating certain randomized algorithms and interactive protocols~\cite{CG88}.

This discussion naturally leads us to wonder whether perfect randomness is essential in different cryptographic primitives, in the sense that the underlying class of sources of randomness allows deterministic extraction of nearly uniformly random bits.
We call such classes of sources \emph{extractable}.
More concretely, the following is our main question.
\begin{question}\label{q:extcrypto}
Does secret sharing, or any of its useful variants such as leakage-resilient or non-malleable secret sharing, \emph{require} access to \emph{extractable} randomness?
\end{question}

This question was first asked by Bosley and Dodis~\cite{BD07} (for $2$-out-of-$2$ secret sharing) and it remains open. 
Bosley and Dodis settled the analogous question for the case of information-theoretic private-key encryption, motivated by a series of (im)possibility results for such schemes in more specific source models~\cite{MP90,DS02,DOPS04}.
More precisely, they showed that encryption schemes using $d$ bits of randomness and encrypting messages of size $b>\log d$ require extractable randomness, while those encrypting messages of size $b<\log d-\log\log d-1$ do not.

As noted in~\cite{DPP06,BD07}, private-key encryption schemes yield $2$-out-of-$2$ secret sharing schemes by seeing the uniformly random key as the left share and the ciphertext as the right share.
Therefore, we may interpret the main result of~\cite{BD07} as settling Question~\ref{q:extcrypto} for the artificial and highly restrictive class of secret sharing schemes where the left share is uniformly random and independent of the secret, and the right share is a deterministic function of the secret and the left share.
No progress has been made on Question~\ref{q:extcrypto} since.

\paragraph{Random-less Reductions for Secret Sharing.} Given that the problem of whether $2$-out-of-$2$ secret sharing requires extractable randomness has been open for 15 years, it is reasonable to consider intermediate problems towards resolving the open question. In a spirit similar to computational complexity, we consider how the question whether $t$ out of $n$ secret sharing requires extractable randomness is related to the same question for a different choice of the parameters $t, n$ i.e.,
\begin{question}\label{q:ssred}
Given $t, n, t', n'$, does the fact that $t$-out-of-$n$ secret sharing require extractable randomness imply that $t'$-out-of-$n'$ secret sharing require extractable randomness?
\end{question}
A natural approach towards resolving this question is to try to construct a $t$-out-of-$n$ secret sharing scheme from a $t'$-out-of-$n'$ secret sharing scheme in a black-box manner without any additional randomness. Intuitively, since we don't have access to any additional randomness, it seems that the most obvious strategy to achieve such reductions is to choose $n$ subsets of the set of $n'$ shares in such a way that any $t$ out of these $n$ subsets contain at least $t'$ out of the original $n'$ shares and any $t-1$ subsets contain at most $t'-1$ of the original $n'$ shares. In particular, there is a trivial reduction when $t = n = 2$ that chooses the first subset to contain the first of the $n'$ shares, and the second subset to contain any $t'-1$ of the remaining shares. This shows the completeness of the extractability of $2$-out-of-$2$ secret sharing with respect to these reductions. Such reductions can be formalized via distribution designs~\cite{SW18}.

\section{Our Results}

In this work, we make progress on both Question~\ref{q:extcrypto} and Question~\ref{q:ssred}. 
Before we proceed to discuss our results, we formalize the notions of an extractable class of randomness sources and threshold secret sharing.
\begin{defn}[Extractable class of sources]
We say a class of randomness sources $\cY$ over $\bits^d$ is \emph{$(\delta,m)$-extractable} if there exists a deterministic function $\ext:\bits^d\to\bits^m$ such that\footnote{We use the notation $X\approx_\delta Y$ to denote the fact that $\Delta(X;Y)\leq \delta$, where $\Delta(\cdot;\cdot)$ corresponds to statistical distance (see Definition~\ref{def:SD}).} $\ext{} (Y)\approx_\delta U_m$ for every $Y\in\cY$, where $U_m$ denotes the uniform distribution over $\bits^m$.
\end{defn}
Note that we may consider the support of all sources in $\cY$ to be contained in some set $\bits^d$ without loss of generality.
Since we will be interested in studying the quality of randomness used by secret sharing schemes, we make the class of randomness sources allowed for a secret sharing scheme explicit in the definition of $t$-out-of-$n$ threshold secret sharing below.
\begin{defn}[Threshold secret sharing scheme]\label{def:thresholdss}
A tuple $(\Share,\Rec,\cY)$ with $\Share:\bitsB\times\bits^d\to\left(\bitsL\right)^n$ and $\Rec:\bits^*\to\bitsB$ deterministic algorithms and $\cY$ a class of randomness sources over $\bits^d$ is a \emph{$(t,n,\eps)$-secret sharing scheme (for $b$-bit messages using $d$ bits of randomness)} if for every randomness source $Y\in\cY$ the following hold:
\begin{enumerate}
    \item If $\cT\subseteq [n]$ satisfies $|\cT|\geq t$ (i.e., $\cT$ is \emph{authorized}), then
    \begin{equation*}
        \Pr_Y[\Rec(\Share(x,Y)_\cT)=x]=1
    \end{equation*}
    for every $x\in\bitsB$;
    
    \item If $\cT\subseteq[n]$ satisfies $|\cT|<t$ (i.e., $\cT$ is \emph{unauthorized}), then for any $x,x'\in\bitsB$ we have
    \begin{equation*}
        \Share(x,Y)_\cT\approx_\eps \Share(x',Y)_\cT,
    \end{equation*}
    where $\Share(x,Y)_\cT$ denotes the shares of parties $i \in \cT$.
\end{enumerate}
\end{defn}

\subsection{Leakage-Resilient $2$-out-of-$2$ Secret Sharing Requires Extractable Randomness.}

As our first contribution, we settle Question~\ref{q:extcrypto} for the important sub-class of \emph{leakage-resilient} $2$-out-of-$2$ secret sharing.
Intuitively, we consider $2$-out-of-$2$ secret sharing schemes with the additional property that the adversary learns almost nothing about the message when they obtain bounded information from each share.
More formally, we have the following definition.
\begin{defn}[Leakage-resilient secret sharing scheme]\label{def:lrsss}
    We say that a tuple $(\Share,\Rec,\cY)$ with $\Share:\bitsB\times\bits^d\to\left(\bitsL\right)^2$ and $\Rec:\bits^*\to\bitsB$ deterministic algorithms and $\cY$ a class of randomness sources over $\bits^d$ is an \emph{$(\eps_1, \eps_2)$-leakage-resilient secret sharing scheme (for $b$-bit messages using $d$ bits of randomness)} if $(\Share,\Rec,\cY)$ is a $(t=2,n=2,\eps_1)$-secret sharing scheme and the following additional property is satisfied: For any two messages $x,x'\in\bitsB$ and randomness source $Y\in\cY$, let $(\Sh_1,\Sh_2)=\Share(x,Y)$ and $(\Sh'_1,\Sh'_2)=\Share(x',Y)$.
    Then, for any leakage functions $f,g:\bitsL\to\bits$ it holds that
    \begin{equation*}
        f(\Sh_1),g(\Sh_2)\approx_{\eps_2}f(\Sh'_1),g(\Sh'_2).
    \end{equation*}
\end{defn}

Leakage-resilient secret sharing has received significant attention recently, with several constructions and leakage models being analyzed~\cite{BDIR21,ADNOPRS19,KMS19,SV19,CGGKLMZ20,LCGSW20,MPSW20}.
Comparatively, Definition~\ref{def:lrsss} considers a significantly weaker notion of leakage-resilience than all works just mentioned.
In particular, we do not require leakage-resilience to hold even when the adversary has full access to one of the shares on top of the leakage.
This means that our results are widely applicable.
Roughly speaking, we prove that every leakage-resilient secret sharing scheme for $b$-bit messages either requires a huge number of bits of randomness, or we can extract several bits of perfect randomness with low error from its underlying class of randomness sources.
More formally, we prove the following.
\begin{thm}\label{thm:final}
    Let $(\Share,\Rec,\cY)$ be an $(\eps_1,\eps_2)$-leakage-resilient secret sharing scheme for $b$-bit messages. (Where the scheme tolerates single bit leakage.)
    Then, either:
    \begin{enumerate}
        \item The scheme uses $d\geq \min\left(2^{\Omega(b)},(1/\eps_2)^{\Omega(1)}\right)$ bits of randomness, or;
        
        \item The class of sources $\cY$ is $(\delta,m)$-extractable with $\delta\leq \max\left(2^{-\Omega(b)},\eps_2^{\Omega(1)}\right)$ and $m=\Omega(\min(b,\log(1/\eps_2)))$. 
        Moreover, if $\Share$ is computable by a $\poly(b)$-time algorithm, then $\cY$ is $(\delta,m)$-extractable by a family of $\poly(b)$-size circuits.
    \end{enumerate}
\end{thm}
An important corollary of Theorem~\ref{thm:final} is that every efficient negligible-error leakage-resilient secret sharing scheme requires extractable randomness with negligible error.
\begin{coro}\label{coro:efficient}
    If $(\Share,\Rec,\cY)$ is an $(\eps_1,\eps_2)$-leakage-resilient secret sharing scheme for $b$-bit messages running in time $\poly(b)$ with $\eps_2=\negl(b)$,\footnote{By $\eps_2=\negl(b)$, we mean that $\eps_2=o(1/b^c)$ for every constant $c>0$ as $b\to\infty$.} it follows that $\cY$ is $(\delta,m)$-extractable with $\delta=\negl(b)$ and $m=\Omega(\min(b,\log(1/\eps_2)))$. 
\end{coro}


\paragraph{Split-state non-malleable coding requires extractable randomness.}
Non-malleable coding, introduced by Dziembowski, Pietrzak, and Wichs~\cite{DPW18}, is another recent notion which has attracted much attention, in particular regarding the \emph{split-state} setting (see~\cite{AO20} and references therein).
Informally, a split-state non-malleable code has the guarantee that if an adversary is allowed to split a codeword in half and tamper with each half arbitrarily but separately, then the tampered codeword either decodes to the same message, or the output of the decoder is nearly independent of the original message.
More formally, we have the following definition.
\begin{defn}[Split-state non-malleable code~\cite{DPW18}]
A tuple $(\Enc,\Dec,\cY)$ with $\Enc:\bitsB\times\bits^d\to(\bits^\ell)^2$ and $\Dec:(\bits^\ell)^2\to\bitsB\cup\{\bot\}$ deterministic algorithms and $\cY$ a class of randomness sources is a \emph{(split-state) $\eps$-non-malleable code} if the following holds for every randomness source $Y\in\cY$:
\begin{enumerate}
    \item $\Pr[\Dec(\Enc(x,Y))=x]=1$ for all $x\in\bitsB$;
    
    \item For tampering functions $f,g:\bits^\ell\to\bits^\ell$, denote by $\Tamp^{f,g}_x$ the tampering random experiment which computes $(L,R)=\Enc(x,Y)$ and outputs $\Dec(f(L),g(R))$.
    Then, for any tampering functions $f$ and $g$ there exists a distribution $D^{f,g}$ over $\bitsB\cup\{\bot,\same\}$ such that
    \begin{equation*}
        \Tamp^{f,g}_x \approx_\eps \Sim^{f,g}_x
    \end{equation*}
    for all $x\in\bitsB$, where $\Sim^{f,g}_x$ denotes the random experiment which samples $z$ according to $D^{f,g}$ and outputs $z$ if $z\neq\same$ and $x$ if $z=\same$.
\end{enumerate}
The notion of non-malleable code in the split-state model is equivalent to the notion of a \textbf{ $2$-out-of-$2$ non-malleable secret sharing scheme}~\cite{GK18}. 
\end{defn}

It is known by~\cite[Lemmas 3 and 4]{AKO17} that every $\eps$-non-malleable coding scheme $(\Enc,\Dec,\cY)$ for $b$-bit messages is also a $(2\eps,\eps)$-leakage-resilient secret sharing scheme, provided $b\geq 3$ and $\eps<1/20$.
Combining this observation with Theorem~\ref{thm:final} yields the following corollary, which states that every split-state non-malleable code either uses a huge number of bits of randomness, or requires extractable randomness with low error and large output length.
\begin{coro}\label{coro:nmc}
    Let $(\Enc,\Dec,\cY)$ be an $\eps$-non-malleable code (i.e., $2$-out-of-$2$ $\eps$-non-malleable secret sharing scheme) for $b$-bit messages with $b\geq 3$ and $\eps<1/20$.
    Then, either:
    \begin{enumerate}
        \item The scheme uses $d\geq \min\left(2^{\Omega(b)},(1/\eps)^{\Omega(1)}\right)$ bits of randomness, or;
        
        \item The class of sources $\cY$ is $(\delta,m)$-extractable with $\delta\leq \max\left(2^{-\Omega(b)},\eps^{\Omega(1)}\right)$ and $m=\Omega(\min(b,\log(1/\eps)))$. 
        Moreover, if $\Enc$ is computable by a $\poly(b)$-time algorithm, then $\cY$ is $(\delta,m)$-extractable by a family of $\poly(b)$-size circuits.
    \end{enumerate}
\end{coro}
As a result, an analogous version of Corollary~\ref{coro:efficient} also holds for split-state non-malleable coding. This resolves Question~\ref{q:extcrypto} for $2$-out-of-$2$ non-malleable secret sharing. 

\subsection{Random-less Reductions for Secret Sharing.}

In this section, we discuss our contribution towards resolving Question~\ref{q:ssred}. We focus on the following complementary scenario:
Suppose we have proved that all $(t,n,\eps)$-secret sharing schemes for $b$-bit messages using $d$ bits of randomness require a $(\delta,m)$-extractable class of randomness sources.
It is then natural to wonder whether such a result can be bootstrapped to conclude that all $(t',n',\eps)$-secret sharing schemes for the same message length $b$ and number of randomness bits $d$ also require $(\delta,m)$-extractable randomness, for different threshold $t'$ and number of parties $n'$.
A natural approach is to set up general \emph{black-box reductions} between different types of secret sharing which, crucially, do not use extra randomness.
In fact, if we can obtain from a $(t',n',\eps)$-secret sharing scheme $(\Share',\Rec',\cY)$ another $(t,n,\eps)$-secret sharing scheme $(\Share,\Rec,\cY)$ for $b$-bit messages which uses the same class of randomness sources $\cY$, then our initial assumption would allow us to conclude that $\cY$ is $(\delta,m)$-extractable.

Remarkably, we are able to obtain the desired reductions for a broad range of parameters by exploiting a connection to the construction of combinatorial objects called \emph{distribution designs}, a term coined by Stinson and Wei~\cite{SW18} for the old technique of devising a new secret sharing scheme by giving multiple shares of the original scheme to each party.
Surprisingly, although these objects have roots going back to early work on secret sharing~\cite{BL88}, they have not been the subject of a general study.
In this work, we obtain general and simple constructions of, and bounds for, distribution designs, which are tight in certain parameter regimes.
We give two examples of reductions we derive from these results.
\begin{coro}[Informal, Restated]\label{coro:inf1}
    If every $(t=2,n,\eps)$-secret sharing scheme for $b$-bit messages using $d$ bits of randomness requires a $(\delta,m)$-extractable class of randomness sources, then so does every $(t',n',\eps)$-secret sharing scheme for $b$-bit messages using $d$ bits of randomness whenever $n\leq \binom{n'}{t'-1}$.
    Moreover, this is the best distribution-design-based reduction possible with $t=2$.
\end{coro}

\begin{coro}[Informal, Restated]\label{coro:inf2}
    If every $(t,n,\eps)$-secret sharing scheme for $b$-bit messages using $d$ bits of randomness requires a $(\delta,m)$-extractable class of randomness sources, then so does every $(t'=n',n',\eps)$-secret sharing scheme for $b$-bit messages using $d$ bits of randomness whenever $n'\geq \binom{n}{t-1}$.
    Moreover, this is the best distribution-design-based reduction possible with $t'=n'$.
\end{coro}

\section{Related Work}

We begin by discussing the results on private-key encryption that led to the work of Bosley and Dodis~\cite{BD07} in more detail.
Early work by McInnes and Pinkas~\cite{MP90} showed that min-entropy sources and Santha-Vazirani sources are insufficient for information-theoretic private-key encryption of even $1$-bit messages.
This negative result was later extended to \emph{computationally} secure private-key encryption by Dodis, Ong, Prabhakaran, and Sahai~\cite{DOPS04}, and was complemented by Dodis and Spencer~\cite{DS02}, who showed that, in fact, non-extractable randomness \emph{is} sufficient for information-theoretic private-key encryption of $1$-bit messages.
Later, the picture was completed by the aforementioned groundbreaking work of Bosley and Dodis~\cite{BD07}.

Besides the results already discussed above for private-key encryption and secret sharing, the possibility of realizing other cryptographic primitives using certain classes of imperfect randomness sources has also been studied.
Non-extractable randomness is known to be sufficient for message authentication~\cite{MW97,DS02}, signature schemes~\cite{DOPS04,ACMPS14}, differential privacy~\cite{DLMV12,DY15,YL18}, secret-key agreement~\cite{ACMPS14}, identification protocols~\cite{ACMPS14}, and interactive proofs~\cite{DOPS04}.
On the other hand, Santha-Vazirani sources are insufficient for bit commitment, secret sharing, zero knowledge, and two-party computation~\cite{DOPS04}, and in some cases this negative result even holds for Santha-Vazirani sources with efficient tampering procedures~\cite{ACMPS14}.

In other directions, the security loss incurred by replacing uniform randomness by imperfect randomness was studied in~\cite{BBNRSSY09,BKMR15}, and the scenario where a perfect common reference string is replaced by certain types of imperfect randomness has also been considered~\cite{CPS07,AORSV20}.
The security of keyed cryptographic primitives with non-uniformly random keys has also been studied~\cite{DY13}.

\section{Preliminaries}
Let us first detail the chapter specific lemmas.

We will use a folklore/standard lemma stemming from a straightforward application of the probabilistic method, which states that, with high probability, a random function extracts almost perfect randomness from a fixed source with sufficient min-entropy.
By a union bound, this result also implies that a random function is a great extractor for all sufficiently small classes of flat sources (and convex combinations thereof), an observation we will exploit later on.
\begin{lemma}\label{lem:probmethod}
Fix an $(n,k)$-source $X$.
Then, for every $\eps>0$ it holds that a uniformly random function $F:\bit^n\to\bit^m$ with $m\leq k-2\log(1/\eps)$ satisfies $F(X)\approx_\eps U_m$ with probability at least $1-2 e^{-\eps^2 2^k}$ over the choice of $F$.
\end{lemma}
\begin{proof}
Fix an $(n,k)$-source $X$ and pick a function $F:\bit^n\to\bit^m$ with $m\leq k-2\log(1/\eps)$ uniformly at random.
It suffices to bound the probability that
\begin{equation*}
    |\Pr[F(X)\in \cT]-\mu(\cT)|\leq \eps
\end{equation*}
holds for every set $\cT\subseteq\bit^m$, where $\mu(\cT)=|\cT|/2^m$ denotes the density of $\cT$.
Fix such a set $\cT$, and let $Z_x=\Pr[X=x]\cdot \mathbf{1}_{F(x)\in \cT}$.
Then, we have $\Pr[F(X)\in \cT]=\sum_{x\in\bit^n}Z_x$ and $\E\left[\sum_{x\in\bit^n}Z_x\right]=\mu(\cT)$.
As a result, since $Z_x\in[0,\Pr[X=x]]$ for all $x\in\bit^n$, Hoeffding's inequality\footnote{The version of Hoeffding's inequality we use here states that if $X_1,\dots,X_N$ are independent random variables and $X_i\in[m_i,M_i]$ for each $i$, then $\Pr\left[\left|\sum_{i=1}^N X_i-\mu\right|>\eps\right]\leq 2\cdot \exp\left(-\frac{2\eps^2}{\sum_{i=1}^N (M_i-m_i)^2}\right)$, where $\mu=\E\left[\sum_{i=1}^N X_i\right]$.} implies that
\begin{align*}
    \Pr\left[\left|\sum_{x\in\bit^n}Z_x-\mu(\cT)\right|>\eps\right]&\leq 2\cdot \exp\left(-\frac{2\eps^2}{\sum_{x\in\bit^n}\Pr[X=x]^2}\right)\\
    &\leq 2\cdot e^{-2\eps^2 2^k}.
\end{align*}
The last inequality follows from the fact that
\begin{equation*}
    \sum_{x\in\bit^n}\Pr[X=x]^2 \leq \max_{x\in\bit^n}\Pr[X=x]\leq 2^{-k},
\end{equation*}
since $X$ is an $(n,k)$-source.
Finally, a union bound over all $2^{2^m}$ sets $\cT\subseteq\bit^m$ shows that the event in question holds with probability at least
\begin{equation*}
    1-2\cdot 2^{2^m}\cdot e^{-2\eps^2 2^k}\geq 1-2 e^{-\eps^2 2^k}
\end{equation*}
over the choice of $F$, given the upper bound on $m$.
\end{proof}

The following extension of Lemma~\ref{lem:probmethod}, stating that a random function condenses weak sources with high probability, will also be useful.
\begin{lemma}\label{lem:randcond}
Fix an $(n,k)$-source $X$.
Then, for every $\eps>0$ it holds that a uniformly random function $F:\bit^n\to\bit^m$ satisfies $F(X)\approx_\eps W$ for some $W$ such that $\minH(W)\geq \min(m,k-2\log(1/\eps))$ with probability at least $1-2 e^{-\eps^2 2^k}$ over the choice of $F$.
\end{lemma}
\begin{proof}
For $m'=\min(m,k-2\log(1/\eps))$, let $F':\bit^n\to\bit^{m'}$ be the restriction of $F$ to its first $m'$ bits.
Then, Lemma~\ref{lem:probmethod} ensures that $F'(X)\approx_\eps U_{m'}$ with probability at least $1-2e^{-\eps^2 2^k}$ over the choice of $F$.
Via a coupling argument, this implies that $F(X)\approx W$ for some $W$ with $\minH(W)\geq m'$.

\end{proof}

\subsection{Secret Sharing}

\begin{defn}[Threshold secret sharing scheme]
A tuple $(\Share,\Rec,\cY)$ with $\Share:\bit^b\times\bit^d\to\left(\bit^\ell\right)^n$ and $\Rec:\bit^*\to\bit^b$ deterministic algorithms and $\cY$ a class of randomness sources over $\bit^d$ is a \emph{$(t,n,\eps)$-secret sharing scheme (for $b$-bit messages using $d$ bits of randomness)} if for every randomness source $Y\in\cY$ the following hold:
\begin{enumerate}
    \item If $\cT\subseteq [n]$ satisfies $|\cT|\geq t$ (i.e., $\cT$ is \emph{authorized}), then
    \begin{equation*}
        \Pr_Y[\Rec(\Share(x,Y)_\cT)=x]=1
    \end{equation*}
    for every $x\in\bit^b$;
    
    \item If $\cT\subseteq[n]$ satisfies $|\cT|<t$ (i.e., $\cT$ is \emph{unauthorized}), then for any $x,x'\in\bit^b$ we have
    \begin{equation*}
        \Share(x,Y)_\cT\approx_\eps \Share(x',Y)_\cT,
    \end{equation*}
    where $\Share(x,Y)_\cT$ denotes the shares of parties $i \in \cT$.
\end{enumerate}
\end{defn}

We consider two natural extensions:
leakage-resilient secret sharing, and non-malleable 
secret sharing.

\begin{defn}[Leakage-resilient secret sharing scheme]
    We say that a tuple $(\Share,\Rec,\cY)$ with $\Share:\bit^b\times\bit^d\to\left(\bit^\ell\right)^2$ and $\Rec:\left(\bit^\ell\right)^2\to\bit^b$ deterministic algorithms and $\cY$ a class of randomness sources over $\bit^d$ is an \emph{$(\eps_1, \eps_2)$-leakage-resilient secret sharing scheme (for $b$-bit messages using $d$ bits of randomness)} if $(\Share,\Rec,\cY)$ is a $(t=2,n=2,\eps_1)$-secret sharing scheme and the following additional property is satisfied: For any two messages $x,x'\in\bit^b$ and randomness source $Y\in\cY$, let $(\Sh_1,\Sh_2)=\Share(x,Y)$ and $(\Sh'_1,\Sh'_2)=\Share(x',Y)$.
    Then, for any leakage functions $f,g:\bit^\ell\to\bit$ it holds that
    \begin{equation*}
        f(\Sh_1),g(\Sh_2)\approx_{\eps_2}f(\Sh'_1),g(\Sh'_2).
    \end{equation*}
\end{defn}

\begin{defn}[Split-state non-malleable code~\cite{DPW18}]
A tuple $(\Enc,\Dec,\cY)$ with $\Enc:\bit^b\times\bit^d\to(\bit^\ell)^2$ and $\Dec:(\bit^\ell)^2\to\bit^b\cup\{\bot\}$ deterministic algorithms and $\cY$ a class of randomness sources is a \emph{(split-state) $\eps$-non-malleable code} if the following holds for every randomness source $Y\in\cY$:
\begin{enumerate}
    \item $\Pr[\Dec(\Enc(x,Y))=x]=1$ for all $x\in\bit^b$;
    
    \item For tampering functions $f,g:\bit^\ell\to\bit^\ell$, denote by $\Tamp^{f,g}_x$ the tampering random experiment which computes $(L,R)=\Enc(x,Y)$ and outputs $\Dec(f(L),g(R))$.
    Then, for any tampering functions $f$ and $g$ there exists a distribution $D^{f,g}$ over $\bit^b\cup\{\bot,\same\}$ such that
    \begin{equation*}
        \Tamp^{f,g}_x \approx_\eps \Sim^{f,g}_x
    \end{equation*}
    for all $x\in\bit^b$, where $\Sim^{f,g}_x$ denotes the random experiment which samples $z$ according to $D^{f,g}$ and outputs $z$ if $z\neq\same$ and $x$ if $z=\same$.
\end{enumerate}
The notion of non-malleable code in the split-state model is equivalent to the notion of a \textbf{ $2$-out-of-$2$ non-malleable secret sharing scheme}~\cite{GK18}. 
\end{defn}

\subsection{Amplifying Leakage-Resilience}
The following lemma states that every secret sharing scheme withstanding $1$ bit of leakage also withstands $t>1$ bits of leakage from each share, at the cost of an increase in statistical error.
\begin{lemma}
        Let $(\Share, \Rec,\cY)$ be an $(\eps_1,\eps_2)$-leakage-resilient secret sharing scheme.
        Then, for all secrets $x,x'\in\bit^b$, randomness source $Y\in\cY$, and functions $f, g : \bit^\ell \to \bit^t$ we have
        \begin{equation*}
            f(\Sh_1), g(\Sh_2) \approx_{\eps'} f(\Sh'_1), g(\Sh'_2)
        \end{equation*}
        with $\eps'=2^{t}\eps_2$, where $(\Sh_1,\Sh_2)=\Share(x,Y)$ and $(\Sh'_1,\Sh'_2)=\Share(x',Y)$.
    \end{lemma}
\begin{proof}
Fix arbitrary secrets $x, x'\in\bit^b$ and a randomness source $Y\in\cY$, and define $(\Sh_1,\Sh_2)=\Share(x,Y)$ and $(\Sh'_1,\Sh'_2)=\Share(x',Y)$.
Suppose that there exist functions $f,g:\bit^\ell\to\bit^t$ such that the distributions $(f(\Sh_1),g(\Sh_2))$ and $(f(\Sh'_1),g(\Sh'_2))$ are $(\eps'=2^t\eps_2)$-far.
Then, the XOR lemma (Lemma \ref{lem:XOR}) implies that there is a non-zero vector $a\in\bit^{2t}$, which we may write as $a=(a^{(1)},a^{(2)})$ for $a^{(1)},a^{(2)}\in\bit^t$, such that the distributions 
\begin{equation*}
    \langle a,(f(\Sh_1),g(\Sh_2))\rangle= \langle a^{(1)}, f(\Sh_1)\rangle+\langle a^{(2)},g(\Sh_2)\rangle
\end{equation*}
and
\begin{equation*}
    \langle a,(f(\Sh'_1),g(\Sh'_2))\rangle= \langle a^{(1)}, f(\Sh'_1)\rangle+\langle a^{(2)},g(\Sh'_2)\rangle
\end{equation*}
are $\eps_2$-far.
Consequently, for $f',g':\bit^\ell\to\bit$ defined as $f'(z)=\langle a^{(1)},f(z)\rangle$ and $g'(z)=\langle a^{(2)},g(z)\rangle$ it holds that
\begin{equation*}
    f'(\Sh_1),g'(\Sh_2)\not\approx_{\eps_2} f'(\Sh'_1),g'(\Sh'_2),
\end{equation*}
contradicting the fact that $(\Share,\Rec,\cY)$ is an $(\eps_1,\eps_2)$-leakage-resilient secret sharing scheme.

\end{proof}

\section{Technical Overview}

\subsection{Leakage-Resilient Secret Sharing Requires Extractable Randomness.}

We present a high-level overview of our approach towards proving Theorem~\ref{thm:final}.
Recall that our goal is to show that if $(\Share,\Rec,\cY)$ is an $(\eps_1,\eps_2)$-leakage-resilient secret sharing for $b$-bit messages using $d$ bits of randomness, then there exists a deterministic function $\ext:\bits^d\to\bits^m$ such that $\ext(Y)\approx_\delta U_m$ for all sources $Y\in\cY$, provided that the number of randomness bits $d$ used is not huge.

Our candidate extractor $\ext$ works as follows on input some $y\in\bits^d$: 
\begin{enumerate}
    \item Compute $(\Sh_1,\Sh_2)=\Share(0^b,y)\in\bits^\ell\times\bits^\ell$;
    
    \item For appropriate leakage functions $f,g:\bits^\ell\to\bits^s$, compute the tuple $(f(\Sh_1),g(\Sh_2))$;
    
    \item For an appropriate function $h:\bits^{2s}\to\bits^m$, output
    \begin{equation*}
        \ext(y)=h(f(\Sh_1),g(\Sh_2)).
    \end{equation*}
\end{enumerate}
The proof of Theorem~\ref{thm:final} follows from an analysis of this candidate construction, and we show the existence of appropriate functions $f$, $g$, and $h$ via the probabilistic method.
Note that the number of sources in $\cY$ may be extremely large.
Consequently, our first step, which is similar in spirit to the first step of the related result for private-key encryption in~\cite{BD07}, is to exploit the leakage-resilience of the scheme in question to show that it suffices to focus on a restricted family to prove the desired result.
More precisely, it suffices to show the existence of functions $f$, $g$, and $h$ as above satisfying
\begin{equation}\label{eq:Zclose}
    h(f(Z_1),g(Z_2))\approx_{\delta'} U_m,
\end{equation}
with $\delta'$ an appropriate error parameter, for all $(Z_1,Z_2)\in \cZ$ defined as
\begin{equation*}
    \cZ=\{\Share(U_b,y):y\in\bits^d\},
\end{equation*}
which contains at most $2^d$ distributions.
Our analysis then proceeds in three steps:
\begin{enumerate}
    \item We show that each $(Z_1,Z_2)\in\cZ$ is close in statistical distance to a convex combination of joint distributions $\closeToPair$ with the property that $\minH(\closeToLeft)+\minH(\closeToRight)$ is sufficiently large for all $i$, where $\minH(\cdot)$ denotes the min-entropy of a distribution;
    
    \item Exploiting the previous step, we prove that if we pick $f$ and $g$ uniformly at random, then with high probability over this choice it holds that the joint distribution $(f(Z_1), g(Z_2))$ is close in statistical distance to a high min-entropy distribution;
    
    \item A well known, standard application of the probabilistic method then shows that a uniformly random function $h$ will extract many perfectly random bits from $(f(Z_1),g(Z_2))$ with high probability over the choice of $h$.
\end{enumerate}
While this proves that there exist functions $f$, $g$, and $h$ such that~\eqref{eq:Zclose} holds for a given $(Z_1,Z_2)\in\cZ$, we need~\eqref{eq:Zclose} to be true simultaneously for all $(Z_1,Z_2)\in\cZ$.
We resolve this by employing a union bound over the at most $2^d$ distributions in $\cZ$.
Therefore, if $d$ is not extremely large, we succeed in showing the existence of appropriate functions $f$, $g$, and $h$, and the desired result follows.
More details can be found in Section~\ref{sec:lrsss}.

\subsubsection{Random-less Reductions for Secret Sharing.}

In this section, we define distribution designs and briefly discuss how they can be used to provide the desired black-box reductions between different types of threshold secret sharing, in particular Corollaries~\ref{coro:inf1} and~\ref{coro:inf2}.
Intuitively, a $(t,n,t',n')$-distribution design distributes shares $(\Sh_1,\Sh_2,\dots,\Sh_{n'})$ of some $(t',n',\eps)$-secret sharing scheme into subsets of shares $\cS_1,\dots,\cS_{n}$, with the property that $(\cS_1,\dots,\cS_n)$ are now shares of a $(t,n,\eps)$-secret sharing scheme.
More formally, we have the following definition, which also appears in~\cite{SW18}.
\begin{defn}[Distribution design]\label{def:distdesign}
We say a family of sets $\cD_1,\cD_2,\dots,\cD_n\subseteq[n']$ is a \emph{$(t,n,t',n')$-distribution design} if for every $\cT\subseteq [n]$ it holds that
\begin{equation*}
    \left|\bigcup_{i\in\cT}\cD_i\right|\geq t'
\end{equation*}
if and only if $|\cT|\geq t$.
\end{defn}
Given a $(t,n,t',n')$-distribution design $\cD_1,\dots,\cD_n\subseteq [n']$, it is clear how to set up a black-box reduction without extra randomness from $(t',n',\eps)$-secret sharing to $(t,n,\eps)$-secret sharing: 
If $(\Share',\Rec',\cY)$ is an arbitrary $(t',n',\eps)$-secret sharing scheme for $b$-bit messages, we can obtain a $(t,n,\eps)$-secret sharing scheme $(\Share,\Rec,\cY)$ for $b$-bit messages by defining
\begin{equation*}
    \Share(x,y)_i = \Share'(x,y)_{\cD_i}
\end{equation*}
for each $i\in[n]$, and
\begin{equation*}
    \Rec(\Share(x,y)_\cT) = \Rec'\left(\Share'(x,y)_{\bigcup_{i\in\cT}\cD_i}\right)
\end{equation*}
for each $\cT\subseteq[n]$.
The following lemma is then straightforward from the definitions of threshold secret sharing and distribution designs, and this construction.
\begin{lem}\label{lem:distred}
If every $(t,n,\eps)$-secret sharing scheme for $b$-bit messages using $d$ bits of randomness requires $(\delta,m)$-extractable randomness and there exists a $(t,n,t',n')$-distribution design, then so does every $(t',n',\eps)$-secret sharing scheme for $b$-bit messages using $d$ bits of randomness.
\end{lem}
Details of our constructions of distribution designs and associated bounds can be found in Section~\ref{sec:reductions}.
The black-box reductions then follow immediately by combining these constructions with Lemma~\ref{lem:distred}.

\subsection{Open Questions}\label{sec:open}

We obtain distribution designs for a wide variety of parameters, but for some of these constructions we could not prove optimality or find a better construction. We leave this as an open question. 
A naturally related question is whether there is an alternative approach to obtain a random-less reduction for secret sharing that does not use distribution designs. 

Finally, we hope this work further motivates research on the main open question of whether $2$-out-of-$2$ secret sharing (or even $t$-out-of-$n$ secret sharing for any $t$ and $n$) requires extractable randomness.

\subsection{Amplifying Leakage-Resilience}

Recall the definition of leakage-resilient secret sharing from Definition~\ref{def:lrsss} already discussed in the beginning of this chapter.
The following lemma states that every secret sharing scheme withstanding $1$ bit of leakage also withstands $t>1$ bits of leakage from each share, at the cost of an increase in statistical error.
\begin{lem}
    \label{lem:amplifying_leakage}
        Let $(\Share, \Rec,\cY)$ be an $(\eps_1,\eps_2)$-leakage-resilient secret sharing scheme.
        Then, for all secrets $x,x'\in\bitsB$, randomness source $Y\in\cY$, and functions $f, g : \bits^\ell \to \bits^t$ we have
        \begin{equation*}
            f(\Sh_1), g(\Sh_2) \approx_{\eps'} f(\Sh'_1), g(\Sh'_2)
        \end{equation*}
        with $\eps'=2^{t}\eps_2$, where $(\Sh_1,\Sh_2)=\Share(x,Y)$ and $(\Sh'_1,\Sh'_2)=\Share(x',Y)$.
    \end{lem}
\begin{proof}
Fix arbitrary secrets $x, x'\in\bitsB$ and a randomness source $Y\in\cY$, and define $(\Sh_1,\Sh_2)=\Share(x,Y)$ and $(\Sh'_1,\Sh'_2)=\Share(x',Y)$.
Suppose that there exist functions $f,g:\bitsL\to\bits^t$ such that the distributions $(f(\Sh_1),g(\Sh_2))$ and $(f(\Sh'_1),g(\Sh'_2))$ are $(\eps'=2^t\eps_2)$-far.
Then, the XOR lemma implies that there is a non-zero vector $a\in\bits^{2t}$, which we may write as $a=(a^{(1)},a^{(2)})$ for $a^{(1)},a^{(2)}\in\bits^t$, such that the distributions 
\begin{equation*}
    \langle a,(f(\Sh_1),g(\Sh_2))\rangle= \langle a^{(1)}, f(\Sh_1)\rangle+\langle a^{(2)},g(\Sh_2)\rangle
\end{equation*}
and
\begin{equation*}
    \langle a,(f(\Sh'_1),g(\Sh'_2))\rangle= \langle a^{(1)}, f(\Sh'_1)\rangle+\langle a^{(2)},g(\Sh'_2)\rangle
\end{equation*}
are $\eps_2$-far.
Consequently, for $f',g':\bits^\ell\to\bits$ defined as $f'(z)=\langle a^{(1)},f(z)\rangle$ and $g'(z)=\langle a^{(2)},g(z)\rangle$ it holds that
\begin{equation*}
    f'(\Sh_1),g'(\Sh_2)\not\approx_{\eps_2} f'(\Sh'_1),g'(\Sh'_2),
\end{equation*}
contradicting the fact that $(\Share,\Rec,\cY)$ is an $(\eps_1,\eps_2)$-leakage-resilient secret sharing scheme.

\end{proof}

\section{Randomness Extraction from Leakage-Resilient Secret Sharing Schemes}\label{sec:lrsss}

In this section, we show that all $2$-out-of-$2$ secret sharing schemes satisfying the weak leakage-resilience requirement from Definition~\ref{def:thresholdss} require extractable randomness with good parameters.
\begin{thm}\label{thm:lrsssextractable}
Given any $\gamma\in(0,1)$, there are absolute constants $c_\gamma,c'_\gamma,c''_\gamma>0$ such that the following holds:
Suppose $(\Share,\Rec,\cY)$ is an $(\eps_1,\eps_2)$-leakage-resilient secret sharing scheme for $b$-bit messages using $d$ bits of randomness.
Then, if $b\geq c_\gamma$ and $d\leq 2^{c'_\gamma b}$ it holds that $\cY$ is $(\delta,m)$-extractable
with $\delta\leq 2^b\eps_2+2^{-c''_\gamma b}$ and $m\geq (1-\gamma)b$.
\end{thm}
We prove Theorem~\ref{thm:lrsssextractable} via a sequence of lemmas by showing the existence of an extractor $\ext:\bits^d\to\bits^m$ for the class $\cY$ with appropriate parameters.
Our construction works as follows: On input $y\in \bits^d$, the extractor $\ext$ computes $(L_y,R_y)=\Share(0^\messageLength,y)$, applies special leakage functions $f,g:\bitsL\to\bits^\messageLength$ to be determined in order to obtain local leakage $(f(L_y),g(R_y))$, and finally outputs $\ext(y)=h(f(L_y),g(R_y))$ for an appropriate function $h:\bits^{2\messageLength}\to\bits^m$.
Our goal is to show that
\begin{equation}\label{eq:goodExt}
    \ext(Y)\approx_\delta U_m
\end{equation}
for all sources $Y\in\cY$.
Similarly in spirit to~\cite{BD07}, our first lemma shows that in order to prove~\eqref{eq:goodExt} we can instead focus on extracting randomness from the family of distributions
\begin{equation*}\label{eq:moddist}
    \cZ=\{\Share(U_\messageLength,y):y\in\bits^d\}.
\end{equation*}

\begin{lem}\label{lem:modExt}
Fix functions $f,g:\bitsL\to\bits^\messageLength$ and $h:\bits^{2\messageLength}\to\bits^m$, and suppose that
\begin{equation}\label{eq:goodExtmod}
    \ext'(Z)=h(f(Z_1),g(Z_2))\approx_{\delta'} U_m
\end{equation}
for all $Z=(Z_1,Z_2)\in\cZ$.
Then, it holds that $\ext$ given by $\ext(y)=h(f(L_y),g(R_y))$, where $(L_y,R_y)=\Share(0^\messageLength,y)$, satisfies
\begin{equation*}
    \ext(Y)\approx_\delta U_m
\end{equation*}
for all $Y\in\cY$ with $\delta=2^{\messageLength}\eps_2+\delta'$.
\end{lem}
\begin{proof}
Lemma~\ref{lem:amplifying_leakage} implies that
\begin{equation*}
    f(L_Y),g(R_Y)\approx_{\eps'} f(L'_Y),g(R'_Y),
\end{equation*}
where $(L'_Y,R'_Y)=\Share(U_\messageLength,Y)$
holds with $\eps'=2^{b}\eps_2$ for all $Y\in\cY$, and so $\ext(Y)\approx_{\eps'} h(f(L'_K),g(R'_K))$.
Since~\eqref{eq:goodExtmod} holds for all $Z\in\cZ$ and $\Share(U_\messageLength,Y)$ is a convex combination of distributions in $\cZ$, it follows that $h(f(L'_Y),g(R'_Y))\approx_{\delta'} U_m$. 
The triangle inequality yields the desired result.

\end{proof}
Given Lemma~\ref{lem:modExt}, we will focus on proving~\eqref{eq:goodExtmod} for appropriate functions $f$, $g$, and $h$ and error $\delta'$ in the remainder of this section.
We show the following lemma, which implies Theorem~\ref{thm:lrsssextractable} together with Lemma~\ref{lem:modExt}.

\begin{lem}\label{lem:fixedKey}
Given any $\gamma\in(0,1)$, there are absolute constants $c_\gamma,c'_\gamma,c''_\gamma>0$ such that if $b\geq c_\gamma$ and $d\leq 2^{c'_\gamma b}$, then there exist functions $f,g:\bits^\ell\to\bitsB$ and $h:\bits^{2b}\to\bits^m$ such that
\begin{equation*}
    \ext'(Z)=h(f(Z_1),g(Z_2))\approx_{\delta'}U_m
\end{equation*}
for all $Z=(Z_1,Z_2)\in\cZ$ with $\delta'\leq 2^{-c''_\gamma \messageLength}$
and $m\geq (1-\gamma)b$.
\end{lem}

The roadmap for the proof ahead is that we are first going to fix a $Z \in \cZ$, and then do the following:
\begin{enumerate}
    \item Justify that $Z = (Z_1, Z_2)$ is statistically close to an appropriate convex combination of distributions with linear min-entropy that suit our purposes. (Lemma \ref{lem:degree})
    
    \item Show that if we pick $f$ and $g$ uniformly at random, then with high probability over this choice it holds that $(f(Z_1), g(Z_2))$ is statistically close to a distribution with decent min-entropy. (Lemma \ref{lem:fgOnDeg})
    
    \item Note that a random function $h$ extracts uniformly random bits from the tuple $(f(Z_1),g(Z_2))$ with high probability, provided that this distribution contains enough min-entropy. A union bound over the $2^d$ distributions in $\cZ$ concludes the argument.
\end{enumerate}

\begin{lem}\label{lem:degree}
    Fix $\beta\in(0,1)$ and an integer $r>0$.
    Then,
    for all $(Z_1, Z_2) \in \cZ$ it holds that $(Z_1, Z_2)$ is $\left(r\cdot 2^{-(1-\beta-1/r)b} \right)$-close to a distribution $D = \sum_{i \in \cI} p_i\cdot \closeToPair$ where for each $i \in \cI\subseteq [r]$ it holds that $\closeToLeft,\closeToRight\in\bits^\ell$, and $\minH(\closeToLeft) \geq \leftEntropy$ and $\minH(\closeToRight|\closeToLeft=\sh_1)\geq \rightEntropy$ for every $\sh_1\in\supp{\closeToLeft}$.
\end{lem}
\begin{proof}
    Fix some $y\in\bits^d$ and set $(Z_1, Z_2)=\Share(U_\messageLength,y)$. 
    It will be helpful for us to see $\Share(\cdot,y)$ as a bipartite graph $G$ with left and right vertex sets $\bitsL$ and an edge between $\sh_1$ and $\sh_2$ if $(\sh_1,\sh_2)\in\supp{Z_1, Z_2}$. 
    Then, $(Z_1, Z_2)$ is the uniform distribution on the $2^\messageLength$ edges of $G$ by the correctness of the scheme.
    For every left vertex $\sh_1\in\bitsL$, we define its neighborhood
\begin{equation*}
    \cA(\sh_1)=\{\sh_2:(\sh_1,\sh_2)\in\supp{Z_1, Z_2}\}
\end{equation*}
and its degree
\begin{equation*}
    \dg(\sh_1)=|\cA(\sh_1)|.
\end{equation*}
Note that $(Z_2|Z_1=\sh_1)$ is uniformly distributed over $\cA(\sh_1)$, and so
\begin{equation*}
    \minH(Z_2|Z_1=\sh_1)=\log\dg(\sh_1).
\end{equation*}

Partition $\supp{Z_1}$ into sets
\begin{equation*}
    \cS_i = \left \{ \sh_1 : 2^{(\frac{i - 1}{r})\messageLength }\leq \dg(\sh_1) < 2^{(\frac{i}{r})\messageLength} \right\}
\end{equation*}
for $i\in[r]$.
With this definition in mind, we can express $(Z_1, Z_2)$ as
\begin{equation*}\label{eq:convexcombrep}
    \sum_{i \in [r]} \Pr[Z_1 \in \cS_i] \ZCondOnSi ,
\end{equation*}
where $\ZCondOnSi$ denotes the distribution $\zPair$ conditioned on the event that $Z_1 \in \cS_i$. 
Call a non-empty set $\cS_i$ \emph{good} if $ \sum_{\sh_1 \in \cS_i} \deg(\sh_1)  \geq 2^{(\beta+1/r)\messageLength}$. Otherwise the set $\cS_i$ is \emph{bad}. 
Let $\cI$ denote the set of indices $i\in[r]$ such that $\cS_i$ is good.
We proceed to show that we can take the target distribution $D$ in the lemma statement to be $D=\sum_{i\in\cI}p_i\cdot\closeToPair$ for
\begin{equation*}
    p_i=\frac{\Pr[Z_1\in\cS_i]}{\Pr[\textrm{$Z_1$ lands on good set}]}
\end{equation*}
with $\closeToPair=(Z_1,Z_2|Z_1\in\cS_i)$ when $i\in\cI$.

To see this, consider the case where $\cS_i$ is good, i.e., we have $\sum_{\sh_1 \in \cS_i} \deg(\sh_1) \geq 2^{(\beta+1/r)\messageLength}$.
For each $\sh_1 \in \cS_i$, we have
\begin{align*}
    \Pr[Z_1 = \sh_1 | Z_1 \in \cS_i] &= \frac{\dg(\sh_1)}{\sum_{s \in \cS_i} \dg(s)}\\
    &\leq \frac{2^{\frac{i}{r}\messageLength}}{ 2^{(\beta+1/r)\messageLength} }\\
    &= 2^{-\leftEntropy}.
\end{align*}
Furthermore, for any $\sh_1 \in \cS_i$ and $\sh_2$ we know that
\begin{align*}
    \Pr[Z_2 = \sh_2 | Z_1 = \sh_1] &\leq 2^{-\rightEntropy}.
\end{align*}
Combining these two observations shows that in this case we have $\minH(Z_1|Z_1\in\cS_i)\geq \leftEntropy$ and $\minH(Z_2|Z_1=\sh_1)\geq \rightEntropy$ for all valid fixings $\sh_1\in\cS_i$.

To conclude the proof, consider $D$ as above, which we have shown satisfies the properties described in the lemma statement.
Noting that $D$ corresponds exactly to $(Z_1,Z_2)$ conditioned on $Z_1$ landing on a good set, we have
\begin{equation*}
    \Delta((Z_1,Z_2);D) \leq \Pr[\textrm{$Z_1$ lands in a bad set}].
\end{equation*}
It remains to bound this probability on the right-hand side.
Assuming the set $\cS_i$ is bad, it holds that $\sum_{\sh_1 \in \cS_i} \deg(\sh_1)< 2^{(\beta+1/r)\messageLength}$. 
Therefore, since $\zPair$ takes on any edge with probability $2^{-\messageLength}$, it holds that $Z_1$ lands in $\cS_i$ with probability at most $2^{-\messageLength} \cdot 2^{(\beta+1/r)\messageLength} = 2^{-(1 - \beta-1/r)\messageLength}$. 
There are at most $r$ bad sets, so by a union bound we have $\Pr[\textrm{$Z_1$ lands in a bad set}]\leq r \cdot 2^{-(1 - \beta-1/r)\messageLength}$.

\end{proof}

\begin{lem}\label{lem:fgOnDeg}
    Fix $\alpha,\beta \in (0, 1)$ and an integer $r$. 
    Then, with probability at least $1-3r\cdot e^{b-\alpha^2 2^{\min(b/r,(\beta-1/r)b)}}$ over the choice of uniformly random functions $f,g:\bits^\ell\to\bitsB$ it holds that $(f(Z_1), g(Z_2))$ is $\left( 2\alpha + r\cdot 2^{-(1-\beta-1/r)b} \right)$-close to a $(2\messageLength, \left( \beta - 1/r\right) \messageLength-4\log(1/\alpha))$-source.
\end{lem}
\begin{proof}
    Suppose we pick functions $f,g:\bits^\ell\to\bitsB$ uniformly at random.
    We begin by expressing $\fgZ$ as
    \begin{equation*}
        \sum_{i \in [r]} \Pr[Z_1 \in \cS_i] \fgZCondOnSi ,
    \end{equation*}
    which by Lemma \ref{lem:degree} is $\left( r\cdot 2^{-(1-\beta-1/r)b} \right)$-close to 
    \begin{equation*}
        \sum_{i \in \cI} \Pr[Z_1 \in \cS_i] \fgcloseToPair.
    \end{equation*}
    We proceed by cases:
    \begin{enumerate}
        \item $\frac{i-1}{r}\geq \beta-1/r$: We know from Lemma~\ref{lem:degree} that $\minH(\closeToRight|\closeToLeft=\sh_1)\geq (\beta-1/r) b$ for all $\sh_1\in\supp{\closeToLeft}$.
        By Lemma~\ref{lem:randcond}, we have
        \begin{equation*}
            (g(\closeToRight)|\closeToLeft=\sh_1)\approx_\alpha V
        \end{equation*}
        for some $V$ with $\minH(V)\geq (\beta-1/r) b-2\log(1/\alpha)$ with probability at least $1-2e^{-\alpha^2 2^{(\beta-1/r) b}}$ over the choice of $g$.
        Since this holds for any valid fixing $\closeToLeft=\sh_1$, we conclude via a union bound over the at most $2^b$ possible fixings that
        \begin{equation*}
            f(\closeToLeft),g(\closeToRight)\approx_\alpha W_i
        \end{equation*}
        for some $W_i$ with $\minH(W_i)\geq (\beta-1/r) b-2\log(1/\alpha)$ with probability at least $1-2e^{b-\alpha^2 2^{(\beta-1/r)b}}$ over the choice of $f$ and $g$.
        
        \item $1/r\leq \frac{i-1}{r}< \beta-1/r$: We know from Lemma~\ref{lem:degree} that $\minH(\closeToLeft)\geq \left(\beta-\frac{i-1}{r}\right)b$ and $\minH(\closeToRight|\closeToLeft=\sh_1)\geq \left(\frac{i-1}{r}\right)b$ for all $\sh_1\in\supp{\closeToLeft}$.
        First, by Lemma~\ref{lem:randcond} we conclude that with probability at least
        \begin{equation*}
            1-2e^{-\alpha^2 2^{\left(\beta-\frac{i-1}{r}\right)b}} \geq 1-2e^{-\alpha^2 2^{b/r}}
        \end{equation*}
        over the choice of $f$ it holds that
        \begin{equation}\label{eq:fclose}
            f(\closeToLeft)\approx_\alpha V_1
        \end{equation}
        for some $V_1$ with $\minH(V_1)\geq (\beta-\frac{i-1}{r})b-2\log(1/\alpha)$.
        Analogously, for every $\sh_1\in\supp{\closeToLeft}$, we can again invoke Lemma~\ref{lem:randcond} to see that with probability at least
        \begin{equation*}
            1-2e^{-\alpha^2 2^{\left(\frac{i-1}{r}\right)b}} \geq 1-2e^{-\alpha^2 2^{b/r}}
        \end{equation*}
        over the choice of $g$, for any $\sh_1\in\supp{\closeToLeft}$ it holds that
        \begin{equation}\label{eq:gclose}
            (g(\closeToRight)|\closeToLeft=\sh_1)\approx_\alpha V_{2,\sh_1}
        \end{equation}
        for some $V_{2,\sh_1}$ with $\minH(V_{2,\sh_1})\geq \left(\frac{i-1}{r}\right)b-2\log(1/\alpha)$.
        By a union bound over the at most $2^b$ possible fixings $\sh_1$, we conclude that~\eqref{eq:gclose} holds simultaneously for all $\sh_1\in\supp{\closeToLeft}$ with probability at least $1-2e^{b-\alpha^2 2^{b/r}}$ over the choice of $g$.
        An additional union bound shows that this holds simultaneously along~\eqref{eq:fclose} with probability at least $1-3e^{b-\alpha^2 2^{b/r}}$ over the choice of $f$ and $g$, which implies that
        \begin{equation*}
            f(\closeToLeft),g(\closeToRight)\approx_{2\alpha} W_i
        \end{equation*}
        for some $W_i$ with
        \begin{multline*}
            \minH(W_i)\geq \left(\beta-\frac{i-1}{r}\right)b-2\log(1/\alpha)+\left(\frac{i-1}{r}\right)b-2\log(1/\alpha) \\=\beta b-4\log(1/\alpha).
        \end{multline*}

        \item $i=1$: In this case, by Lemma~\ref{lem:degree} we know that $\minH(\closeToLeft)\geq \beta b$.
        Therefore, Lemma~\ref{lem:randcond} implies that $f(\closeToLeft)\approx_\alpha V_1$ for some $V_1$ such that $\minH(V_1)\geq \beta b-2\log(1/\alpha)$ with probability at least $1-2e^{-\alpha^2 2^{\beta b}}\geq 1-2e^{-\alpha^2 2^{b/r}}$.
        This implies that $f(\closeToLeft),g(\closeToRight)\approx_\alpha W_i$ for some $W_i$ with $\minH(W_i)\geq \beta b-2\log(1/\alpha)$.
    \end{enumerate}
    Finally, a union bound over the at most $r$ indices $i\in\cI$ yields the desired statement.
    
\end{proof}

We are now ready to prove Lemma~\ref{lem:fixedKey} with the help of Lemma~\ref{lem:fgOnDeg}.
\begin{proof}[Proof of Lemma~\ref{lem:fixedKey}]
Fix some $\gamma\in(0,1)$.
Then, we set $\beta=1-\gamma/2>1-\gamma$, $\alpha=2^{-cb}$ for a sufficiently small constant $c>0$, and $r>0$ a sufficiently large integer so that
\begin{equation}\label{eq:choice1}
    1-\gamma\leq \beta - 1/r - 6c
\end{equation}
and
\begin{equation}\label{eq:choice2}
    1/r+6c \leq \frac{\min(\beta,1-\beta)}{100}.
\end{equation}
According to Lemma~\ref{lem:fgOnDeg}, we know that for any given $Z=(Z_1,Z_2)\in\cZ$ it holds that $(f(Z_1),g(Z_2))$ is $(2\alpha+r\cdot 2^{-(1-\beta-1/r)b})$-close to some $(2b,(\beta-1/r)b-4\log(1/\alpha))$-source $W$ with probability at least $1-3r\cdot e^{b-\alpha^2 2^{\min(b/r,(\beta-1/r)b)}}$ over the choice of $f$ and $g$.

Let $m=(1-\gamma)b$ and pick a uniformly random function $h:\bits^{2b}\to\bits^m$.
Then, since $m\leq \minH(W)-2\log(1/\alpha)$ by~\eqref{eq:choice1}, Lemma~\ref{lem:probmethod} implies that $h(W)\approx_\alpha U_m$, and hence
\begin{equation}\label{eq:hclose}
    h(f(Z_1),g(Z_2))\approx_{3\alpha + r\cdot 2^{-(1-\beta-1/r)b}} U_m,
\end{equation}
with probability at least
\begin{align*}
    1-&2e^{-\alpha^2 2^{(\beta-1/r)b-4\log(1/\alpha)}}-3r\cdot e^{b-\alpha^2 2^{\min(b/r,(\beta-1/r)b)}}\\
    &\geq 1-5r\cdot e^{b-\alpha^2 2^{\min(b/r,(\beta-1/r)b)-4\log(1/\alpha)}}
\end{align*}
over the choice of $f$, $g$, and $h$, via a union bound.

Now, observe that from~\eqref{eq:choice2}, if $b\geq c_\gamma$ for a sufficiently large constant $c_\gamma>0$, it follows that
\begin{equation*}
    5r\cdot e^{b-\alpha^2 2^{\min(b/r,(\beta-1/r)b)-4\log(1/\alpha)}} \leq 2^{-2^{2c'_\gamma b}}
\end{equation*}
for some constant $c'_\gamma>0$.
Moreover, under~\eqref{eq:choice2} we also have that
\begin{equation*}
    \delta':=3\alpha + r\cdot 2^{-(1-\beta-1/r)b}\leq 2^{-c''_\gamma b}
\end{equation*}
for some constant $c''_\gamma>0$.
Finally, a union bound over the $2^d$ distributions in $\cZ$ shows that~\eqref{eq:hclose} holds simultaneously for all $Z\in\cZ$ with probability at least $1-2^{d-2^{2c'_\gamma b}}$.
Consequently, if $d\leq 2^{c'_\gamma b}$ it follows that there exist functions $f$, $g$, and $h$ such that~\eqref{eq:hclose} holds for all $Z\in\cZ$ with the appropriate error $\delta'$ and output length $m$.

\end{proof}

\subsection{The Main Result}\label{sec:mainres}

We now use Theorem~\ref{thm:lrsssextractable} to obtain the main result of this section. 
\begin{thm}[First part of Theorem~\ref{thm:final}, restated]
    Suppose $(\Share,\Rec,\cY)$ is an $(\eps_1,\eps_2)$-leakage-resilient secret sharing scheme for $b$-bit messages.
    Then, either:
    \begin{itemize}
        \item The scheme uses $d\geq \min\left(2^{\Omega(b)},(1/\eps_2)^{\Omega(1)}\right)$ bits of randomness, or;
        
        \item The class of sources $\cY$ is $(\delta,m)$-extractable with $\delta\leq \max\left(2^{-\Omega(b)},\eps_2^{\Omega(1)}\right)$ and $m=\Omega(\min(b,\log(1/\eps_2)))$. 
    \end{itemize}
\end{thm}
\begin{proof}
Given the scheme $(\Share,\Rec,\cY)$ from the theorem statement, let $b'=\min\left(b,\left\lceil\frac{\log(1/\eps_2)}{100}\right\rceil\right)$ and consider the modified scheme $(\Share',\Rec',\cY)$ for $b'$-bit messages obtained by appending $0^{b-b'}$ to every $b'$-bit message and running the original scheme $(\Share,\Rec,\cY)$.
Applying Theorem~\ref{thm:lrsssextractable} to $(\Share',\Rec',\cY)$ we conclude that either $\Share'$, and hence $\Share$, uses at least
\begin{equation*}
    2^{\Omega(b')}=\min\left(2^{\Omega(b)},(1/\eps_2)^{\Omega(1)}\right)
\end{equation*}
bits of randomness, or $\cY$ is $(\delta,m)$-extractable with
\begin{equation*}
    \delta\leq 2^{-\Omega(b')}=\max\left(2^{-\Omega(b)},\eps_2^{\Omega(1)}\right)
\end{equation*}
and $m=\Omega(b')=\Omega(\min(b,\log(1/\eps_2)))$.

\end{proof}

\subsection{Efficient Leakage-Resilient Secret Sharing Requires Efficiently Extractable Randomness}\label{sec:efficient}

In this section, we prove the remaining part of Theorem~\ref{thm:final}.
We show that every low-error leakage-resilient secret sharing scheme $(\Share,\Rec,\cY)$ for $b$-bit messages where $\Share$ is computed by a $\poly(b)$-time algorithm admits a low-error extractor for $\cY$ computable by a family of $\poly(b)$-size circuits.
Similarly to~\cite[Section 3.1]{BD07}, this is done by replacing the uniformly random functions $f$, $g$, and $h$ in the proof of Theorem~\ref{thm:lrsssextractable} by \emph{$t$-wise independent functions}, for an appropriate parameter $t$.

We say that a family of functions $\cF_t$ from $\bits^p$ to $\bits^q$ is \emph{$t$-wise independent} if for $F$ sampled uniformly at random from $\cF_t$ it holds that the random variables $F(x_1),F(x_2),\dots,F(x_t)$ are independent and uniformly distributed over $\bits^q$ for any distinct $x_1,\dots,x_t\in\bits^p$.
There exist $t$-wise independent families of functions $\cF_t$ such that every $f\in\cF_t$ can be computed in time $\poly(b)$ and can be described by $\poly(b)$ bits whenever $p$, $q$, and $t$ are $\poly(b)$~\cite{Dod00,TV00,BD07}.
Therefore, since $\Share$ admits a $\poly(b)$-time algorithm, it suffices to show the existence of functions $f$, $g$, and $h$ belonging to appropriate $\poly(b)$-wise independent families of functions such that $\ext(Y)=h(f(\Sh_1),g(\Sh_2))$ is statistically close to uniform, where $(\Sh_1,\Sh_2)=\Share(0^b,Y)$, for every source $Y\in\cY$ (the advice required to compute $\ext$ would be the description of $f$, $g$, and $h$).
We accomplish this with the help of some auxiliary lemmas.
The first lemma states a standard concentration bound for the sum of $t$-wise independent random variables.
\begin{lem}[\protect{\cite[Theorem 5]{Dod00}, see also \cite[Lemma 2.2]{BR94}}]\label{lem:dodisconc}
Fix an even integer $t\geq 2$ and suppose that $X_1,\dots,X_N$ are $t$-wise independent random variables in $[0,1]$.
Let $X=\sum_{i=1}^N X_i$ and $\mu=\E[X]$.
Then, it holds that
\begin{equation*}
    \Pr[|X-\mu|\geq \eps\cdot \mu]\leq 3\left(\frac{t}{\eps^2 \mu}\right)^{t/2}
\end{equation*}
for every $\eps<1$.
\end{lem}
We can use Lemma~\ref{lem:dodisconc} to derive analogues of Lemmas~\ref{lem:probmethod} and~\ref{lem:randcond} for $t$-wise independent functions.
\begin{lem}\label{lem:randtwise}
Suppose $f:\bits^p\to\bits^q$ is sampled uniformly at random from a $2t$-wise independent family of functions with $q\leq k - \log t - 2\log(1/\eps)-5$ and $t\geq q$, and let $Y$ be a $(p,k)$-source.
Then, it follows that
\begin{equation*}
    f(Y)\approx_\eps U_q
\end{equation*}
with probability at least $1-2^{-t}$ over the choice of $f$.
\end{lem}
\begin{proof}
    Fix a $(p,k)$-source $Y$ and suppose $f:\bits^p\to\bits^q$ is sampled from a family of $2t$-wise independent functions.
    Note that
    \begin{equation*}
        \Delta(f(Y);U_q)=\frac{1}{2}\sum_{z\in\bits^q}|\Pr[f(Y)=z]-2^{-q}|.
    \end{equation*}
    For each $y\in\bits^p$ and $z\in\bits^q$, consider the random variable\\
    $W_{y,z}=\Pr[Y=y]\cdot \mathbf{1}_{\{f(y)=z\}}$.
    Then, we may write
    \begin{equation*}
        \Delta(f(Y);U_q)=\frac{1}{2}\sum_{z\in\bits^q}\left|\sum_{y\in\bits^p}W_{y,z}-2^{-q}\right|.
    \end{equation*}
    Note that the $W_{y,z}$'s are $2t$-wise independent, $\E[\sum_{y\in\bits^n}W_{y,z}]=2^{-q}$, and that $2^k\cdot W_{y,z}\in[0,1]$.
    Therefore, an application of Lemma~\ref{lem:dodisconc} with the random variables $(2^k\cdot W_{y,z})_{y\in\bits^p,z\in\bits^q}$ shows that
    \begin{equation*}
        \Pr\left[\left|\sum_{y\in\bits^p}W_{y,z}-2^{-q}\right|>2\eps\cdot 2^{-q}\right]\leq 3\left(\frac{t\cdot 2^q}{2\eps^2 2^k}\right)^t.
    \end{equation*}
    Therefore, a union bound over all $z\in\bits^q$ shows that $f(Y)\approx_\eps U_q$ fails to hold with probability at most $3\cdot 2^q\cdot 2^{-t}\left(\frac{t\cdot 2^{q}}{\eps^2\cdot 2^k}\right)^{t}\leq 2^{-t}$ over the choice of $f$, where the inequality follows by the upper bound on $q$.
    
\end{proof}
The proof of the following lemma is analogous to the proof of Lemma~\ref{lem:randcond}, but using Lemma~\ref{lem:randtwise} instead of Lemma~\ref{lem:probmethod}.
\begin{lem}\label{lem:randcondtwise}
Suppose $f:\bits^p\to\bits^q$ is sampled uniformly at random from a $2t$-wise independent family of functions with $t\geq q$, and let $Y$ be a $(p,k)$-source.
Then, it follows that $f(Y)\approx_\eps W$ for some $W$ such that $\minH(W)\geq \min(q,k-\log t-2\log(1/\eps)-5)$ with probability at least $1-2^{-t}$ over the choice of $f$.
\end{lem}

Following the reasoning used in the proof of Theorem~\ref{thm:lrsssextractable} but sampling $f,g:\bits^\ell\to\bits^b$ and $h:\bits^b\to\bits^m$ from $2t$-wise independent families of functions with $t=100\max(b,d)=\poly(b)$, and using Lemmas~\ref{lem:randtwise} and~\ref{lem:randcondtwise} in place of Lemmas~\ref{lem:probmethod} and~\ref{lem:randcond}, respectively, yields the following result analogous to Theorem~\ref{thm:lrsssextractable}.
Informally, it states that efficient low-error leakage-resilient secret sharing schemes require low-complexity extractors for the associated class of randomness sources.
\begin{thm}\label{thm:efflrsss}
    There exist absolute constants $c,c'>0$ such that the following holds for $b$ large enough:
    Suppose $(\Share,\Rec,\cY)$ is an $(\eps_1,\eps_2)$-leakage-resilient secret sharing for $b$-bit messages using $d$ bits of randomness such that $\Share$ is computable by a $\poly(b)$-time algorithm.
    Then, there exists a deterministic extractor $\ext:\bits^d\to\bits^m$ computable by a family of $\poly(b)$-size circuits
    with output length $m\geq c\cdot b$ such that
    \begin{equation*}
        \ext(Y)\approx_\delta U_m
    \end{equation*}
    with $\delta=2^b\eps_2+2^{-c'\cdot b}$ for every $Y\in\cY$.
\end{thm}

Finally, replacing Theorem~\ref{thm:lrsssextractable} by Theorem~\ref{thm:efflrsss} in the reasoning from Section~\ref{sec:mainres} yields the remaining part of Theorem~\ref{thm:final}.

\subsection{An Extension to the Setting of Computational Security}

In this work we focus on secret sharing schemes with information-theoretic security.
However, it is also natural to wonder whether our result extends to secret sharing schemes satisfying a reasonable notion of computational security.
Indeed, a slight modification to the argument used to prove Theorem~\ref{thm:final} also shows that computationally-secure efficient leakage-resilient secret sharing schemes require randomness sources from which one can efficiently extract bits which are pseudorandom (i.e., computationally indistinguishable from the uniform distribution).
We briefly discuss the required modifications in this section.
For the sake of exposition, we refrain from presenting fully formal definitions and theorem statements.

First, we introduce a computational analogue of Definition~\ref{def:lrsss}.
We say that $(\Share,\Rec,\cY)$ is a \emph{computationally secure leakage-resilient secret sharing scheme (for $b$-bit messages)} if the scheme satisfies Definition~\ref{def:lrsss} except that the leakage-resilience property is replaced by the following computational analogue: ``For any leakage functions $f, g : \bits^\ell \to \bits$ computed by $\poly(b)$-sized circuits and any two secrets $x,x'\in\bits^b$, it holds that any adversary computable by $\poly(b)$-sized circuits cannot distinguish between the distributions $(f(\Sh_1),g(\Sh_2))$ and $(f(\Sh'_1),g(\Sh'_2))$ with non-negligible advantage (in some security parameter $\lambda$), where $(\Sh_1,\Sh_2)=\Share(x)$ and $(\Sh'_1,\Sh'_2)=\Share(x')$.'' 


Using this definition, the exact argument we used to prove Theorem~\ref{thm:final} combined with a modified version of Lemma~\ref{lem:modExt} then shows that we can extract bits which are \emph{computationally indistinguishable} from the uniform distribution using the class of randomness sources used to implement such a computationally-secure leakage-resilient secret sharing scheme.
In fact, the proof of Theorem~\ref{thm:final} only uses the leakage-resilience property of the secret sharing scheme in the proof of Lemma~\ref{lem:modExt}.
The remaining lemmas only make use of the correctness property of the scheme, which remains unchanged in the computational analogue of Definition~\ref{def:lrsss}.
Crucially, as shown in Section~\ref{sec:efficient}, we can construct the functions $f$, $g$, and $h$ so that they are computed by $\poly(b)$-sized circuits assuming that the sharing procedure is itself computable by $\poly(b)$-sized circuits.
Therefore, the following computational analogue of Lemma~\ref{lem:modExt}, which suffices to conclude the proof of the computational analogue of Theorem~\ref{thm:final}, holds:
``Suppose that there are functions $f, g : \bits^\ell \to \bits$ and a function 
$h : \bits^{2b} \to \bits^m$ computable by $\poly(b)$-sized circuits such that 
\begin{equation*}
    h(f(Z_1), g(Z_1)) \approx_{\delta} \unif{m}
\end{equation*}
for $\delta = \negl(\lambda)$ and for all
$(Z_1, Z_2)$ in $\mathcal{Z}$.
Then, it holds that no adversary computable by $\poly(b)$-sized circuits can distinguish $\ext(Y)$ from a uniformly random string with $Y \in \mathcal{Y}$, where $\ext(Y) = h(f(L_Y), g(R_Y))$ and $(L_y, R_y) = \Share(0^b, Y)$.''

\section{Random-less Reductions for Secret Sharing}\label{sec:reductions}


In this section, we study black-box deterministic reductions between different types of threshold secret sharing.
Such reductions from $(t',n',\eps)$-secret sharing schemes to $(t,n,\eps)$-secret sharing schemes (for the same message length $b$ and number of randomness bits $d$) would allow us to conclude that if all these $(t,n,\eps)$-secret sharing schemes require a $(\delta,m)$-extractable class of randomness sources, then so do all $(t',n',\eps)$-secret sharing schemes.
We provide reductions which work over a large range of parameters and prove complementary results showcasing the limits of such reductions.
As previously discussed,
our starting point for devising black-box reductions is the notion of a \emph{distribution design} as formalized by Stinson and Wei~\cite{SW18} (with roots going back to early work on secret sharing~\cite{BL88}), which we defined in Definition~\ref{def:distdesign}.
As stated in Lemma~\ref{lem:distred}, the existence of a $(t,n,t',n')$-distribution design yields the desired reduction from $(t',n',\eps)$-secret sharing to $(t,n,\eps)$-secret sharing.
Therefore, we focus directly on the study of distribution designs in this section.


We begin with a naive construction.
\begin{thm}\label{thm:naivered}
There exists a $(t,n,t',n')$-distribution design whenever $t'\geq t$ and $n'\geq n+(t'-t)$.
In particular, if every $(t,n,\eps)$-secret sharing scheme for $b$-bit messages and using $d$ bits of randomness requires a $(\delta,m)$-extractable class of randomness sources, then so does every $(t',n',\eps)$-secret sharing scheme for $b$-bit messages using $d$ bits of randomness whenever $t'\geq t$ and $n'\geq n+(t'-t)$.
\end{thm}
\begin{proof}
Consider the $(t,n,t',n')$-distribution design $\cD_1,\dots,\cD_n$ obtained by setting $\cD_i=\{i\}\cup\{n'-(t'-t)+1,n'-(t'-t)+2,\dots,n'\}$, which is valid exactly when the conditions of the theorem are satisfied.

\end{proof}

The following result shows the limits of distribution designs, and will be used to show the optimality of our constructions when $t=2$ or $t'=n'$.
\begin{thm}\label{thm:existdist}
A $(t,n,t',n')$-distribution design exists only if $\binom{n'}{t'-1}\geq\binom{n}{t-1}$ and $t'\geq t$.
\end{thm}
\begin{proof}
Consider an arbitrary $(t,n,t',n')$-distribution design $\cD_1,\cD_2,\dots,\cD_n$.
First, note that it must be the case that all the $\cD_i$'s are non-empty.
This implies that we must have $t'\geq t$.
Second, to see that $\binom{n'}{t'-1}\geq \binom{n}{t-1}$, consider all $\binom{n}{t-1}$ distinct subsets $\cT\subseteq[n]$ of size $t-1$, and denote $\cD_\cT=\bigcup_{i\in\cT}\cD_i$.
By the definition of distribution design, it must hold that
\begin{equation*}
    \left|\cD_\cT\right|\leq t'-1.
\end{equation*}
Consider now modified sets $\widehat{\cD_\cT}$ obtained by adding arbitrary elements to $\cD_\cT$ so that $|\widehat{\cD_\cT}|=t'-1$.
Then, from the definition of distribution design, for any two distinct subsets $\cT,\cT'\subseteq [n]$ of size $t-1$ it must be the case that
\begin{equation*}
    \left|\widehat{\cD_\cT}\cup\widehat{\cD_{\cT'}}\right|\geq t'.
\end{equation*}
This implies that $\widehat{\cD_\cT}\neq \widehat{\cD_{\cT'}}$ for all distinct subsets $\cT,\cT'\subseteq[n]$ of size $t-1$, which can only hold if $\binom{n'}{t'-1}\geq \binom{n}{t-1}$.

\end{proof}

We now show that Theorem~\ref{thm:existdist} is tight for a broad range of parameters.
In particular, when $t=2$ or $t'=n'$ we are able to characterize exactly under which parameters a $(t,n,t',n')$-distribution design exists.
\begin{thm}
There exists a $(t=2,n,t',n')$-distribution design if and only if $n\leq \binom{n'}{t'-1}$.
In particular, if every $(t=2,n,\eps)$-secret sharing scheme for $b$-bit messages using $d$ bits of randomness requires $(\delta,m)$-extractable randomness, then so does every $(t',n',\eps)$-secret sharing scheme for $b$-bit messages using $d$ bits of randomness whenever $n\leq\binom{n'}{t'-1}$.
\end{thm}
\begin{proof}
Note that the condition $n\leq \binom{n'}{t'-1}$ implies that we can take $\cD_1,\dots,\cD_n$ to be distinct subsets of $[n']$ of size $t'-1$, and so $|\cD_i\cup \cD_j|\geq t'$ for any distinct indices $i$ and $j$.
The reverse implication follows from Theorem~\ref{thm:existdist}.

\end{proof}

\begin{thm}
There exists a $(t,n,t'=n',n')$-distribution design if and only if $n'\geq\binom{n}{t-1}$.
In particular, if every $(t,n,\eps)$-secret sharing scheme for $b$-bit messages using $d$ bits of randomness requires $(\delta,m)$-extractable randomness, then so does every $(n',n',\eps)$-secret sharing scheme for $b$-bit messages using $d$ bits of randomness whenever $n'\geq\binom{n}{t-1}$.
\end{thm}
\begin{proof}
We show that a $(t,n,n',n')$-distribution design exists whenever $n'=\binom{n}{t-1}$, which implies the desired result.
Let $\cP$ denote the family of all subsets of $[n]$ of size $t-1$, and set $n'=|\cP|=\binom{n}{t-1}$ (we may use any correspondence between elements of $\cP$ and integers in $[n']$).
Then, we define the set $\cD_i\subseteq \cP$ for $i\in[n]$ to contain all elements of $\cP$ except the subsets of $[n]$ which contain $i$.
We argue that $\cD_1,\dots,\cD_n$ is a distribution design with the desired parameters.
First, observe that for any distinct indices $i_1,i_2,\dots,i_{t-1}\in[n]$ it holds that
\begin{equation*}
    \bigcup_{j=1}^{t-1}\cD_{i_j}=\cP\setminus\{\{i_1,i_2,\dots,i_{t-1}\}\}.
\end{equation*}
On the other hand, since $\{i_1,\dots,i_{t-1}\}\in\cD_{i_t}$ for any index $i_t\neq i_1,\dots,i_{t-1}$, it follows that $\bigcup_{j=1}^t\cD_{i_j}=\cP$, as desired.

The reverse implication follows from Theorem~\ref{thm:existdist}.

\end{proof}

\subsection{Distribution Designs from Partial Steiner Systems}

In this section, we show that every partial Steiner system is also a distribution design which beats the naive construction from Theorem~\ref{thm:naivered} for certain parameter regimes.
Such set systems have been previously used in seminal constructions of pseudorandom generators and extractors~\cite{NW94,Tre01}, and are also called combinatorial designs.
\begin{defn}[Partial Steiner system]
We say a family of sets $\cD_1,\dots,\cD_n\subseteq[n']$ is an \emph{$(n,n',\ell,a)$-partial Steiner system} if it holds that $|\cD_i|=\ell$ for every $i\in[n]$ and $|\cD_i\cap\cD_j|\leq a$ for all distinct $i,j\in[n]$.
\end{defn}
The conditions required for the existence of a partial Steiner system are well-understood, as showcased in the following result from~\cite{EFF85,NW94,Tre01}, which is nearly optimal~\cite{Rod85,RRV02}.
\begin{lem}[\cite{EFF85,NW94,Tre01}]\label{lem:existSteiner}
Fix positive integers $n$, $\ell$, and $a\leq \ell$.
Then, there exists an $(n,n',\ell,a)$-partial Steiner system for every integer $n'\geq e\cdot n^{1/a}\cdot \frac{\ell^2}{a}$.
\end{lem}
Noting that every partial Steiner system with appropriate parameters is also a distribution design, we obtain the following theorem.
\begin{thm}\label{thm:distfromSteiner}
    Fix an integer $a\geq 1$.
    Then, there exists a $(t,n,t',n')$-distribution design whenever $t'\geq t^2+\frac{at(t-1)^2}{2}$ and $n'\geq \frac{e n^{1/a}}{a}\cdot\left(1+\frac{t'}{t}+\frac{a(t-1)}{2}\right)^2$.
\end{thm}
\begin{proof}
Fix an integer $a\geq 1$ and an $(n,n',\ell,a)$-partial Steiner system $\cD_1,\dots,\cD_n\subseteq [n']$ with $\ell=\left\lceil\frac{t'}{t}+\frac{a(t-1)}{2}\right\rceil$.
By Lemma~\ref{lem:existSteiner} and the choice of $\ell$, such a partial Steiner system is guaranteed to exist whenever $n'$ satisfies the condition in the theorem statement.
We proceed to argue that this partial Steiner system is also a $(t,n,t',n')$-distribution design.
First, fix an arbitrary set $\cT\subseteq[n]$ of size $t-1$.
Then, we have
\begin{equation*}
    \left| \cD_\cT\right|\leq \ell(t-1)\leq t'-1,
\end{equation*}
where the rightmost inequality holds by our choice of $\ell$ and the condition on $t'$ and $t$ in the theorem statement.
Second, fix an arbitrary set $\cT\subseteq[n]$ of size $t$.
Then, it holds that
\begin{align*}
    \left| \cD_\cT\right|&\geq \ell+(\ell-a)+(\ell-2a)+\cdots+(\ell-a(t-1))\\
    &=\ell\cdot t-\frac{at(t-1)}{2}\\
    &\geq t',
\end{align*}
where the last equality follows again from our choice of $\ell$ and the condition on $t'$ and $t$ in the theorem statement.

\end{proof}

When $n$ is sufficiently larger than $t$ and $t'$ and $t'$ is sufficiently larger than $t$, the parameters in Theorem~\ref{thm:distfromSteiner} cannot be attained by the naive construction from Theorem~\ref{thm:naivered}, which always requires choosing $t'\geq t$ and $n'\geq n$.
For example, if $t^3\leq t'\leq C t^3$ for some constant $C\geq 1$ then we can choose $a=2$, in which case we have
\begin{equation}\label{eq:condtprime}
    t^2+\frac{at(t-1)^2}{2}\leq t^3\leq t'.
\end{equation}
Moreover, it holds that
\begin{align}
    \frac{e n^{1/a}}{a}\cdot\left(1+\frac{t'}{t}+\frac{a(t-1)}{2}\right)^2 &\leq \frac{e \sqrt{n}}{2}\cdot\left(Ct^2+t\right)^2\nonumber\\
    &\leq 2eC^2 \sqrt{n} t^4.\label{eq:boundnprime}
\end{align}
Combining~\eqref{eq:condtprime} and~\eqref{eq:boundnprime} with Theorem~\ref{thm:distfromSteiner}, 
we obtain the following example result showing it is possible to improve on Theorem~\ref{thm:naivered} in some parameter regimes.
\begin{coro}
    Suppose $t^3\leq t'\leq Ct^3$ for some constant $C\geq 1$.
    Then, there exists a $(t,n,t',n')$-distribution design for any $n'\geq 2eC^2\sqrt{n}t^4$.
    In particular, if $t\leq n^{1/9}$ and $n$ is large enough, we may choose $n'$ significantly smaller than $n$.
\end{coro}

\chapter{Non-malleable Extraction, and {Privacy Amplification}}
In this chapter (based on the paper \cite{ACO23}) we turn our attention to non-malleable cryptography. Expanding again on adversaries that may not respect the boundaries on what is information or states are accessible during the design of cryptographic primitives, while some adversaries may choose to be passive by only collecting information, others may take on a more active role by tampering with public communication between parties, execution states, or the computation of the primitives themselves. Indeed, there are certain real life attacks that work by flipping RAM bits (and therefore tamper with states/memory) on targeted machines \cite{KDKFLLWLM14}, or heating up devices to cause hardware faults that influence execution of the protocol \cite{BDL01}. To address this, non-malleable cryptography addresses adversaries that are allowed to actively tamper or ``mall'' certain states or stored information in an attempt to influence the outcome of the protocol in a meaningful manner. Here are a few examples in non-malleable cryptography:
\emph{Non-malleable encryption} was firstly introduced by Dolev, Dwork, and Naor in \cite{DDN91} where the adversary is allowed to actively tamper with or ``mall'' ciphertexts into other ciphertexts that can be decrypted to related plaintexts. Boaz Barak gave a non-malleable commitment scheme in \cite{Bar02} which allows a party to commit to a value to eventually reveal it where an adversary is allowed to tamper with messages used in the communication. Dziembowski, Pietrzak, and Wichs in \cite{DPW18} introduce non-malleable codes, which are an extension of error-correcting codes, where one can think of the error as being adversarially chosen, and the decoding algorithm must either detect potential tampering or output something unrelated to the initially encoded message. We note that the aforementioned primitives are extentions of pre-existing primitives, namely: public-key encryption, commitment schemes, and error-correcting codes.

Along similar lines, we turn our attention to studying extractors, and their non-malleable extensions. The problem of constructing efficient two-source extractors for low min-entropy sources with negligible error has been an important focus of research in pseudorandomness for more than 30 years, with fundamental connections to combinatorics and many applications in computer science. The first non-trivial construction was given by Chor and Goldreich~\cite{CG88} who showed that the inner product function is a low-error two-source extractor for $n$-bit sources with min-entropy $(1/2+\gamma)n$, where $\gamma>0$ is an arbitrarily small constant. A standard application of the probabilistic method shows that (inefficient) low-error two-source extractors exist for polylogarithmic min-entropy. While several attempts were made to improve the construction of~\cite{CG88} to allow for sources with smaller min-entropy, the major breakthrough results were obtained after almost two decades. Raz~\cite{R05} gave an explicit low-error two-source extractor where one of the sources must have min-entropy $(1/2+\gamma)n$ for an arbitrarily small constant $\gamma>0$, while the other source is allowed to have logarithmic min-entropy. In an incomparable result, Bourgain~\cite{Bou05} gave an explicit low-error two-source extractor for sources with min-entropy $(1/2-\gamma)n$, where $\gamma>0$ is a small constant. An improved analysis by Lewko~\cite{Lew19} shows that Bourgain's extractor can handle sources with min-entropy $4n/9$.

\paragraph{((Seeded) non-malleable extractors.)} The problem of privacy amplification against active adversaries was first considered by Maurer and Wolf~\cite{MW97}. In a breakthrough result, Dodis and Wichs~\cite{DW09} introduced the notion of seeded non-malleable extractors as a natural tool towards achieving a privacy amplification protocol in a minimal number of rounds, and with minimal entropy loss. Roughly speaking, the output of a seeded non-malleable extractor with a uniformly random seed $Y$, and a source $X$ with some min-entropy independent of $Y$, should look uniformly random to an adversary who can tamper the seed, and obtain the output of the non-malleable extractor on a tampered seed. 

More precisely, we require that
\[
    \bigg(\mathbf{nmExt}(X,Y),\mathbf{nmExt}(X,g(Y)), Y \bigg)
\approx_\eps \bigg(U_m, \mathbf{nmExt}(X,g(Y)), Y\;\bigg),
\]
where $X$ and $Y$ are independent sources with $X$ having sufficient min-entropy and $Y$ uniformly random, $g$ is an arbitrary tampering function with no fixed points, $U_m$ is uniform over $\bit^m$ and independent of $X, Y$, and $\approx_\eps$ denotes the fact that the two distributions are $\eps$-close in statistical distance (for small $\eps$). 

Prior works have also studied seeded extractors with weaker non-malleability guarantees such as look-ahead extractors~\cite{DW09} or affine-malleable extractors~\cite{AHL16}, and used these to construct privacy amplification protocols. 

\paragraph{(Non-malleable two-source extractors.)}
A natural strengthening of both seeded non-malleable extractors, and two-source extractors are two-source \emph{non-malleable} extractors.  Two-source non-malleable extractors were introduced by Cheraghchi and Guruswami~\cite{CG17}. Roughly speaking, a function $\nmExt:\bit^n\times\bit^n\to\bit^m$ is said to be a non-malleable extractor if the output of the extractor remains close to uniform (in statistical distance), even conditioned on the output of the extractor inputs correlated with the original sources.
In other words, we require that
\[
    \bigg(\nmExt(X,Y),\nmExt(f(X),g(Y)),Y\bigg)
    \approx_\eps \bigg(U_m, \nmExt(f(X),g(Y)),Y\bigg) \;.
\]
where $X$ and $Y$ are independent sources with enough min-entropy, $f, g$ are arbitrary tampering functions such that one of $f, g$ has no fixed points. 

The original motivation for studying efficient two-source non-malleable extractors stems from the fact that they directly yield explicit split-state non-malleable codes~\cite{DPW18} (provided the extractor also supports efficient preimage sampling). 

The first constructions of non-malleable codes~\cite{DKO13,ADL18} relied heavily on the (limited) non-malleability of the inner-product two-source extractor. Subsequent improved constructions of non-malleable codes in the split-state model relied on both the inner-product two-source extractor~\cite{ADKO15a,AO20}, and on more sophisticated constructions of the two-source non-malleable extractors~\cite{CGL20,Li17,L20}.  Soon after they were introduced, non-malleable extractors have found other applications such as non-malleable secret sharing~\cite{GK18,ADNOPRS19}.

\paragraph{(Connections, and state-of-the-art constructions.)} As one might expect, the various notions of extractors mentioned above are closely connected to each other. Li~\cite{Li12} obtained the first connection between seeded non-malleable extractors and two-source extractors based on inner products. This result shows that an improvement of Bourgain's result would immediately lead to better seeded non-malleable extractors, and a novel construction of seeded non-malleable extractors with a small enough min-entropy requirement and a small enough seed size would immediately lead to two-source extractors that only require small min-entropy. However,~\cite{Li12} could only obtain seeded non-malleable extractors for entropy rate above $1/2$. 

In yet another breakthrough result,~\cite{CGL20} obtained a sophisticated construction of seeded non-malleable extractors for polylogarithmic min-entropy. Additionally, they showed that similar techniques can also be used to obtain two-source non-malleable extractors. This immediately led to improved privacy amplification protocols and improved constructions of non-malleable codes in the split-state model. Building on this result, in a groundbreaking work, Chattopadhyay and Zuckerman~\cite{CZ19} gave a construction of two-source extractors with polylogarithmic min-entropy and polynomiallly small error. All of these results have subsequently been improved in~\cite{Li16,BDT17,C17,Li17,L20}. We summarize the parameters of the best known constructions of seeded extractors, two-source extractors, seeded non-malleable extractors, and two-source non-malleable extractors alongside those of our construction in Table~\ref{table:main}. We note here that all prior constructions of two-source non-malleable extractors required both sources to have almost full min-entropy. A recent result~\cite{GSZ21} has not been included in this table since it constructs a weaker variant of a non-malleable two-source extractor (that does not fulfil the standard definition) that is sufficient for their application to network extraction. Even if one is willing to relax the definition to that in~\cite{GSZ21}, the final parameters of our two-source non-malleable extractor are better!

The research over the past few years has shown that non-malleable two-source extractors, seeded non-malleable extractors, two-source extractors, non-malleable codes, and privacy amplification protocols are strongly connected to each other in the sense that improved construction of one of these objects has led to improvements in the construction of others. Some results have made these connections formal by transforming a construction of one object into a construction of another object. For instance, in addition to the connections already mentioned, Ben-Aroya et al.\ \cite{BACDLTS18} adapt the approach of~\cite{CZ19} to show explicit \emph{seeded} non-malleable extractors with improved seed length lead to explicit low-error two-source extractors for low min-entropy.

Also,~\cite{AORSS20} showed that some improvement in the parameters of non-malleable two-source extractor constructions from~\cite{CGL20,Li17,L20} leads to explicit low-error two-source extractors for min-entropy $\delta n$ with a very small constant $\delta>0$.

\begin{table}[]
    \centering 
    \begin{tabular}{|l|l|l|l|l|l|l|}
    \hline
    Citation                                                                                        & Left Rate                                  & Right Rate                                                & Non-malleability                                          \\\hline
       \textbf{Seeded} \\\hline
        \begin{tabular}[c]{@{}l@{}}\cite{RRV02} Theorem 1\end{tabular}                                & $\text{polylog}(n) / n$                                 & $1$                                  & None                                                 \\\hline
    \begin{tabular}[c]{@{}l@{}}\cite{GUV09} Theorem 4.17\end{tabular}                             & $\log (n) / n$                                 & $1$                                  & None                                                 \\\hline
        \textbf{Seeded, Non-malleable} \\\hline
        \cite{Li12}                                                                                    & $1/2 - \gamma$                             & $1$                                      & Right source                                      \\\hline
    \cite{CRS14}                                                                                    & $1/2 + \gamma$                             & $1$                                      & Right source                                      \\\hline
    \begin{tabular}[c]{@{}l@{}}\cite{DLWZ14} Theorem 1.4\end{tabular}                             & $1/2 + \gamma$                             & $1$                                           & Right source               \\\hline
       \cite{CGL20} & $\log^2 n/n$                     & $1$                                    & Right-source                               \\\hline
    \begin{tabular}[c]{@{}l@{}}\cite{Li17} Theorem 6.2\end{tabular}                               & $\log(n)/n$                                 & $1$                                    & Right source       \\\hline
    \begin{tabular}[c]{@{}l@{}}\cite{L20}\end{tabular}                              & $\log(n)/n$                               & $1$                                              & Right-source        \\\hline     
    \textbf{Two-source} \\\hline
    \cite{CG88}                                                                                     & $1/2$                                      & $1/2$                                                     & None                                               \\\hline
    \cite{Bou05}                                                                                    & $1/2 - \gamma$                             & $1/2 - \gamma$                                            & None                                               \\\hline
    \cite{R05}                                                                                                                & $\log(n)/n$                 & $1/2 + \gamma$                               & None                                      \\\hline

    \textbf{Two-source, Non-malleable} \\\hline
    \cite{CGL20} & $1-\frac{1}{n^\gamma}$                     & $1-\frac{1}{n^\gamma}$                                    & Two-sided                               \\\hline
      \begin{tabular}[c]{@{}l@{}}\cite{Li17}\end{tabular}                              & $(1-\gamma)$                               & $(1-\gamma)$                                              & Two-sided        \\\hline     
    \begin{tabular}[c]{@{}l@{}}\cite{L20} Theorem 1.11\end{tabular}                              & $(1-\gamma)$                               & $(1-\gamma)$                                              & Two-sided        \\\hline     
    \color{red}{This Work }                                                                                                         &\color{red}{ $\text{polylog}(n) / n$     }              & \color{red}{ $4/5 + \gamma$}                    & \color{red}{Two-sided} \\\hline
\end{tabular}
\caption{In the table, we assume that the left source has length $n$, and $\gamma$ is a very small universal constant that has a different value for different results. Most of the constructions two-source non-malleable extractors including ours allow for $t$-time tampering at the cost of a higher min-entropy requirement. In particular (as described in Remark \ref{multitamper1}, \ref{multitamper2}, and \ref{multitamper3}), for our extractor we require the left source to have min-entropy rate $\text{polylog}(n)/n$, and the right source has min-entropy rate  $(1-\frac{1}{2t+3})$.}
\label{table:main}
\end{table}

Parameters for each extractor were chosen such that the error is $2^{-\kappa^c}$ and the output length is $\Omega(\kappa)$ for some constant $c$, where $\kappa$ is the amount of entropy in the left source.

\paragraph{(Best of all worlds.)}

Notice that the seeded non-malleable extractor, and the two-source extractors can be seen as special case of a two-source non-malleable extractor. With this view, the known constructions of negligible error (non-malleable) two-source extractors can be broadly classified in three categories:
\begin{itemize}
\item Constructions where one source has min-entropy rate about $1/2$, the other source can have small min-entropy rate, but the extractor doesn't guarantee non-malleability.
    \item Constructions where one source is uniform, and the other can have small min-entropy rate, and the extractor guarantees non-malleability when the uniform source is tampered.
    \item Constructions where both sources have entropy rate very close to $1$ and the extractor guarantees non-malleability against the tampering of both sources. 
\end{itemize}

The main focus of this work is the question whether we can have one construction that subsumes all the above constructions. 

\begin{question}
Is there an explicit construction of a two-source non-malleable extractor which requires two sources of length $n_1$ and $n_2$, and min-entropy requirement $c n_1$ (for some constant $c < 1$), and $\poly \log n_2$, respectively, that guarantees non-malleability against the tampering of both sources, and for which the error is negligible? In particular, can we obtain a construction with parameters suitable for application to privacy amplification with tamperable memory~\cite{AORSS20}?
\end{question}

In this work, we make progress towards answering this question.

\paragraph{(Applications of two-source non-malleable extractors.)} Two-source non-malleable  extractors have in the recent years attracted a lot of attention, and have very quickly become fundamental objects in cryptosystems involving communication channels that cannot be fully trusted.  As we discussed earlier, two-source non-malleable extractors have applications in the construction of non-malleable codes, and in constructing two-source extractors. The other primary applications of two-source non-malleable extractors include non-malleable secret sharing~\cite{GK18,ADNOPRS19}, non-malleable commitments~\cite{GPR16}, network extractors~\cite{GSZ21}, and privacy amplification~\cite{CKOS19,AORSS20}.  

In particular, in~\cite{AORSS20}, the authors introduce an extension of privacy amplification (PA) against active adversaries where, Eve as the active adversary is additionally allowed to \emph{fully corrupt} the internal memory of one of the honest parties, Alice and Bob, before the execution of the protocol.  Their construction required two-source non-malleable extractors with one source having a small entropy rate $\delta$ (where $\delta$ is a constant close to $0$). Since no prior construction of two-source non-malleable extractor satisfied these requirements, the authors constructed such extractors under computational assumptions and left the construction of the information-theoretic extractor with the desired parameters as an open problem. Our construction in this work resolves this open problem. We do not include here the details of the PA protocol due to space constraints. We refer the reader to~\cite{AORSS20} for the PA protocol.

\paragraph{(Subsequent work.)} Li, inspired by our work and that of~\cite{GSZ21}, in \cite{L23} gives a two-source non-malleable extractor construction
with $\frac{2}{3}$-rate entropy in one source and $\frac{\log(n)}{n}$-rate entropy in the other. Based on the proof sketch in~\cite{L23}, the key idea of the construction and proof seems similar, the fundamental difference being the use of an  correlation breaker with advice instead of a collision resistant extractor.

\section{Chapter Organisation and Our Results}
We build two-source non-malleable extractors, with one source having polylogarithmic min-entropy, and the other source having min-entropy rate $0.81$. We introduce collision-resistant extractors, and extend and improve efficiency of the privacy amplification protocol from~\cite{AORSS20}. The following is a roadmap of this chapter.  
\begin{itemize}
    \item In Section~\ref{sec:overview}, we give an overview of our technical details. 
    
\item In Section~\ref{sec:generic-reduction}, we give a generic transformation that, takes in (1) a non-malleable two-source extractor which requires sources with high min-entropy, and (2) a two-source extractor which requires sources with smaller min-entropy and an additional collision-resistance property, and constructs a two-source non-malleable extractor with min-entropy requirement comparable to (but slightly worse) that of the two-source extractor used by the construction. 

\item In Section~\ref{sec:coll-res-generic}, we give a generic transformation that converts any seeded extractor (two-source extractor where one of the source is uniformly distributed)  to a collision-resistant seeded extractor with essentially the same parameters.

\item In Section~\ref{sec:coll-res-raz}, we show that the two-source extractor from~\cite{R05} is collision resistant.

\item In Section~\ref{sec:fullynmseeded}, we apply our generic transformation from Section~\ref{sec:coll-res-generic} to the seeded extractor from~\cite{RRV02} to obtain a collision-resistant seeded extractor. We then use the generic transformation from Section~\ref{sec:generic-reduction} along with the non-malleable extractor from~\cite{L20} to obtain a two-source non-malleable extractor, where one of the source is uniform and the other has  min-entropy polylogarithmic in the length of the sources.

\item In Section~\ref{sec:nmraz}, we apply the generic transformation from Section~\ref{sec:generic-reduction} to the non-malleable extractor from Section~\ref{sec:fullynmseeded}, and the two-source extractor from~\cite{R05} to obtain a two-source non-malleable extractor where one source is required to have polylogarithmic min-entropy and the source is required to have min-entropy rate greater than $0.8$.

\item Applications:
\begin{itemize}
\item In Section~\ref{sec:ratehalfnm}, we use a generic transformation from~\cite{AKOOS21} to obtain a non-malleable two-source extractor where the length of the output is $1/2 - o(1)$ times the length of the input. Notice that via the probabilistic method, it can be shown that the output length of this construction is optimal. \footnote{The main drawback of this construction compared to the construction from Section~\ref{sec:nmraz} is that this is not a strong two-source non-malleable extractor, and hence cannot be used in most applications.}

\item In Section~\ref{App:PA}, we sketch the details of the privacy amplification protocol that uses our non-malleable two-source extractor.  We extend the protocol by \cite{AORSS20} to obtain a secret of optimal size while  maintaining security against a memory tampering adversary.
\end{itemize}
\end{itemize}

\section{Preliminaries}

\begin{lemma}\label{lma:randDistinguisher}
    Let $X, Y$ be random variables. Further let $f_{\mathcal{I}}$ be a family of functions $f$ indexed by set $\mathcal{I}$ and let $S$ be a random variable supported on $\mathcal{I}$ that is independent of both $X$ and $Y$. Then $f_S$ can be thought of as a randomised function such that $f_S(x) = f_s(x)$ with probability $\Pr[S = s]$.

    Then it holds that:
    \begin{equation*}
        \dist{f_S(X)}{f_S(Y)} \leq \dist{X}{Y}.
    \end{equation*}
\end{lemma}

\begin{lemma}[Lemma 4 of \cite{DDV10}, Lemma 9 of \cite{ADKO15a}]\label{lma:reveal}
    Let $A, B$ be independent random variables and consider a sequence $\blocks{V}{i}$ of random variables, where for some function $\phi$, $V_i = \phi_i(C_i) = \phi(\blocks{V}{i-1}, C_i)$ with each $C_i \in \{A, B\}$. Then $A$ and $B$ are independent conditioned on $V_1, \ldots, V_i$. That is, $\mutualInfoCond{A}{B}{\blocks{V}{i}} = 0$.
\end{lemma}

\begin{definition}\label{defn:biasedAgainstTests}
    Call a sequence of variables $\blocks{Z}{N}$ $(k, \eps)$-biased against linear tests if for any non-empty $\tau \subseteq [N]$ such that $\lvert \tau \rvert \leq k$, $\lvert \Pr[\bigoplus_{i \in \tau} Z_i = 0] - \frac{1}{2} \rvert \leq \eps$.
\end{definition}

\begin{lemma}[Theorem 2 of \cite{AGHP02}]\label{lma:linearTests}
    Let $N = 2^t - 1$ and let $k$ be an odd integer. Then it is possible to construct $N$ random variables $Z_i$ with $i \in [N]$ which are $(k, \eps)$-biased against linear tests using a seed of size at most $2 \lceil \log(1/\eps) + \log \log N + \log k \rceil + 1$ bits.
\end{lemma}

\subsection{Rejection Sampling for Extractors}
In this section we present two lemmas that use rejection sampling to lower the entropy requirement for strong two-source extractors and their collision resistance.

We first define a sampling algorithm $\samp$ that given
a flat distribution $Y' \givenBy (n, k)$, tries to approximate some distribution $Y \givenBy (n, k - \delta)$ (with $supp(Y) \subseteq supp(Y')$). Letting $d = \max_{y \in supp(Y)} \left \{ \frac{\Pr[Y = y]}{\Pr[Y' = y]} \right \}$:

\begin{equation*}
    \samp(y) = \begin{cases}
        y, & w.p.\ \frac{\Pr[Y = y]}{d \cdot \Pr[Y' = y]}\\
        \bot, &else 
    \end{cases}
\end{equation*}

\begin{lemma}\label{lma:rejectionSample}
    The probability $\samp(Y') = y$ is $\frac{\Pr[Y = y]}{d}$ and furthermore, the probability that $\samp(Y') \neq \bot$ is $\frac{1}{d}$. Consequently, the distribution $\samp(Y')$ conditioned on the event that $\samp(Y') \neq \bot$ is identical to $Y$.
\end{lemma}

\begin{proof}
    Letting $\samp$ and $d$ be defined as above, then:
    \begin{equation*}
        \Pr[\samp(Y') = y] = \frac{1}{d} \frac{\Pr[Y = y]}{\Pr[Y' = y]} \cdot \Pr[Y' = y] = \frac{\Pr[Y = y]}{d}
    \end{equation*}

    Then it follows that:
    \begin{equation*}
        \Pr[\samp(Y') \neq \bot] = \sum_{y} \Pr[\samp(Y) = y] = \sum_{y} \frac{\Pr[Y = y]}{d} = \frac{1}{d}
    \end{equation*}

    Thus, conditioned on the event that $\samp(Y') \neq \bot$, $\samp(Y')$ is the distribution $Y$.
    \begin{align*}
        \Pr[\samp(Y') = y | \samp(Y') \neq \bot] &= \frac{\Pr[\samp(Y') \neq \bot |\samp(Y') = y ] \Pr[\samp(Y') = y]}{\Pr[\samp(Y') \neq \bot]} \\&= \Pr[Y = y]
    \end{align*}
\end{proof}

\paragraph{Lowering the Entropy Requirement for Strong Two-Source Extractors.}
We will use rejection sampling now to show that 
we can lower entropy requiements for extractors,
at the cost of increased statistical distance in its output from the full uniform distribution.

\begin{lemma}\label{lma:extReduction}
    Let $\extractorParam{\ext}{(n_1, k_1)}{(n_2, k_2)}{m}{\eps}$ be a strong two-source extractor using input distributions $X$ and $Y'$. Then letting $Y \givenBy (n_2, k_2 - \delta)$:
    
    \begin{equation*}
        \distCond{\ext(X, Y)}{\unif{m}}{Y} \leq 2^\delta \eps
    \end{equation*}
\end{lemma}

\begin{proof}
    Assume by contradiction that there exists a distribution
    $Y \givenBy (n, k -\delta)$ for which $\distCond{\ext(X, Y)}{\unif{m}}{Y} > 2^\delta \eps$, i.e. there exists a distinguisher $A : \bit^m \to \bit$ such that $\lvert \Pr[A(\ext(X, Y), Y) = 1] - \Pr[A(\unif{m}, Y) = 1] \rvert > 2^\delta \eps$. We want to use this fact to 
    create a distinguisher $D$ that distinguishes $\ext(X, Y')$
    from $\unif{m}$ \emph{for some distribution $Y'\givenBy (n, k)$}. Note that 
    $Y$ can be expressed as a convex combination
    of $(n, k - \delta)$ flat distributions,  i.e. $Y = \sum_i \alpha_i Y_i$.
    We define $Y'$ in the following way:
    $Y'$ is a convex combination of flat 
    distributions $Y'_i$ where each $Y'_i$
    is some $(n, k)$ flat distribution such that
    $supp(Y_i) \subseteq supp(Y'_i)$. We first note
    that for all $y \in supp(Y)$:
    \begin{equation*}
        \frac{Pr[Y = y]}{Pr[Y' = y]} = \frac{\sum_i \alpha_i Pr[Y_i = y] }{ \sum_i \alpha_i Pr[Y'_i = y] }
        \leq \frac{2^{-k + \delta}}{2^{-k}} \leq 2^\delta
    \end{equation*}
    Furthermore, note that $Y'$ has min-entropy $k$. To see this, note that for any $y \in supp(Y)$:
    \begin{equation*}
        \Pr[Y' = y] = \sum_i \alpha_i \Pr[Y'_i = y]
        \leq \sum_i \alpha_i 2^{-k}
        \leq 2^{-k}
    \end{equation*}
    
    Let 
    $\samp$ be a rejection sampler that on input distribution $Y$, samples for $Y'$.
    Now, $D$ is defined as follows:
    \begin{equation*}
        D(Z, Y') = \begin{cases}
                        A(Z, Y') &,\ \text{if }\samp(Y') \neq \bot\\
                        1 &,\ w.p.\ \frac{1}{2},\ \text{if }\samp(Y') = \bot\\
                        0 &,\ else
                    \end{cases}
    \end{equation*}
    Note that by Lemma \ref{lma:rejectionSample}, $\Pr[\samp(Y') \neq \bot] \geq \frac{1}{2^\delta}$ and $\samp(Y')$ is identical to $Y$ conditioned on the event that $\samp(Y') = \bot$. Then the advantage that $D$ distinguishes between $\ext(X, Y')$ and $\unif{m}$ given $Y'$ is given as:
    \begin{align*}
        &\lvert \Pr[D(\ext(X, Y'), Y') = 1] - \Pr[D(\unif{m}, Y') = 1] \rvert\\
        &\geq \Pr[\samp(Y) \neq \bot] \lvert \Pr[A(\ext(X, Y), Y) = 1]
         - \Pr[A(\unif{m}, Y) = 1] \rvert\\
        &>\frac{1}{2^\delta} 2^\delta \eps = \eps
    \end{align*}
    Which in turn implies that $\distCond{\ext(X, Y')}{\unif{m}}{Y'} > \eps$, which implies the desired contradiction.
\end{proof}

\paragraph{Lowering the Entropy Requirement for Collision Resistance in Extractors.}
Similarly, it will be useful for us to first show that
we can lower the entropy requirement (again, at the cost 
of increased distance in the output from the uniform 
distribution) 
\begin{lemma}\label{lma:colReduction}
    Let $\extractorParam{\ext}{(n_1, k_1)}{(n_2, k_2)}{m}{\eps}$ be a strong two-source extractor using input distributions $X$ and $Y'$ that has collision probability $\collisionError$. Then letting $Y \givenBy (n_2, k_2 - \delta)$ and $f$ be any fixed-point-free function:
    
    \begin{equation*}
        \Pr[\ext(X, Y) = \ext(f(X), Y)] \leq 2^\delta \collisionError
    \end{equation*}
\end{lemma}
\begin{proof}
    For the sake of contradiction, $Y$ be any $(n, k - \delta)$ distribution for which
    the collision probability is at least 
    $2^{\delta} \cdot \collisionError$.
    
    Note that 
    $Y$ can be expressed as a convex combination
    of $(n, k - \delta)$ flat distributions, i.e. $Y = \sum_i \alpha_i Y_i$.
    We define $Y'$ in the following way:
    $Y'$ is a convex combination of flat 
    distributions $Y'_i$ where each $Y'_i$
    is some $(n, k)$ flat distribution such that
    $supp(Y_i) \subseteq supp(Y'_i)$. We first note
    that for all $y \in supp(Y)$:
    \begin{align*}
        \frac{Pr[Y = y]}{Pr[Y' = y]} &= \frac{\sum_i \alpha_i Pr[Y_i = y] }{ \sum_i \alpha_i Pr[Y'_i = y] } \\
        &\leq \frac{2^{-k + \delta}}{2^{-k}} \leq 2^\delta
    \end{align*}
    Furthermore, note that $Y'$ has min-entropy $k$. To see this, note that for any $y \in supp(Y)$:
    \begin{equation*}
        \Pr[Y' = y] = \sum_i \alpha_i \Pr[Y'_i = y]
        \leq \sum_i \alpha_i 2^{-k}
        \leq 2^{-k}
    \end{equation*}

    Let $\samp$ be a rejection sampler that on input distribution
    $Y'$, samples for $Y$. By the collision resilience property of $\ext$, it follows that:
    \begin{align*}
        \collisionError &\geq \Pr[ \ext(X, Y') = \ext(f(X), Y') ]\\
                        &\geq \Pr[ \ext(X, Y) = \ext(f(X), Y) | \samp(Y) \neq \bot ] \Pr[ \samp(Y) \neq \bot ]\\
                        &= \Pr[ \ext(X, Y) = \ext(f(X), Y)] 2^{-\delta}
    \end{align*}
\end{proof}

\subsection{Extractors}
\begin{definition}[Two Source Non-malleable Extractor]\label{defn:twosourceNMext}
    Call $\extractorParam{\ext}{(n_1, k_1)}{(n_2, k_2)}{m}{\eps}$ a \emph{two source non-malleable extractor} if additionally for any pair of functions $f : \bit^{n_1} \to \bit^{n_1}$, $g : \bit^{n_2} \to \bit^{n_2}$ such at least one of $f, g$ is fixed-point-free\footnote{A function $f$ is said to be fixed-point-free if for any $x$, $f(x) \neq x$}, the following holds:

    \begin{equation*}
        \distCond{\ext(X, Y)}{\unif{m}}{\ext(f(X), g(Y))} \leq \eps
    \end{equation*}

 Additionally, we call the extractor $\ext$ a \emph{right strong non-malleable two-source extractor} if:
    \begin{equation*}
        \distCond{\ext(X, Y)}{\unif{m}}{\ext(f(X), g(Y)),Y} \leq \eps \;,
    \end{equation*}
    and 
    we call the extractor $\ext$ a \emph{left strong non-malleable two-source extractor} if:
    \begin{equation*}
        \distCond{\ext(X, Y)}{\unif{m}}{\ext(f(X), g(Y)),X} \leq \eps \;,
    \end{equation*}
\end{definition}

\begin{definition}[(Fully) Non-malleable Seeded Extractor]
    Call $\extractorParam{\ext}{(n_1, k_1)}{(n_2, n_2)}{m}{\eps}$ a \emph{non-malleable seeded extractor} if additionally for some fixed-point-free function $g : \bit^{n_2} \to \bit^{n_2}$, the following holds:
    
    \begin{equation*}
        \distCond{\ext(X, Y)}{\unif{m}}{\ext(X, g(Y))} \leq \eps
    \end{equation*}

    A natural strengthening of a non-malleable seeded extractor is to consider a pair of tampering functions on both its inputs rather than on just the seed. Thus call a $\ext$ a \emph{fully non-malleable seeded extractor} if additionally for some pair of fixed-point-free functions $g : \bit^{n_2} \to \bit^{n_2}$, and $f : \bit^{n_1} \to \bit^{n_1}$, the following holds:
    
    \begin{equation*}
        \distCond{\ext(X, Y)}{\unif{m}}{\ext(f(X), g(Y))} \leq \eps
    \end{equation*}
\end{definition}

One useful thing to note is that the extractor remains non-malleable even if the functions $f, g$ are randomised with shared coins (independent of $X$ and $Y$).
\begin{lemma}\label{lma:randTamp}
    Let $\ext$ be a two source non-malleable extractor for $(n, k)$-sources $X, Y$ with output length $m$ and error $\eps$. Let $f_S, g_S$ random functions over the shared randomness of $S$ independent of $X$ and $Y$ such that for all $s \in \supp{S}$, at at least one of $f_s$ or $g_s$ is fixed-point-free. Then
    \begin{equation*}
        \distCond{\ext(X, Y)}{\unif{m}}{\ext(f_S(X), g_S(Y))} \leq \eps
    \end{equation*}
\end{lemma}
\begin{proof}
    Let $X \givenBy (n, k)$ and $Y \givenBy (n, k)$ be independent sources. Let $f_S, g_S : \bit^n \to \bit^n$ be fixed-point-free random functions over the randomness of $S$ which is independent of $X$ and $Y$.    
    \begin{align*}
        &\distCond{\ext(X, Y)}{\unif{m}}{\ext(f_S(X), g_S(Y))} \\
        &= \sum_{a, b} \lvert \Pr[\ext(X, Y) = a, \ext(f_S(X), g_S(Y)) = b] - \Pr[\unif{m}, \ext(f_S(X), g_S(Y)) = b] \rvert\\
        &= \sum_{a, b} \lvert \sum_s \Pr[S = s] \Pr[\ext(X, Y) = a, \ext(f_S(X), g_S(Y)) = b | S = s]\\
        & \hspace*{2em} - \sum_s \Pr[S = s] \Pr[\unif{m}, \ext(f_S(X), g_S(Y)) = b | S = s]\rvert\\
        &= \sum_{a, b} \sum_s \Pr[S = s] \lvert \Pr[\ext(X, Y) = a, \ext(f_S(X), g_S(Y)) = b | S = s] \\
        & \hspace*{2em}- \Pr[\unif{m}, \ext(f_S(X), g_S(Y)) = b | S = s]\rvert\\
        &= \sum_{a, b} \sum_s \Pr[S = s] \lvert \Pr[\ext(X, Y) = a, \ext(f_s(X), g_s(Y)) = b] \\
        & \hspace*{2em} - \Pr[\unif{m}, \ext(f_s(X), g_s(Y)) = b]\rvert\\
        &= \sum_s \Pr[S = s] \sum_{a, b} \lvert \Pr[\ext(X, Y) = a, \ext(f_s(X), g_s(Y)) = b] \\
        & \hspace*{2em}- \Pr[\unif{m}, \ext(f_s(X), g_s(Y)) = b]\rvert\\
        &= \sum_s \Pr[S = s] \distCond{\ext(X, Y)}{\unif{m}}{\ext(f_s(X), g_s(Y))}\\
        &\leq \sum_s \Pr[S = s] \cdot \eps \\
        &= \eps
    \end{align*}
    Note that $S$ is independent of $X$ and $Y$ and thus $\ext(X, Y)$ is independent of $S$. Now for a fixed $s$, $f_s$ and $g_s$ are fixed functions. So the last inequality follows as $\ext$ is a two source non-malleable extractor.
\end{proof}

\subsection{Privacy Amplification}
The following two definitions are taken verbatim from \cite{AORSS20}.
\begin{definition}[Protocol against memory-tampering active adversaries]
    An {\emph{$(r,\ell_1,k_1,\ell_2,\linebreak[1] k_2,m)$}-protocol against memory-tampering active adversaries} is a protocol between Alice and Bob, with a man-in-the-middle Eve, that proceeds in $r$ rounds.
    Initially, we assume that Alice and Bob have access to random variables $(W,A)$ and $(W,B)$, respectively, where $W$ is an $(\ell_1,k_1)$-source (the \emph{secret}), and $A$, $B$ are $(\ell_2,k_2)$-sources (the \emph{randomness tapes}) independent of each other and of $W$. The protocol proceeds as follows:
    \begin{description}
        \item In the first stage, Eve submits an arbitrary function $F:\{0,1\}^{\ell_1}\times\{0,1\}^{\ell_2}\to \{0,1\}^{\ell_1}\times\{0,1\}^{\ell_2}$ and chooses one of Alice and Bob to be corrupted, so that either $(W,A)$ is replaced by $F(W,A)$ (if Alice is chosen), or $(W,B)$ is replaced by $F(W,B)$ (if Bob is chosen).
    
   \item  In the second stage, Alice and Bob exchange messages $(C_1,C_2,\dots,C_r)$ over a non-authenticated channel, with Alice sending the odd-numbered messages and Bob the even-numbered messages, and Eve is allowed to replace each message $C_i$ by $C'_i$ based on $(C_1,C'_1,\dots,C_{i-1},C'_{i-1},C_i)$ and independent random coins, so that the recipient of the $i$-th message observes $C'_i$.
    Messages $C_i$ sent by Alice are deterministic functions of $(W,A)$ and $(C'_2,C'_4,\dots,C'_{i-1})$, and messages $C_i$ sent by Bob are deterministic functions of $(W,B)$ and $(C'_1,C'_3,\dots,C'_{i-1})$.
    
 \item    In the third stage, Alice outputs $S_A\in\{0,1\}^m\cup\{\bot\}$ as a deterministic function of $(W,A)$ and $(C'_2,C'_4,\dots)$, and Bob outputs $S_B\in\{0,1\}^m\cup\{\bot\}$ as a deterministic function of $(W,B)$ and $(C'_2,C'_4,\dots)$.
\end{description}
\end{definition}

Furthermore, we will use the following 
explicit constructions of extractors in this chapter.
We will need the following constructions of extractors. 
\begin{lemma}[Theorem 6.9 of \cite{L20}]\label{lma:li}
    There exists a constant $0 < \gamma < 1$ and an explicit two-source non-malleable extractor $\extractorParam{\liExt}{(n, (1-\gamma)n)}{(n, (1-\gamma)n)}{\Omega(n)}{\liError}$ such that $\liError = 2^{-\Omega(n \frac{\log \log n}{\log n})}$.
\end{lemma}

\begin{lemma}[Theorem 2 of \cite{RRV02}]\label{lma:tre}
    For every $n, k$ there exists an explicit strong seeded extractor $\extractorParam{\treExt}{(n, k)}{(d, d)}{\Omega(k)}{\eps}$ such that $d = O(\log^2(n)\log(1/\eps))$.
\end{lemma}

\begin{lemma}[Theorem 1 of \cite{R05}]\label{lma:raz}
    For any $n_1, n_2, k_1, k_2, m$ and any $0 < \delta < \frac{1}{2}$ such that:
    \begin{enumerate}
        \item $k_1 \geq 5\log(n_2 - k_2)$
        \item $n_2 \geq 6 \log n_2 + 2\log n_1$,
        \item $k_2 \geq (\frac{1}{2} + \delta) \cdot n_2 + 3\log n_2 + \log n_1$,
        \item $m = \Omega(\min \{ n_2, k_1 \})$,
    \end{enumerate}
    there exists a strong two-source extractor $\extractorParam{\razExt}{(n_1, k_1)}{(n_2, k_2)}{m}{\eps}$, such that $\eps = 2^{-\frac{3m}{2}}$.
\end{lemma}

\begin{definition}[Privacy amplification protocol against memory-tampering active adversaries]
    An \emph{$(r,\ell_1,k_1,\ell_2,k_2,m,\eps,\delta)$-privacy amplification protocol against memory-tampering active adversaries} is an $(r,\ell_1,k_1,\ell_2,k_2,m)$-protocol against memory-tampering active adversaries with the following additional properties:
    \begin{itemize}
        \item \textbf{ If Eve is passive:} In this case, $F$ is the identity function and Eve only wiretaps.
        Then, $S_A=S_B\neq \bot$ with $S_A$ satisfying
        \begin{equation}
            S_A,C\approx_\eps U_m,C,
        \end{equation}
        where $C=(C_1,C'_1,C_2,C'_2,\dots,C_r,C'_r)$ denotes Eve's view.
        
        \item \textbf{ If Eve is active:} Then, with probability at least $1-\delta$ either $S_A=\bot$ or $S_B=\bot$ (i.e., one of Alice and Bob detects tampering), or $S_A=S_B\neq\bot$ with $S_A$ satisfying~\eqref{eq:sa}.
    \end{itemize}
\end{definition}

\section{Technical overview}
\label{sec:overview}

\subsection{Collision Resistant Extractors}
At the core of our non-malleable extractor compiler is a new object we call a \emph{collision resistant extractor}. An extractor is an object that takes as input two sources of randomness $X$ and $Y$ (in case of the seeded extractors $Y$ but uniform) and guarantees that, as long as $X$ and $Y$ are independent and have sufficient min-entropy, the output $\ext(X,Y)$ will be uniform (even given $Y$ \footnote{This property is often referred to as strong extraction \label{footstrong}}). A \emph{collision resistant extractor} $\innerExt$ has the added property that for all fixed-point-free functions $f$ (i.e. $f(x)\neq x$ for all $x$) the probability that $\innerExt(X,Y)=\innerExt(f(X),Y))$ is negligible \footnote{This notion might somewhat resemble various non-malleability notions, however in case of the non-malleability one would expect $\innerExt(f(X),Y))$ to be independent of $\innerExt(X,Y)$, here we only expect that those two outputs don't collide}. 

Readers might notice the resemblance to the collision resistant hashing families and the leftover hash lemma. The leftover hash lemma states that if the probability that $h(x_0,Y)=h(x_1,Y)$ is sufficiently small then $h(.,.)$ is an extractor. Obremski and Skorski (\cite{OS18}) showed that the inverse is almost true --- there exists a `core' of inputs on which every extractor has to fulfill the small collision probability property. This inverse leftover hash lemma is sadly not constructive and not efficient (the description of the core might be exponential), and thus we are unable to use it to obtain an efficient \emph{collision resistant extractor}.

We show that Raz's extractor (\cite{R05}) is a \emph{collision resistant extractor} with essentially the same parameters. We obtain this result by carefully modifying the original proof. The proof techniques are similar and we do not discuss the details in this section. One other aspect to note is that while Cohen, Raz, and Segev did indeed show that the same extractor is non-malleable. Their exact construction actually has Raz use $\log(n)$ fully uniform bits as a seed, and the non-malleability
requirement is in the seed. Our result differs, in the sense that we show Raz on input with $\log(n)$ min-entropy 
is collision-resistant --- we weaken the guarantee from non-malleability to collision-resistance, in exchange for also
improving the requirement from fully-uniform to just min-entropic.

We also show a generic transform that turns any seeded extractor (a two-source extractor where one source is uniform) into a \emph{collision resistant extractor} with a slight increase in the size of the seed.

\section{General Compiler for Seeded Extractors}

We first construct a collision-resistant extractor $h$ with a short output based on the Nisan-Widgerson generator \cite{NW94} or Trevisan's extractor \cite{RRV02}. Given the input $X$ and the seed $Z$, function $h$ will output $\hat{X}(Z_1)\concat\hat{X}(Z_2)\concat \cdots \concat\hat{X}(Z_t)$ where $\ecc$ is an error-correcting code of appropriate minimum distance, and $a \circ b$ denotes the concatenation of $a$ and $b$, $\hat{X} = \ecc(X)$, and $Z=Z_1\concat Z_2\concat \cdots \concat Z_t$, and $\hat{X}(Z_i)$ denotes $Z_i$-th bit of $\hat{X}$. Proof that this is an extractor follows directly from Nisan-Widgerson generator properties, while the collision resistance follows from the large distance of the error-correcting code.

We can now use any seeded extractor and the collision resistant extractor mentioned above to obtain a collision resistant seeded extractor with output size comparable to the seeded extractor.  Consider seeded extractors that take as input a random source $X$ and a short but uniform source $S$ and output $\ext(X,S)$ which is uniform (even given $S$ \footref{footstrong}).
Let us require on input a slightly longer uniform seed $S\concat Z$ (where $\concat$ denotes concatenation), and consider the following extractor: $\innerExt(X, S\concat Z)=\ext(X,S)\concat h(X,Z)$, where $h$ is either a collision resistant hash function or a collision resistant extractor. 

The proof follows quite easily. Function $h$ ensures that collisions indeed happen with negligible probability, the only thing left to show is that $\innerExt(X,S\concat Z)$ is uniform. 
First notice that by the definition the seeded extractor $\ext(X,S)$ is uniform, so we only have to show that $h(X,Z)$ is uniform even given $\ext(X,S)$. Observe that $Z$ is uniform and independent given $X,S$, so it suffices to show that $X$ has some remaining entropy given $\ext(X,S),S$, then $h(X,Z)$ will be uniform (either by leftover hash lemma, if $h$ is a collision resistant hash function, or by the definition of collision resistant extractor). This last step can be ensured simply by setting $\ext$ to  extract fewer bits than the entropy of $X$, thus a slight penalty in the parameters. 
Also notice that $h$ above can be a fairly bad extractor in terms of the rate or the output size and seed size. We can make the output and the seed of $h$ very small since we only need it to help provide collision resistance and thus the parameters of $\innerExt$ will be dominated by the parameters of $\ext$.


\subsection{Our Non-Malleable Extractor Compiler}

Our compiler takes as an input two objects, one is a collision resistant extractor (as discussed in the previous section), the other object is a strong two-source non-malleable extractor. A right-strong~\footnote{Notice that unlike many results in the literature, we need to distinguish between left strong and right strong for our extractor since the construction is inherently not symmetric.} non-malleable extractor gives the guarantee that $\ext(X,Y)$ is uniform even given $\ext(f(X),g(Y))$ and $Y$ (or $X$ in case of a left-strong non-malleable extractor) for any tampering functions $f,g$ where at least one of them are fixed-point-free. When we refer to a non-malleable extractor as strong without specifying if it's left-strong or right-strong we mean that the non-malleable extractor is both left-strong and right-strong. 
The construction is as follows: For a collision resistant extractor $\innerExt$, and a strong non-malleable extractor $\outerExt$ we consider following extractor:
\begin{equation}
    \twonmext(X, Y_\ell \concat Y_r) := \outerExt( Y_\ell \concat Y_r,  \innerExt(X, Y_\ell) ) \;.
\end{equation}
We will show that $\twonmext$ inherits the best of both worlds --- strong non-malleability of $\outerExt$ and the good entropy requirements of $\innerExt$.

There are two main issues to handle:
\paragraph{(Issue of the independent tampering.)}
Notice that the definition of the non-malleable extractor guarantees that $\ext(X,Y)$ is uniform given $\ext(X',Y')$ only if the sources are tampered independently (i.e. $X'$ is a function of only $X$, and $Y'$ is a function of only $Y$).

To leverage the non-malleability of $\outerExt$, we need to ensure that the tampering $X\rightarrow X'$ and $Y_\ell\concat Y_r \rightarrow Y'_\ell\concat Y'_r$ translates to the independent tampering of $Y_\ell\concat Y_r \rightarrow Y'_\ell \concat Y'_r$ and $\innerExt(X,Y_\ell) \rightarrow \innerExt(X',Y'_\ell)$. The problem is that both tamperings depend on $Y_\ell$. To address this issue we will simply reveal $Y_\ell$ and $Y'_\ell$ (notice that $Y'_\ell$ can depend on $Y_r$ thus revealing $Y_\ell$ alone is not sufficient). Once $Y_\ell=y_\ell$ and $Y'_\ell=y'_\ell$ are revealed (and therefore fixed) the tampering $y_\ell\concat Y_r \rightarrow y'_\ell \concat Y'_r$ and $\innerExt(X,y_\ell) \rightarrow \innerExt(X',y'_\ell)$ becomes independent since right tampering depends only on $X$, which is independent of $Y_\ell\concat Y_r$ and remains independent of $Y_r$ even after we reveal $Y_\ell$ and $Y'_\ell$ (this extra information only lowers the entropy of $Y_r$).

\paragraph{(Issue of the fixed points (or why we use collision resistance).)} Non-malleable extractors guarantee that $\ext(X,Y)$ is uniform given $\ext(X',Y')$ if and only if $(X,Y) \neq (X',Y')$. 

The issue in our compiler is clear: If $Y_\ell \concat Y_r$ do not change, and $X$ is tampered to be $X'\neq X$ but $\innerExt(X',Y_\ell)=\innerExt(X,Y_\ell)$ then 
\begin{align*}
    \twonmext(X, Y_\ell \concat Y_r) &= \outerExt( Y_\ell \concat Y_r,  \innerExt(X, Y_\ell) ) \\
    &= \outerExt( Y_\ell \concat Y_r,  \innerExt(X', Y_\ell) ) \\&
    =\twonmext(X', Y_\ell \concat Y_r)\;.
\end{align*}
To mitigate this problem, we require $\innerExt$ to be collision resistant, which means the probability that $\innerExt(X,Y_\ell)=\innerExt(X',Y_\ell)$ is negligible thereby resolving this issue. It is also possible to use $\innerExt$ without the collision resilience property, this gives a weaker notion of non-malleable extractor as was done in~\cite{GSZ21} .

\paragraph{(Is $\twonmext$ strong?)} Here we briefly argue that if $\outerExt$ is strong (i.e. both left and right strong) then $\twonmext$ will also be strong. To argue that compiled extractor is  left-strong, we notice that revealing $X$ on top of $Y_\ell$ and $Y'_\ell$ (which we had to reveal to maintain independence of tampering) translates to revealing $\innerExt(X,Y_\ell)$ which reveals right input of $\outerExt$ (revealing of $Y_\ell$ and $Y'_\ell$ is irrelevant since $Y_r$ maintains high enough entropy). As for the right-strongness, revealing $Y_r$ on top of $Y_\ell$ and $Y'_\ell$ translates to revealing of the left input of $\outerExt$, notice that $\innerExt(X,Y_\ell)$ remains uniform given $Y_\ell$ by the strong extraction property of $\innerExt$. 

$ $\\
For our construction, we will apply the compiler twice. First, we will use a collision resistant seeded extractor and the Li's extractor \cite{L20}. This gives us a strong non-malleable extractor $\fnmext$ for the first source with poly-logarithmic entropy, and the second source being uniform. We will refer to this object as a \emph{fully non-malleable seeded extractor}. We emphasize that this object is stronger than the seeded non-malleable extractor since it guarantees non-malleability for both sources. Then, we will then apply our compiler to Raz's extractor \cite{R05} and $\fnmext$ which will produce an extractor $\betterTwoNMExt$ that is a strong non-malleable extractor for the first source with poly-logarithmic entropy and the second source with entropy rate\footnote{Entropy rate is a ratio of min-entropy of the random variable to its length: $\frac{\minEnt{X}}{|X|}$}  $0.8$.

\subsubsection{Compiling Seeded Extractor with Li's Extractor}
In this section we will apply our compiler to the collision resistant seeded extractor $\crTreExt$ and strong non-malleable extractor $\liExt$ from \cite{L20}, yielding the following construction: 
\begin{equation}
    \fnmext(X,Y_\ell\concat Y_r)=\liExt(Y_\ell\concat Y_r, \crTreExt(X,Y_\ell)).
\end{equation}
The extractor $\liExt(\text{0.99}, \text{0.99})$ requires both sources to have a high entropy rate of $~99\%$\footnote{This is a simplification, formally speaking there exist a constant $\delta$ such that sources are required to have entropy rate above $1-\delta$. The reader may think of $\delta$ as $0.01$.}, while the extractor $\crTreExt(\text{poly-log}, \text{uniform})$ requires first source to have poly-logarithmic entropy, and the second source to be uniform. 
Let us analyse the entropy requirements of the extractor $\fnmext$:
Since part of the construction is $\crTreExt(X,Y_\ell)$ we require $Y_\ell$ to be uniform, which means that whole $Y_\ell \concat Y_r$ has to be uniform. On the other hand $X$ has to only have a poly-logarithmic entropy. The output of $\crTreExt(X,Y_\ell)$ will be uniform which will fulfill the $0.99$ entropy rate requirement of $\liExt$. There is a small caveat: While $Y_\ell \concat Y_r$ is uniform one has to remember that we had to reveal $Y_\ell$ and $Y'_\ell$ to ensure independent tampering, therefore we only have to make sure that $Y_\ell$ is very short so $Y_\ell \concat Y_r$ will have over $0.99$ entropy rate even given $Y_\ell$ and $Y'_\ell$. This is possible since $\crTreExt$ requires only a very short seed length. Thus we get that $\fnmext(\text{poly-log},\text{uniform})$ requires first source to have poly-logarithmic entropy, while the second source is uniform, and non-malleability is guaranteed for both sources.

\subsubsection{Compiling Raz's Extractor with the Above}
Now we will compile Raz's extractor \cite{R05} with above obtained $\fnmext$. The result will be:
\begin{equation}
    \betterTwoNMExt(X,Y_\ell\concat Y_r)=\fnmext(Y_\ell\concat Y_r, \razExt(X,Y_\ell)).
\end{equation}
As we discussed above $\fnmext(\text{poly-log},\text{uniform})$ requires first source to have poly-logarithmic entropy, while the second source has to be uniform, $\razExt(\text{poly-log},0.5)$ requires first source to have poly-logarithmic entropy while the second source has to have over $0.5$ entropy rate. Therefore we require $Y_\ell$ to have an entropy rate above $0.5$ and it is sufficient if $X$ has poly-logarithmic entropy. As for requirements enforced by $\fnmext$, since the output of $\razExt$ will be uniform we only have to check if $Y_\ell\concat Y_r$ has poly-logarithmic entropy given $Y_\ell$ and $Y'_\ell$. Given that $Y'_\ell$ can not lower the entropy of $Y_r$ by more than its size $|Y'_\ell|$ we have two equations:
\begin{align*}
    &\minEnt{Y_r} > |Y_\ell| \\
    &\minEnt{Y_\ell} > 0.5 |Y_\ell|
\end{align*}
which implies
\begin{align*}
    &\minEnt{Y_\ell\concat Y_r} > 2|Y_\ell| \\
    &\minEnt{Y_\ell\concat Y_r} > |Y_r| + 0.5 |Y_\ell|
\end{align*}
which asserts that $\frac{\minEnt{Y_\ell\concat Y_r}}{|Y_\ell\concat Y_r|} > 0.8$. Therefore $\betterTwoNMExt(\text{poly-log}, 0.8)$ requires first source to have poly-logarithmic entropy, while second source has to have entropy rate above $0.8$.

Finally notice that $\razExt$ has a relatively short output (shorter than both inputs) but that is not a problem since $\fnmext$ can have its first input much longer than the second input. We can adjust the output size of $\crTreExt$ to accommodate the input size requirements of $\liExt$ (this extractor requires both inputs to have the same length). We stress however that taking into consideration all inputs requirements both in terms of entropy and in terms of sizes is not trivial and our construction is tuned towards seeded-extractors and the Raz's extractor.

\section{A Generic Construction of a Two-Source Non-Malleable Extractor}
\label{sec:generic-reduction}
In this section we present a generic construction that transforms a non-malleable two-source extractor $\outerExt$ into another non-malleable two-source extractor with a much smaller entropy rate requirement via a two-source extractor. We note that the below theorem statement depends on $\outerExt$ and $\innerExt$. We will subsequently give an explicit construction for $\innerExt$.

\begin{theorem}
\label{thm:main_reduction}
 For any integers $n_1, n_2, n_3, n_4, k_1, k_2, k_3, k_4, m$ and \\
 $\outerError, \innerError, \collisionError > 0$, $n_4 < n_1$, given an efficient construction of
 \begin{itemize}
    \item a strong non-malleable extractor  $\extractorParam{\outerExt}{(n_1, k_1)}{(n_2, k_2)}{m}{\outerError}$,
    \item a right strong two-source extractor $\extractorParam{\innerExt}{(n_3, k_3)}{(n_4, k_4)}{n_2}{\innerError}$ that is $\collisionError$-collision resistant,
\end{itemize}
then for any integers $k_1^*, k_2^*$, $\eps, \tau > 0$ that satisfy the following conditions, there is an efficient construction of a left and right strong non-malleable two-source extractor $\extractorParam{\twonmext}{(n_3, k_1^*)}{(n_1, k_2^*)}{m}{\eps}$. \[k_1^* \ge k_3 \;,\]
\[k_2^* \ge \log 1/\tau + \max\left(k_4 + (n_1 - n_4), k_1 + 2 n_4 \right)   \;,\]
and 
\[
\eps \le 3\tau + 3 \outerError + 2 \innerError + 2 \sqrt{\collisionError}\;.
\]
\end{theorem}
\begin{proof}
Our construction is as follows: Given inputs $x \in \bit^{n_3}$ and $y = y_\ell \concat y_r$, where $y_\ell \in \bit^{n_4}$, and $y_r \in \bit^{n_1-n_4}$ our extractor is defined as:
\begin{equation}
    \twonmext(x, y) := \outerExt( y_\ell \concat y_r,  \innerExt(x, y_\ell) ) \;.
\end{equation}

Let $f:\bit^{n_3} \to \bit^{n_3}$ and $g:\bit^{n_1} \to \bit^{n_1}$. For any $y \in \bit^{n_1}$, by $g(y)_\ell$ we denote the $n_4$ bit prefix of $g(y)$. We assume that $f$ does not have any fixed points. The proof for the case when $g$ not having any fixed points is similar (in fact, simpler) as we explain later. 

\paragraph{(Right strongness.)} We first prove that our non-malleable extractor is right strong. 

\begin{claim}\label{clm:CloseToUnif}
 Let $\tY$ be a random variable with min-entropy $k_2^* - \log 1/\tau$ and is independent of $X$.   Consider the randomized function $\modifiedTampering$ that given $a, b, c$, samples $\innerExt(f(X), c)$ conditioned on $\innerExt(X, b) = a$, i.e., \[\modifiedTampering : a, b, c \mapsto \innerExt(f(X), c)_{| \innerExt(X, b) = a}\;.\], then it holds that:

\begin{equation}
    \distCond{
        \begin{array}{c}
            \innerExt(X, \tY_\ell)\\
            \outerExt( \tY_\ell \concat \tY_r,  \innerExt(X, \tY_\ell) )\\
            T
        \end{array}
    }{
        \begin{array}{c}
            \unif{d}\\\outerExt( \tY_\ell \concat \tY_r,  \unif{d} )\\
            T'
        \end{array}
    }{      
         \begin{array}{c}
            \tY_r\\
            \tY_\ell\\
            g(\tY)_\ell
             \end{array}
            } \leq \innerError \;
    \end{equation}
    , with $T = \outerExt( g(\tY),  \innerExt(f(X), g(\tY)_\ell) )$ and 
 $T' = \outerExt( g(\tY),  \modifiedTampering(\unif{\innerOutputLength}, \tY_\ell, g(\tY)_\ell) )$.

\end{claim}
\begin{proof}
 We have that $\minEnt{X} \geq k_1^* \ge k_3$ and $\minEnt{\tY_\ell} \geq k_2^* - \log 1/\tau - |\tY_r| = k_2^* - \log 1/\tau -  (n_1 - n_4) \ge k_4$, and $X, \tY_\ell$ are independently distributed. It follows that $\distCond{\innerExt(X, \tY_\ell)}{\unif{\innerOutputLength}}{\tY_\ell} \leq \innerError$. Then,  Lemma~\ref{lma:randDistinguisher} implies that
    \begin{equation*}
    \distCond{\innerExt(X, \tY_\ell)}{\unif{\innerOutputLength}}
    {\begin{array}{c}
    \tY_\ell \\ \tY_r \\ g(\tY)_\ell
     \end{array}} \le \innerError \;.
    \end{equation*}

Observing that since $\tY_r$ is independent of $\innerExt(f(X), g(\tY)_\ell), \innerExt(X, \tY_\ell)$ given $\tY_\ell, g(\tY)_\ell$, we have that the tuple $\innerExt(X, \tY_\ell), \tY_\ell, \tY_r,  \modifiedTampering(\innerExt(X, \tY_\ell), \tY_\ell, g(\tY)_\ell)$ is identically distributed as $\innerExt(X, \tY_\ell), \tY_\ell, \tY_r,  \innerExt(f(X), g(\tY)_\ell)$.   Again applying Lemma \ref{lma:randDistinguisher}, we get the desired statement.
 \end{proof}
    
    Now, let $\cY_0$ be the set of $y$ such that $g(y)_\ell = y_\ell$, and $\cY_1$ be the set of all $y$ such that $g(y)_\ell \neq y_\ell$ (in other words, $\cY_0$ contains all the fixed-points of $g$, and $\cY_1$ is the complement set). Also, let $\cY_{0,0}$ be the set of all $y \in \cY_0$ such that $\Pr[C(X,y_\ell) = C(f(X), y_\ell)] \le \sqrt{\collisionError}$, and $\cY_{0,1} = \cY_0 \setminus \cY_{0,0}$.  
    
    \begin{claim}\label{cl:Y1}
        If $\Pr[Y \in \cY_1] \ge \tau$, then:
        \begin{equation}
    \distCond{
        \begin{array}{c}
            \outerExt( \tY_\ell \concat \tY_r,  \innerExt(X, \tY_\ell) )
        \end{array}
    }{
        \begin{array}{c}
           U_m
        \end{array}
    }{
        \begin{array}{c}
            T \\ \tY_\ell\\ \tY_r
        \end{array}
    } \leq 
    \begin{array}{c}
    \innerError + \outerError
    \end{array}
     \;,
    \end{equation}
    
    where $\tY = Y|_{Y \in \cY_1}$, 
    $T = \outerExt( g(\tY),  \innerExt(f(X), g(\tY)_\ell) )$. 
    \end{claim}
    \begin{proof}
        Notice that conditioned on $Y$ being in $\cY_1$, $g$ does not have a fixed point. Thus, since $U_{n_2}$ is independent of $\tY_r$ given $\tY_\ell, g(\tY)_\ell$, and $H_\infty(U_{n_2}) = n_2 \ge k_2$, $H_\infty(\tY_r|\tY_\ell, g(\tY_r)) \ge k_2^* - \log 1/\tau - 2n_4 \ge k_1$, by the definition of a strong non-malleable extractor, we have that (letting $V = \outerExt( g(\tY),  T_{f,g}(U_{n_2}, \tY_\ell, g(\tY)_\ell) )$)
        
    \[
    \distCond{
        \begin{array}{c}
            \outerExt( \tY_\ell \concat \tY_r,  U_{n_2} )\\
            V
        \end{array}
    }{
        \begin{array}{c}
           U_m\\
            V
        \end{array}
    }{
        \begin{array}{c}
        \tY_\ell\\ \tY_r
        \end{array}
    } \le \outerError \;. 
    \]
    Furthermore, from Claim~\ref{clm:CloseToUnif} and Lemma~\ref{lma:randDistinguisher}, we get that:
    \[
    \distCond{
        \begin{array}{c}
            \outerExt( \tY_\ell \concat \tY_r,  U_{n_2} )\\
            V
        \end{array}
    }{
        \begin{array}{c}
        \outerExt( \tY_\ell \concat \tY_r,  \innerExt(X, \tY_\ell ))\\
            T
        \end{array}
    }{
        \begin{array}{c}
        \tY_\ell\\ \tY_r
        \end{array}
    } \le \innerError \;. 
    \]
    The desired statement follows from triangle inequality.
    \end{proof}
    
    Similarly, we prove the following claim:
    
  \begin{claim}
        If $\Pr[Y \in \cY_{0,0}] \ge \tau$, then 
        \begin{equation}
    \distCond{
        \begin{array}{c}
            \outerExt( \tY_\ell \concat \tY_r,  \innerExt(X, \tY_\ell) )
        \end{array}
    }{
        \begin{array}{c}
           U_m
        \end{array}
    }{
        \begin{array}{c}
        T \\ \tY_\ell\\ \tY_r
        \end{array}
    } \leq 
    \begin{array}{c}
        \outerError  \\
          + 2\innerError \\
          + \sqrt{\collisionError}
    \end{array} \;,
    \end{equation}
    where $\tY = Y|_{Y \in \cY_{0,0}}, T =  \outerExt( g(\tY),  \innerExt(f(X), g(\tY)_\ell) )$. 
    \end{claim}
    \begin{proof}
        Notice that  the probability that $\innerExt(X, \tY) = \innerExt(f(X), g(\tY)_\ell)$ is at most $\sqrt{\collisionError}$. Thus, by Claim~\ref{clm:CloseToUnif}, the probability that $U_{n_2} = T_{f,g}(U_{n_2}, \tY_\ell, g(\tY)_\ell)$ is at most $\sqrt{\collisionError} + \innerError$.   Also, since $U_{n_2}$ is independent of $\tY_r$ given $\tY_\ell, g(\tY)_\ell$, and $H_\infty(U_{n_2}) = n_2 \ge k_2$, $H_\infty(\tY_r|\tY_\ell, g(\tY_r)) \ge k^* - \log 1/\tau - 2n_4 \ge k_2$, by the definition of a strong non-malleable extractor, we have that (letting $V = \outerExt( g(\tY),  T_{f,g}(U_{n_2}, \tY_\ell, g(\tY)_\ell) )$):
    \begin{equation*}
    \distCond{
        \begin{array}{c}
            \outerExt( \tY_\ell \concat \tY_r,  U_{n_2} )
        \end{array}
    }{
        \begin{array}{c}
           U_m
        \end{array}
    }{\begin{array}{c}
        \tY_\ell\\ \tY_r \\ V
        \end{array}} \le {\begin{array}{c}
             \outerError +
             \innerError + 
              \sqrt{\collisionError}
        \end{array}}   \;. 
    \end{equation*}
    
    Furthermore, from Claim~\ref{clm:CloseToUnif} and Lemma~\ref{lma:randDistinguisher}, we get that:
    \[
    \distCond{
        \begin{array}{c}
            \outerExt( \tY_\ell \concat \tY_r,  U_{n_2} )\\
            V
        \end{array}
    }{
        \begin{array}{c}
        \outerExt( \tY_\ell \concat \tY_r,  \innerExt(X, \tY_\ell ))\\
        T
        \end{array}
    }{\begin{array}{c}
        \tY_\ell\\ \tY_r
        \end{array}} \le \innerError \;. 
    \]
    The desired statement follows from triangle inequality.
    \end{proof}
    We now show that $Y \in \cY_{0,1}$ with small probability.
    
    \begin{claim}
    \label{claim:Y01}
       \[ \Pr[Y \in \cY_{0,1}] \le \tau + \sqrt{\collisionError}\;.\]
    \end{claim}
    \begin{proof}
        If $\Pr[Y \in \cY_0] < \tau$, then the statement trivially holds. So, we assume $\Pr[Y \in \cY_0] \ge \tau$. Let $\tY = Y|_{Y \in \cY_0}$  Then $H_\infty(\tY) \ge k_2^* - \log 1/\tau - (n_1 - n_4) \ge k_4$. Since $\mathbf{C}$ is collision-resistant, we have that 
        \begin{align*}
            \collisionError &\ge \Pr[\innerExt(X, \tY_\ell) = \innerExt(f(X), g(\tY)_\ell)] 
             \Pr[\tY \in \cY_{0,1}] \cdot \sqrt{\collisionError} \\
             &\ge \Pr[Y \in \cY_{0,1}] \cdot \sqrt{\collisionError} \;.
        \end{align*}
    \end{proof}
    We now conclude the proof of right strongness of our non-malleable extractor as follows. We shorthand         $\twonmext( X, Y), Y, \twonmext( f(X),  g(Y))$ by $\phi(X,Y)$, and $U_m, Y, \twonmext( f(X),  g(Y))$ by $\psi(X,Y)$.
    \begin{align*}
        \dist{
            \phi( X, Y)
    }{
        \psi(X,Y)
    } &\le \Pr[Y \in \cY_{0,1}] \\
    &\;\;\;\;\;\;\;\;\;+ \Pr[Y \in \cY_1] \cdot \dist{
            \phi( X, Y)|_{Y \in \cY_1}
    }{
        \psi(X,Y)|_{Y \in \cY_1}
    } \\
    &\;\;\;\;\;\;\;\;\; + \Pr[Y \in \cY_{0,0}] \cdot \dist{
            \phi( X, Y)|_{Y \in \cY_{0,0}}
    }{
        \psi(X,Y)|_{Y \in \cY_{0,0}}
    } \\
    &\le (\tau + \outerError + \innerError)\\
    & \;\;\;\;\;\;\;\;\; + (\tau+2\outerError + \innerError+\sqrt{\collisionError}) \\
    & \;\;\;\;\;\;\;\;\;  + (\tau+\sqrt{\collisionError}) \\
    &= 3\tau + 3 \outerError + 2 \innerError + 2 \sqrt{\collisionError} \;. 
    \end{align*}
    
    Note that we assumed that $f$ does not have fixed points. On the other hand, if $g$ does not have fixed points then a simpler proof works that does not need to partition the domain into $\cY_{0,0}, \cY_{0,1}, \cY_1$. Since the first source for the non-malleable extractor $\outerExt$, we can conclude the statement similar to Claim~\ref{cl:Y1} with $Y$ instead of $\tY$. 
    
    \paragraph{(Left strongness.)} 
 The proof of left strongness is nearly the same (the statistical distance statements include $X$ instead of $Y_r$), but we include it here for completeness. 

\begin{claim}\label{clm:CloseToUnif_2}
 Let $\tY$ be a random variable with min-entropy $k^* - \log 1/\tau$ and is independent of $X$.   Consider the randomized function $S$ that given $a, b$, samples $X$ conditioned on $\innerExt(X, b) = a$, i.e., \[S : a, b \mapsto X|_{ \innerExt(X, b) = a}\;.\] Then:

     \begin{equation}
    \distCond{
        \begin{array}{c}
            \innerExt(X, \tY_\ell) \\ X
        \end{array}
    }{
            \begin{array}{c}
            \unif{d},\\S(U_{n_2}, \tY_\ell)
             \end{array}
    }{
            \begin{array}{c}
        \tY_\ell\\ \tY_r
        \end{array}
    } \leq \innerError \;.
    \end{equation}
\end{claim}
\begin{proof}
 We have that $\minEnt{X} \geq k_1^* \ge k_3$ and $\minEnt{\tY_\ell} \geq k_2^* - \log 1/\tau - |\tY_r| = k_2^* - \log 1/\tau -  (n_1 - n_4) \ge k_4$, and $X, \tY_\ell$ are independently distributed. It follows that $\distCond{\innerExt(X, \tY_\ell)}{\unif{\innerOutputLength}}{\tY_\ell} \leq \innerError$. Then, using Lemma~\ref{lma:randDistinguisher} and 
observing that since $\tY_r$ is independent of $X$ given $\tY_\ell$, we have that  $\innerExt(X, \tY_\ell), \tY_\ell, \tY_r,  S(\innerExt(X, \tY_\ell), \tY_\ell)$ is identically distributed as $\innerExt(X, \tY_\ell), \tY_\ell, \tY_r, X$, we get the desired statement.
 \end{proof}
    
    Now, let $\cY_0$ be the set of $y$ such that $g(y)_\ell = y_\ell$, and $\cY_1$ be the set of all $y$ such that $g(y)_\ell \neq y_\ell$. Also, let $\cY_{0,0}$ be the set of all $y \in \cY_0$ such that $\Pr[C(X,y_\ell) = C(f(X), y_\ell)] \le \sqrt{\collisionError}$, and $\cY_{0,1} = \cY_0 \setminus \cY_{0,0}$.  
    
    \begin{claim}\label{cl:Y1_2}
        If $\Pr[Y \in \cY_1] \ge \tau$, then:
        \begin{equation}
     \distCond{
        \begin{array}{c}
            \outerExt( \tY_\ell \concat \tY_r,  \innerExt(X, \tY_\ell) )
        \end{array}
    }{
        \begin{array}{c}
           U_m
        \end{array}
    }{
        \begin{array}{c}
        T \\
        \tY_\ell\\ g(\tY)_\ell\\ X
        \end{array}
    } \leq 2\innerError + \outerError  \;,
    \end{equation}
    where $\tY = Y|_{Y \in \cY_1}, T = \outerExt( g(\tY),  \innerExt(f(X), g(\tY)_\ell) )$. 
    \end{claim}
    \begin{proof}
        Notice that conditioned on $Y$ being in $\cY_1$, $g$ does not have a fixed point. Thus, since $U_{n_2}$ is independent of $\tY_r$ given $\tY_\ell, g(\tY)_\ell$, and $H_\infty(U_{n_2}) = n_2 \ge k_2$, $H_\infty(\tY_r|\tY_\ell, g(\tY_r)) \ge k^* - \log 1/\tau - 2n_4 \ge k_1$, by the definition of a strong non-malleable extractor, we have that (letting $V = \outerExt( g(\tY), \innerExt(f(S(U_{n_2}, \tY_\ell)), g(\tY)_\ell))$:
        
    \[
     \distCond{
         \begin{array}{c}
            \outerExt( \tY_\ell \concat \tY_r,  U_{n_2} )
        \end{array}
    }{
        \begin{array}{c}
           U_m
        \end{array}
    }{
    \begin{array}{c}
           U_{n_2} \\
          V\\
          \tY_\ell \\ g(\tY)_\ell
    \end{array}} \le \outerError \;,
    \]
    Furthermore, by applying Claim~\ref{clm:CloseToUnif} and Lemma~\ref{lma:randDistinguisher} twice, we get that:
     \[
    \distCond{
            \begin{array}{c}
                 U_m\\ S(U_{n_2}, \tY_\ell) \\ V         
            \end{array}
           }{
           \begin{array}{c}
           U_m\\ X \\ T
           \end{array}
           }{\begin{array}{c}
                 \tY_\ell  \\
                 g(\tY)_\ell
            \end{array}} \le \innerError \;. 
    \]
    and also
     \[
    \distCond{
    \begin{array}{c}
           \outerExt(\tY_\ell \concat \tY_r, U_{n_2}) \\ S(U_{n_2}, \tY_\ell) \\ 
           V \end{array}
           }{
           \begin{array}{c}
           \outerExt(\tY_\ell \concat \tY_r, \innerExt(X, \tY_\ell))\\ X \\ T 
           \end{array}}{
            \begin{array}{c}
                 \tY_\ell  \\
                 g(\tY)_\ell
            \end{array}
           } \le \innerError \;. 
    \]
    The desired statement follows from triangle inequality.
    \end{proof}
    
    Similarly, we prove the following claim.
    
  \begin{claim}
        If $\Pr[Y \in \cY_{0,0}] \ge \tau$, then
   \begin{equation}
    \distCond{
        \begin{array}{c}
            \outerExt( \tY_\ell \concat \tY_r,  \innerExt(X, \tY_\ell) )
        \end{array}
    }{
        \begin{array}{c}
           U_m
        \end{array}
    }{
        \begin{array}{c}
        \tY_\ell\\ g(\tY)_\ell\\ X \\ T 
        \end{array}
    } \leq 
    \begin{array}{c}
         \outerError +  \\
         3\innerError +\\ 
         \sqrt{\collisionError}
    \end{array}
       \;,
    \end{equation}
    where $\tY = Y|_{Y \in \cY_{0,0}}, T = \outerExt( g(\tY),   \innerExt(f(X), g(\tY)_\ell)$. 
    \end{claim}
    \begin{proof}
        Notice that  the probability that $\innerExt(X, \tY) = \innerExt(f(X), g(\tY)_\ell)$ is at most $\sqrt{\collisionError}$. Thus, by Claim~\ref{clm:CloseToUnif}, the probability that $U_{n_2} = \innerExt(S(U_{n_2}, \tY_\ell), g(\tY)_\ell)$ is at most $\sqrt{\collisionError} + \innerError$.   Also, since $U_{n_2}$ is independent of $\tY_r$ given $\tY_\ell, g(\tY)_\ell$, and $H_\infty(U_{n_2}) = n_2 \ge k_2$, $H_\infty(\tY_r|\tY_\ell, g(\tY_r)) \ge k^* - \log 1/\tau - 2n_4 \ge k_1$, by the definition of a strong non-malleable extractor, we have that \\
        (letting $V = \outerExt( g(\tY), \innerExt(f(S(U_{n_2}, \tY_\ell)), g(\tY)_\ell))$):
    \[
     \distCond{
        \begin{array}{c}
            \outerExt( \tY_\ell \concat \tY_r,  U_{n_2} )
        \end{array}
    }{
        \begin{array}{c}
           U_m
        \end{array}
    }{
    \begin{array}{c}
    U_{n_2}\\
    V\\
    \tY_\ell\\
    g(\tY)_\ell
    \end{array}}
    \le 
    \begin{array}{cc}
         \outerError &  \\
         + \innerError & \\
         + \sqrt{\collisionError}
    \end{array}\;.
    \]
    Furthermore, by applying Claim~\ref{clm:CloseToUnif} and Lemma~\ref{lma:randDistinguisher} twice, we get that:
     \[
    \distCond{
            \begin{array}{c}
                 U_m\\ S(U_{n_2}, \tY_\ell) \\ V
            \end{array}
             }{
             \begin{array}{c}
             U_m\\ X \\ T
             \end{array}
              }{
              \begin{array}{c}
                   \tY_\ell\\ g(\tY)_\ell
              \end{array}
              } \le \innerError \;. 
    \]
    and:
     \[
    \distCond{
    \begin{array}{c}
           \outerExt(\tY_\ell \concat \tY_r, U_{n_2}) \\ S(U_{n_2}, \tY_\ell) \\ V 
           \end{array}
           }{
           \begin{array}{c}
           \outerExt(\tY_\ell \concat \tY_r, \innerExt(X, \tY_\ell)) \\ X \\ T
           \end{array}}{\begin{array}{c}
                 \tY_\ell  \\
                 g(\tY)_\ell
            \end{array}} \le \innerError \;. 
    \]
    The desired statement follows from triangle inequality.
    \end{proof}
    We then conclude the proof of right strongness of our non-malleable extractor exactly as we obtained left strongness. 
\end{proof}

\begin{remark}\label{multitamper1}
    We remark that one can apply the above compiler to multi-tampering non-malleable extractors as a $\outerExt$. Briefly speaking $t$-tampering non-malleable extractor guarantees that extraction output remains uniform even given not one but $t$ tampering outputs:
    \begin{equation*}
        \distCond{E(X, Y)}{\unif{m}}{E(f_1(X), g_1(Y)),...,E(f_t(X), g_t(Y))} \leq \eps.
    \end{equation*}
    As a result, compiled extractor will also be $t$-tamperable.
    The proof is almost identical, there are only two differences:
    \begin{enumerate}
        \item To ensure the reduction to split state tampering it is not sufficient to reveal $\tY_\ell$ and $g_1(\tY)_\ell$, but also all other tamperings: $g_2(\tY)_\ell, \ldots, g_t(\tY)_\ell$. This will have an impact of the calculations of entropy requirement.
        \item Notice that when considering the collision resistance adversary has now $t$ chances instead of $1$, but since the attempts are non-adaptive we can easily bound the collision probability by $t\cdot \collisionError$, this impacts the error calculations.
    \end{enumerate}
\end{remark}

\section{Collision resistance of Extractors}
\label{sec:coll-res}
\subsection{Generic Collision Resistance for Seeded Extractors}
\label{sec:coll-res-generic}
\begin{lemma}\label{lma:CRSeededExt}
    Let $\extractorParam{\ext}{(n, k)}{(d, d)}{m}{\eps}$ be a strong seeded extractor. Then there exists a strong seeded extractor $\extractorParam{\crTreExt}{(n, k)}{(d + z, d + z)}{m - 2\log(1/\collisionError)}{\eps + \treError + \sqrt{\collisionError}}$ with collision probability $\collisionError$ and $z = O(\log(1/\collisionError)\log^2(\log(1/\collisionError))\log(1/\treError))$.
\end{lemma}
\begin{proof}
    We will first mention \cite{RRV02}'s construction of $\treExt$. The aforementioned construction uses an error correcting code and a weak design, defined respectively as below:

    \begin{lemma}[Error Correcting Code, Lemma 35 of \cite{RRV02}]
        For every $n \in \bN$, and $\delta > 0$, there exists a code $\ecc : \bit^n \to \bit^{\hat{n}}$ where $\hat{n} = poly(n, 1/\delta)$ such that for $x, x' \in \bit^n$ with $x \neq x'$, it is the case that $\ecc(x)$ and $\ecc(x')$ disagree in at least $(\frac{1}{2} - \delta) \hat{n}$ positions.
    \end{lemma}

    \begin{definition}[Weak Design, Definition 6 of \cite{RRV02}]
        A family of sets $\blocks{\cS}{m} \subseteq [d]$ is a weak $(\ell, \rho)$-design if:
        \begin{enumerate}
            \item For all $i$, $\lvert \cS_i \rvert = \ell$;
            \item For all $i$,
            \begin{equation*}
                \sum_{j < i} 2^{\lvert \cS_i \cap \cS_j \rvert} \leq \rho \cdot (m - 1).
            \end{equation*}
        \end{enumerate}
    \end{definition}
    In particular, any family of disjoint sets $\blocks{\cS}{m} \subseteq [d]$ with $\lvert \cS_i \rvert = \ell$ is trivially a weak design as well.

    Extractor $\treExt$ operates in the following way: $X$ is firstly evaluated on an error correcting code $\ecc$ to obtain $\hat{X}$. Then viewing seed bits $Z$ as $Z_1 \concat Z_2 \concat \ldots \concat Z_d$, then the $i^{th}$ bit of $\treExt(X, Z)$ is given as the $(Z_{|\cS_i})^{th}$ bit of $\hat{X}$ where $Z_{|\cS_i}$ is understood to specify an $\ell$-bit index $Z_{j_1} \concat Z_{j_2} \concat \ldots \concat Z_{j_\ell}$ for $\cS_i = \{ j_1, j_2, \ldots, j_\ell \}$. In short, the output is given as:

    \begin{equation*}
        \treExt(X, Z) = \hat{X}(Z_{|\cS_1}) \concat \hat{X}(Z_{|\cS_2}) \concat \cdots \concat \hat{X}(Z_{|\cS_m}).
    \end{equation*}

    The modification is to truncate the output of $\ext(X, S)$ by $t = \frac{5}{2}\log(1/\collisionError)$ bits, and then treating $Z$ as $\frac{4t}{5}$ blocks of $\ell = O(\log^2(t)\log(1/\treError))$ many bits, we concatenate the output with $\frac{4t}{5}$ bits. In short, the output is given as:

    \begin{equation*}
        \crTreExt(X, S \concat Z)_i = \begin{cases}
                                        \ext(X, S)_i &,\text{if } i \leq m - t\\
                                        \hat{X}(Z_{i - (m - t)}) &, \text{if } i > m - t
                                    \end{cases}
    \end{equation*}
    where $Z_j$ denotes the $j^{th}$ block of $Z$.

    To show that $\crTreExt$ is indeed a strong extractor, note that $S$ and $Z$ are independent and furthermore by Lemma \ref{lem:avgminH} $\avgCondMinEnt{X}{\ext(X, S),S} \geq k - m + t \geq t$. Instantiating $\crTreExt$ with a family of disjoint sets, an error correcting code $\ecc$ with minimum distance $(\frac{1}{2} - \frac{\treError}{4m}) \hat{n}$ for inputs of min-entropy $(t, \frac{4t}{5})$ and seed length $O( \log(1/\collisionError)\log^2(t) \log(1/\treError))$, Lemma \ref{lma:avgCaseMinExtraction} implies that:

    \begin{equation*}
        \distCond{ \ext(X, S)  \concat \treExt(X, Z) }{ \ext(X, S), \unif{\Omega(t)} }{S, Z} \leq \treError + 2^{-\frac{t}{5}}
    \end{equation*}
    which in turn yields us:
    \begin{equation*}
        \distCond{ \ext(X, S) \concat \treExt(X, Z) }{ \unif{m - O(t)} }{S, Z} \leq \eps + \treError + 2^{-\frac{t}{5}} = \eps + \treError + \sqrt{\collisionError}
    \end{equation*}

    As for the collision probability, note that for any $x$ and fixed-point-free function $f$:
    \begin{align*}
        \Pr[\crTreExt(x, S \concat Z) = \crTreExt(f(x), S \concat Z)] 
        &\leq \Pr\big[\forall i, \ecc(x)(Z_i) = \ecc(f(x))(Z_i)\big]\\
        &\leq \left(\frac{1}{2} + \frac{\treError}{4m}\right)^{2\log(1/\collisionError)}\\
        &\leq \collisionError
    \end{align*}

    Since this bound holds for all possible values $x$, it follows that it holds for any random variable $X$ as well.
\end{proof}

An instantiation that will suit our purpose will be to use Trevisan's extractor $\treExt$ as $\ext$. Then for any $n, k$, we have $\extractorParam{\crTreExt}{(n, k)}{(d, d)}{\Omega(k)}{3\treError}$ with $d = O( \log^2(n) \log(1/\treError)) + O(\log(1/\treError) \log^2(\log(1/\treError))\log(1/\treError)) = O(\log^2(n)\log^2(1/\treError) )$ such that $\eps = \treError$ andwith collision probability $\collisionError = (\treError)^2 < 2^{-\Omega(k)}$. 

\subsection{Collision Resistance of the Raz Extractor}
\label{sec:coll-res-raz}
\begin{lemma}\label{lma:CRRaz}
    For any $n_1, n_2, k_1, k_2, m$ and any $0 < \delta < \frac{1}{2}$ such that:
    \begin{enumerate}
        \item $k_1 \geq 12\log(n_2 - k_2) + 15$,
        \item $n_2 \geq 6 \log n_2 + 2\log n_1 + 4$,
        \item $k_2 \geq (\frac{1}{2} + \delta) \cdot n_2 + 3\log n_2 + \log n_1 + 4$,
        \item $m = \Omega(\min \{ n_2, k_1 \})$,
    \end{enumerate}
    there exists a strong two-source extractor $\extractorParam{\razExt}{(n_1, k_1)}{(n_2, k_2)}{m}{\eps}$, such that $\eps = 2^{-\frac{3m}{2}}$ with collision probability $2^{-m + 1}$.
\end{lemma}

\begin{proof}
We will show that the two-source extractor by Raz satisfies the collision resistant property. We first recap \cite{R05}'s construction. Given independent sources $X \givenBy (n_1, k_1)$ and $Y \givenBy (n_2, k_2)$, $\razExt(X, Y)$ uses $Y$ as seed (using Lemma~\ref{lma:linearTests}) to construct $m \cdot 2^{n_2}$ many $0$-$1$ random variables $Z_{(i, X)}(Y)$ with $i \in [m]$ and $x \in \bit^{n_1}$, where random variables are $(t', \eps)$-biased for $t' \geq t \cdot m$.

The idea is to generate a sequence random variables are $\eps$-biased for tests of size $2tm$, and then the probability of collision can be bounded in a similar manner as the proof that function is a two-source extractor. Define $\gamma_i(X, Y) = (-1)^{Z_{i, X}(Y)}$ and let $f$ be any fixed-point-free function. Furthermore let $t' \geq 2 \cdot m t$ for some value of $t$ such that the set of random variables $Z_{(i, x)}(Y)$ are $(t', \eps)$-biased. The idea will be to show that we can leverage the $(t, \eps)$-biasedness to show that with high probability over the choice of $X$, for each $i \in [m]$, the probability of the extractor colliding on the $i^{th}$ bit is close to $1/2$. Then we use the Lemma \ref{lma:XORLemma} to argue that overall the probability of colliding on all bits is small. 

More formally, define $\gamma_i(X, Y) = \bE\left[ (-1)^{Z_{i, X}(Y)} \right]$, and let $f$ be any fixed-point free function. We will first bound $\lvert \gamma_i(X, Y) \rvert$. 

\begin{claim}[Claim 3.2 in \cite{R05}]\label{clm:invokeEpsBias}
    For any $i \in [m]$, any $r \in [t']$ and any set of distinct values $\blocks{x}{r} \in \bit^{n_1}$:
    \begin{equation*}
        \sum_{y \in \bit^{n_2}} \prod^r_{j = 1} (-1)^{Z_{i, x_j}(y)} \leq 2^{n_2} \cdot \eps
    \end{equation*}
\end{claim}

\begin{proof}
    Since $Z_{i, x}$ are $(t', \eps)$-biased:
    \begin{align*}
        \sum_{y \in \bit^{n_2}} \prod^r_{j = 1} (-1)^{Z_{i, x_j}(y)} &= \sum_{y \in \bit^{n_2}} (-1)^{\bigoplus_{j} Z_{i, x_j}(y)}\\
        &= 2^{n_2} \sum_{y \in \bit^{n_2}} \Pr[\unif{n_2} = y] (-1)^{\bigoplus_{j} Z_{i, x_j}(y)}\\
        &= 2^{n_2} (-1)^{\bigoplus_{j} Z_{i, x_j}(\unif{n_2})} \leq 2^{n_2} \cdot \eps
    \end{align*}
\end{proof}

\begin{claim}\label{clm:pointwiseBiased}
    Letting $Z_{(i, x)}(Y)$ be $(2t, \eps)$-biased, $\Pr[ Z_{(i, x)}(Y) = Z_{(i, f(x))}(Y) ] = \Pr[ Z_{(i, x)}(Y) \oplus Z_{(i, f(x))}(Y) = 0] \leq \frac{1}{2} + \eps'$ where:
    \begin{equation*}
        \eps' = 2^{(n_2 - k_2) / t} \cdot \left( \eps^{1/t} + (2t) \cdot 2^{-\frac{k_1}{3}} \right)
    \end{equation*}  
\end{claim}
\begin{proof}
    Let $t$ be some even positive integer, then consider $\left( \gamma(X, Y) \gamma(f(X), Y) \right)^t$. By Jensen's inequality we can bound the term as:
    \begin{align*}
        \left( \gamma(X, Y) \gamma(f(X), Y) \right)^t &=  \left( \frac{1}{2^{k_1 + k_2}} \sum_{(x, y) \in \supp{X, Y}} \ (-1)^{Z_{(i, x)}(y) \oplus Z_{(i, f(x))}(y)} \right)^t \\
          &\leq  \left(\frac{1}{2^{k_2}}\right) \sum_{y \in \supp{Y}} \left( \frac{1}{2^{k_1}} \sum_{x \in \supp{X}} (-1)^{Z_{(i, x)}(y) \oplus Z_{(i, f(x))}(y)}  \right)^t \\
      &\leq  \left(\frac{1}{2^{k_2}}\right) \sum_{y \in \bit^{n_2}} \left( \frac{1}{2^{k_1}} \sum_{x \in \supp{X}} (-1)^{Z_{(i, x)}(y) \oplus Z_{(i, f(x))}(y)}  \right)^t \\
        &=  \left(\frac{1}{2^{k_2 + k_1 \cdot t}}\right) \sum_{\blocks{x}{t} \in \supp{X}} \sum_{y \in \bit^{n_2}}  \prod_{j = 1}^t  (-1)^{Z_{(i, x_j)}(y) \oplus Z_{(i, f(x_j))}(y)}  \\
    \end{align*}
  
    Then we partition the summands (based on $\blocks{x}{t}$) into two categories: (1) When the values $\blocks{x}{t}, \functionBlocks{x}{t}{f}$ has at least one unique value $x$ that does not otherwise occur in $\blocks{x}{t}$ and $\functionBlocks{x}{t}{f}$ or else (2) when the every value in $\blocks{x}{t}, \functionBlocks{x}{t}{f}$ occurs at least twice.
    
    (1) In the first case, Claim \ref{clm:invokeEpsBias} implies the respective summands can be bounded by $2^{n_1} \cdot \eps$ and there are at most $2^{k_1 \cdot t}$ many of these summands. (2) In the latter case, we will bound the sum using the following claim:

    \begin{claim}
        If $\blocks{x}{t}, \functionBlocks{x}{t}{f}$ are such that every value occurs at least twice and $f(x_i) \neq x_i$ for all $i \in [t]$, then there exists a subset of indices $\cS \subseteq [t]$ such that $\lvert \cS \rvert \leq \frac{2}{3} t $ and $\{ \blocks{x}{t} \} \subseteq \{ x : s \in \cS \} \cup \{ f(x) : s \in \cS \}$.
    \end{claim}

    \begin{proof}
        Define set $\cA$ to contain the values of \blocks{x}{t} that occur at least twice within \blocks{x}{t}. Define $\cS_\cA$ be the set of indices of the first occurrence of each value in $\cA$, and furthermore define $\cB$ to be $\{ \blocks{x}{t} \} \setminus \{ x_j, f(x_j) : j \in \cS_\cA \}$. Then if $\lvert \cA \rvert = \ell$, $\lvert \cB \rvert = r \leq t - 2\ell$. Let $\cB = \{b_1, \ldots, b_r\}$.
        
        Since each $\blocks{x}{t}, \functionBlocks{x}{t}{f}$ has that every value occurs twice, and $b_i$ for any $i \in [r]$ does not occur in $\{x, f(x) \: : \: x \in \cS_\cA\}$, it implies that $\blocks{b}{r} \in \cB$ must be a fixed-point-free permutation of $\functionBlocks{\cB}{r}{f}$. Thus, the permutation $f$ defines a disjoint union of cycles over the set $\cB$. Define $\cS_\cB$ to be the set that for each such cycle includes every alternate element. More precisely, for each such cycle, say $(b_{i_1}, \ldots, b_{i_q})$ with \[
        f(b_{i_1}) = b_{i_2}, f(b_{i_2}) = b_{i_3}, \ldots, f(b_{i_{q -1}}) = b_{i_q}, f(b_{i_q}) = b_{i_1} \;, \] we include $b_{i_1}, b_{i_3}, \ldots, b_{i_{1 + 2 \lfloor (q-1)/2\rfloor}}$ in the set $\cS_\cB$. Then $\cS = \cS_\cA \cup \cS_\cB$ satisfy the desired condition. Also, \[|\cS_\cB| \le r \max_{q \in \mathbb{N} \setminus \{1\}} \frac{\lceil q/2 \rceil}{q} \le \frac{2r}{3}\;,\] since $\frac{\lceil q/2 \rceil}{q}$ is $1/2$ when $q$ is even, and $(q+1)/2q$ when $n$ is odd, and hence is maximized for $q = 3$.  Thus, 
        \[
        |S| \le \ell + \frac{2r}{3} \le \ell + \frac{2(t-2\ell)}{3} = \frac{2t}{3} - \frac{\ell}{3} \le \frac{2t}{3} \;,
        \]
        as needed.
    \end{proof}

    To obtain the bound on the number of summands in the case (2), note that there are $\binom{2^{k_1}}{\frac{2}{3}t }$ possible sets $\cS$, and for each set, there are $\left( \frac{4t}{3}\right)^t$ possible sequences that satisfy Case 2. In each such case, we bound the summand by $2^{n_2}$. Combining the two cases, we get that:

    \begin{align*}
        \left( \gamma(X, Y) \gamma(f(X), Y) \right)^t &\leq \left(\frac{1}{2^{k_2 + k_1 \cdot t}}\right) \sum_{\blocks{x}{t} \in \supp{X}} \sum_{y \in \bit^{n_2}}  \prod_{j = 1}^t  (-1)^{Z_{(i, x_j)}(y) \oplus Z_{(i, f(x_j))}(y)}  \\
        &\leq \left(\frac{1}{2^{k_2 + k_1 \cdot t}}\right) \left( 2^{k_1 \cdot t} 2^{n_2} \cdot \eps + 2^{n_2}\binom{2^{k_1}}{ \frac{2}{3}t }  \left( \frac{4t}{3}\right)^t \right)\\&\leq \left(\frac{1}{2^{k_2 + k_1 \cdot t}}\right) \left( 2^{k_1 \cdot t} 2^{n_2} \cdot \eps + 2^{n_2} (2t)^t \cdot 2^{-\frac{k_1}{3}t} \right)\\
        \lvert \gamma(X, Y) \gamma(f(X), Y) \rvert &\leq 2^{(n_2 - k_2) / t} \cdot \left( \eps^{1/t} + (2t) \cdot 2^{-\frac{k_1}{3}} \;. \right)
    \end{align*}
\end{proof}

Now that we have shown that for any coordinate $i \in [m]$, the probability the extractor collides on the $i^{th}$ bit is at most $\frac{1}{2} + \eps'$, we wish to invoke the Lemma \ref{lma:XORLemma} to argue that the probability the extractor collides on all the coordinates is small.

Define $\tau \subseteq [m]$, and consider the set of random variables\\ $\left \{ \bigoplus_{i \in \tau} Z_{i, x}(Y) \oplus \bigoplus_{i \in \tau} Z_{i, f(x)}(Y) : x \in \bit^{n_1} \right\}$. Since $\lvert \tau \rvert \leq m$, the set of random variables is $\eps$-biased for linear tests of size up to $\frac{2t'}{m}$, and hence $\bigoplus_{i \in \tau} Z_{i, x}(Y) \oplus \bigoplus_{i \in \tau} Z_{i, f(x)}(Y)$ is $\eps'$-biased by Claim \ref{clm:pointwiseBiased}. Then by the Lemma \ref{lma:XORLemma}, since this holds for any $\tau \subseteq [m]$, the sequence $( Z_{1, X}(Y) \oplus Z_{1, f(X)}(Y),\ldots, Z_{m, X}(Y) \oplus Z_{m, f(X)}(Y))$ is $\eps' \cdot 2^{\frac{m}{2}}$-close to $\unif{m}$. It follows that, the probability of collision is at most:
\begin{equation*}
    2^{-m} + \eps' \cdot 2^{\frac{m}{2}} = 2^{-m} + 2^{\frac{m}{2}} \cdot 2^{(n_2 - k_2) / t} \cdot \left( \eps^{1/t} + (2t) \cdot 2^{-\frac{k_1}{3}} \right) \;.
\end{equation*}

We now bound the probability of collision based on our choice of parameters. Recall that Lemma \ref{lma:linearTests} asserts that we can construct $m \cdot 2^{n_2}$ many variables $Z_{(i, x)}$ that are $(t', \eps)$-biased using $2 \lceil \log(1/\eps) + \log \log (m 2^{n_2})   + \log(t') \rceil = 2 \lceil \log(1/\eps) + \log \log (m 2^{n_2}) + \log(2mt) \rceil$ random bits. Set $\eps = 2^{-r}$ where $r = \frac{1}{2} n_2 + 3\log n_2 + \log n_1$, $n_2 \geq 16$ and $k_1 \geq 64$. We then bound the probability separately depending on $k_1$'s value relative to $4(n_2 - k_2)$.

\textbf{If $k_1 \leq 4(n_2 - k_2)$:} Choose $t$ to be the smallest even integer such that $t \geq \frac{8 (n_2 - k_2)}{k_1}$. Then $t \leq n_2 - k_2$, or else that would imply that $k_1 \leq 8$. Then it follows that:
\begin{equation*}
    \frac{8(n_2 - k_2)}{k_1} \leq t \leq \frac{16(n_2 - k_2)}{k_1} \leq \frac{8 n_2}{k_1}
\end{equation*}

Using the inequality above:
\begin{align*}
    2^{(n_2 - k_2) / t} \cdot \left( \eps^{1/t} + (2t) \cdot 2^{-\frac{k_1}{3}} \right)
    &\leq 2^{(n_2 - k_2 - r) / t} + \frac{32(n_2 - k_2)}{k_1} 2^{-\frac{k_1}{3}}\\
    &\leq 2^{-\delta n_2 / t} + \frac{32(n_2 - k_2)}{k_1} 2^{-\frac{k_1}{3}}
    \\
    &\leq 2^{-\delta n_2 / t} + 2^{-\frac{k_1}{3} + \frac{k_1}{12}}\\ 
    &\leq 2^{-\delta \frac{k_1}{8}} + 2^{-\frac{k_1}{4}}\\ &\leq 2^{-\delta \frac{k_1}{8} + 1}\\
\end{align*}

\textbf{Otherwise, if $k_1 > 4(n_2 - k_2)$:} Set $t = 2$. Then:

\begin{align*}
    2^{(n_2 - k_2) / 2} \cdot \left( \eps^{1/2} + 4 \cdot 2^{-\frac{k_1}{3}} \right) &= 2^{(n_2 - k_2 - r) / 2} + 2^{(n_2 - k_2 )/2} \cdot 4 \cdot 2^{-\frac{k_1}{3}}\\
    &\leq 2^{-\delta n_2 / 2} + 2^{(n_2 - k_2 )/2} \cdot 4 \cdot 2^{-\frac{k_1}{3}}\\
    &\leq 2^{-\delta n_2 / 2} + 2^{-\frac{k_1}{8}}
\end{align*}

Choosing $m \leq \delta \min\{ \frac{n_2}{4}, \frac{k_1}{16} \} - 1$, we get that the collision probability is at most $2^{-m} + 2^{\frac{m}{2} - 2m - 1} \leq 2^{-m + 1}$.
\end{proof}

\section{A Fully Non-malleable Seeded Extractor}
\label{sec:fullynmseeded}
In this section, we will use $\crTreExt$ as $\innerExt$ and $\liExt$ as $\outerExt$ for Theorem \ref{thm:main_reduction} with the following instantiations:

\begin{enumerate}
    \item $\crTreExt$ is an extractor given by $[(n_x, k_x), (s, s) \mapsto d \sim \treError]$ for \\
    $s = O(\log^2(n_x)\log^2(1/\treError)$, and $d = \Omega(k_x)$, with collision probability $\left(\frac{\treError}{3}\right)^2$.
    \item $\liExt$ is an extractor given by $[(d, (1-\gamma)d), (d, (1-\gamma)d) \mapsto m \sim \liError]$ for some constant $\gamma$, $m = \Omega(d)$, and $\liError = 2^{-d \left(\frac{\log \log d}{\log d} \right)}$.
\end{enumerate}
with $\collisionError = 2^{-(k_x)^c}$ for some $c < \frac{1}{2}$. It follows that $s = o(d)$.

\begin{theorem}\label{thm:FNMSExt}
    For any $n_x, k_x$, there exists a fully non-malleable seeded extractor $\extractorParam{\fnmext}{(n_x, k_x)}{(s + d, s + d)}{m}{\fnmError}$ with $m = \Omega(d)$, $d < k_x$, $s = O(\log^2(n_x)\log^2(\treError))$, $\fnmError < 10\treError$ with $\treError = 2^{-(\frac{k_x}{2})^c}$ for some $c < \frac{1}{2}$.
\end{theorem}
\begin{proof}
    It suffices to show that for our choice of parameters, the entropy requirements of $\crTreExt$ (from Lemma \ref{lma:CRSeededExt}) and $\liExt$ (from Lemma \ref{lma:li}) are met for Theorem \ref{thm:main_reduction}.

    Setting input parameters $n_3 = n_x$, $k^*_1 = k_x$, $n_4 = k_4 = s$, $k^*_2 = s + d$, and extractor parameters $n_1 = n_2 = d$, $k_1 = k_2 = (1-\gamma)d$, $k_3 = k_x$, note that indeed $k_1^* \geq k_3$. Furthermore, 
    \begin{align*}
        k^*_2 &= s + d = s + k_4 + n_1 - n_4\\
        k^*_2 &= d + s \geq \left(\frac{\gamma}{2}\right)d + (1-\gamma)d + 2s \;.
    \end{align*}

    And thus by our choice of $s$, $\fnmError \leq 3\cdot 2^{-(\frac{k_x}{2})^{2c}} + 7\treError < 10 \treError$ with $\treError = 2^{-(\frac{k_x}{2})^c}$ for some $c < \frac{1}{2}$.
\end{proof}
    
It will also be useful in the subsequent subsection that we relax the entropy requirement of this extractor.
\begin{theorem}\label{thm:relaxedFNMSExt}
    For any $n_x, k_x$, there exists a fully non-malleable seeded extractor $\extractorParam{\fnmext}{(n_x, k_x)}{(s + d, s + d - 1)}{m}{\fnmError}$ with $m = \Omega(k_x)$, $d < k_x$, $s = O(\log^2(n_x)\log^2(\treError))$, $\fnmError < 12\treError$ with $\treError = 2^{-(\frac{k_x}{2})^c}$ for some $c < \frac{1}{2}$.
\end{theorem}
    
\begin{proof}
    By Lemma \ref{lma:extReduction} and Lemma \ref{lma:colReduction}, $\crTreExt$ can also be viewed as\\ $\extractorParam{\crTreExt}{(n_x, k_x)}{(s, s - 1)}{\Omega(k_x)}{2 \treError}$ with collision probability\\ ${2 \collisionError = 2\left(\frac{\treError}{3} \right)^2 \leq \treError^2}$. 
    
    For a similar choice of parameters: $n_3 = n_x$, $k^*_1 = k_x$, $n_4 = s$, $k_4 = s - 1$ $k^*_2 = s + d$, and extractor parameters $n_1 = n_2 = d$, $k_1 = k_2 = (1-\gamma)d$, $k_3 = k_x$, note that indeed $k_1^* \geq k_3$. Furthermore, 
    \begin{align*}
        k^*_2 &= s + d - 1 = s + s - 1 + d - s = s + k_4 + n_1 - n_4\\
        k^*_2 &= s + d - 1 \geq \left(\frac{\gamma}{2}\right)d + (1-\gamma)d + 2s - 1
    \end{align*}

    And thus by our choice of $s$, $\fnmError \leq 3\cdot 2^{-(\frac{k_x}{2})^{2c}} + 9\treError < 12 \treError$ with $\treError = 2^{-(\frac{k_x}{2})^c}$ for some $c < \frac{1}{2}$.
\end{proof}
\begin{remark} \label{multitamper2}
    As we have already mentioned in the Remark \ref{multitamper1}, we can use $t$-tamperable extractor like \cite{Li17}. As a result our $\fnmext$ will be $t$-tamperable non-malleable extractor with negligible error. One only has to make sure that $|y_\ell|<\frac{\gamma \cdot n }{t+1}$, which follows from the first point in the Remark \ref{multitamper1}, where $(1-\gamma)\cdot n$ is the entropy requirement from \cite{Li17} extractor. The error one obtains is therefore at least $2^{-\Omega(n/ \log n)}+2^{-\Omega(\frac{\gamma \cdot n}{t+1} - \log^2(n))}+t\cdot 2^{-\Omega (\frac{\gamma \cdot n}{t+1})}\geq 2^{-\Omega(n^c)}$, for $c<1$ depending on $t$ only. Please notice that the entropy requirements for this extractor do not change.
\end{remark}

\section{A Two-Source Non-malleable Extractor}
\label{sec:nmraz}
In this section, we will use $\razExt$ as $\innerExt$ and $\fnmext$ as $\outerExt$ from Theorem \ref{thm:relaxedFNMSExt} with the following instantiations:

\begin{enumerate}
    \item $\extractorParam{\razExt}{(n_x, k_x)}{(\YLeftLength, k_\ell)}{d}{\razError{d}}$ with $d = \Omega(\min\{k_x, k_\ell\})$ and collision probability $2^{-d + 1}$.
    \item $\extractorParam{\fnmext}{(n_y, \tau \cdot d)}{(d, d - 1)}{m}{\fnmError}$ is a two-source non-malleable extractor for some $0 < \tau < 1$, $m = \Omega(d)$, and $\fnmError < 12 \cdot \treError$ with $\treError < 2^{-\Omega((m)^c)}$ for some $c < \frac{1}{2}$.
\end{enumerate}

\begin{theorem}\label{thm:betterTwoNMExt}
    There exists a two source non-malleable seeded extractor $\twonmext : [(n_x, k_x), (n_y, k_y) \mapsto m \sim \twonmError]$, and $m = \Omega( \min\{ n_y, k_x \} )$, such that:

    \begin{enumerate}
        \item $k_x \geq 12 \log(n_y - k_y) + 15$,
        \item $n_y \geq 30 \log(n_y) + 10\log(n_x) + 20$,
        \item $k_y \geq (\frac{4}{5} + \gamma)n_y + 3\log(n_y) + \log(n_x) + 4$,
        \item $\twonmError \leq 3 \cdot 2^{- \frac{9\gamma}{10} n_y} + 40 \cdot \treError$ where $\treError = 2^{-\Omega(d^c)}$ with $c < \frac{1}{2}$.
    \end{enumerate}
\end{theorem}

\begin{proof}
For any given $Y \givenBy (n_y, k_y)$, we treat it as $Y = \YLeft \concat \YRight$ where $\lvert \YLeft \rvert = n_\ell$ and $\lvert \YRight \rvert = n_r$. 

The extractor $\extractorParam{\razExt}{(n_x, k_x)}{(\YLeftLength, k_\ell)}{d}{\razError{d}}$ from Lemma $\ref{lma:CRRaz}$ requires the following conditions:
\begin{enumerate}
    \item $k_x \geq 12\log(\YLeftLength - k_\ell) + 15$
    \item $\YLeftLength \geq 6 \log \YLeftLength + 2\log n_x + 4$,
    \item $k_\ell \geq (\frac{1}{2} + \gamma) \cdot \YLeftLength + 3\log \YLeftLength + \log n_x + 4$,
    \item $d \leq \gamma \min\{ \frac{n_\ell}{4}, \frac{k_x}{16} \} - 1$
\end{enumerate}
for some $0 < \gamma < \frac{1}{2}$.

Setting $\YLeftLength = (\frac{2}{5} - \gamma) n_y$ (and consequently $n_r = (\frac{3}{5} + \gamma)n_y$), we first show that indeed the input requirements for $\razExt$ are met. Note that 
\begin{align*}
    (n_y - k_y) - (n_\ell - k_\ell) = n_y - k_y - (n_\ell - (k_y - n_r)) = 0
\end{align*}
which implies that:
\begin{align*}
    k_x \geq 12 \log(n_y - k_y) + 15 = 12 \log(n_\ell - k_\ell) + 15
\end{align*}
Next:
\begin{align*}
    n_\ell \geq \frac{1}{5} n_y \geq 6 \log(n_y) + 2\log(n_x) + 4 \geq 6 \log(n_\ell) + 2\log(n_x) + 4
\end{align*}
And lastly:
\begin{align*}
    k_\ell \geq k_y - n_r &= \left(\frac{4}{5} + \gamma\right)n_y + 3\log(n_\ell) + \log(n_x) + 4 - \left(\frac{3}{5} + \gamma\right) n_y\\
    &= \left(\frac{1}{5}\right)n_y + 3\log(n_\ell) + \log(n_x) + 4\\
    &= \left(\frac{1}{5}\right) \left(\frac{1}{0.4 - \gamma}\right) n_\ell + 3\log(n_\ell) + \log(n_x) + 4\\
    &\geq \left(\frac{1}{2} + \frac{5\gamma}{4}\right) n_\ell + 3\log(n_\ell) + \log(n_x) + 4\\
\end{align*}

Setting input parameters $n_3 = n_x$, $k_1^* = k_x$, $n_1 = n_y$, $k_2^* = (\frac{4}{5} + \gamma) n_y$, and extractor parameters $n_4 = n_\ell$, $k_4 = k_\ell$, $n_1 = n_y$, $k_1 = \tau \cdot d$, $n_2 = d$, $k_2 = d - 1$ for some $0 < \tau < 1$, we get that $k_1^* \geq k_3$. Furthermore:
\begin{align*}
    k_2^* - k_4 - n_1 + n_4 &= k_y  - k_\ell - n_y + n_\ell = k_y - k_\ell - \left(\frac{3}{4} + \gamma\right) n_y \\
    &\geq \left( \frac{1}{5} + \gamma \right) n_y - \left( \frac{1}{2} + \gamma \right)\left( \frac{2}{5} - \gamma \right)n_y\\
    &= \left( \frac{11}{10} \gamma + \gamma^2 \right) n_y
\end{align*}
and:
\begin{align*}
    k_2^* - k_1 - 2n_4 &= k_2^* - \tau \cdot d - 2n_\ell \geq \gamma n_y - \tau \gamma \frac{n_\ell}{4} \geq \frac{9\gamma}{10} n_y 
\end{align*}

Thus, by Theorem \ref{thm:main_reduction} it follows that $\extractorParam{\twonmext}{(n_3, k_1^*)}{(n_1, k_2^*)}{m}{ \twonmError }$ is a strong non-malleable extractor with error:
\begin{align*}
    \twonmError &\leq 3 \cdot 2^{- \frac{9\gamma}{10} n_y} + 36 \cdot \treError + 2 \cdot 2^{-\frac{3}{2} d} + 2 \sqrt{2^{-d + 1}}\\
                &\leq 3 \cdot 2^{- \frac{9\gamma}{10} n_y} + 40 \cdot \treError
\end{align*}
where $\treError = 2^{-\Omega(d^c)}$ with $c < \frac{1}{2}$.
\end{proof}
\begin{remark}\label{multitamper3}
    As we noted in Remark \ref{multitamper1} we can use multi-tampering extractor from Remark \ref{multitamper2}, and obtain a $t$-tamperable non-malleable extractor. The error of such extractor remains negligible. Entropy requirements change due to first point from Remark \ref{multitamper1}: One source can have poly-logarithmic entropy, while the other requires entropy rate $(1-\frac{1}{2t+3})$. 
\end{remark}

\section{A Two-Source Non-malleable Extractor With Rate $\frac{1}{2}$}
\label{sec:ratehalfnm}
In \cite{AKOOS21} the authors give a compiler that turns any left-strong non-malleable extractor into a non-malleable extractor with optimal output rate of $\frac 12$. The construction looks as follows:
\begin{equation*}
    \twonmext^*(X,Y)=\sext(X, \twonmext(X,Y)),
\end{equation*}
where $\sext$ is a seeded extractor from \cite{GUV09} with output size equal $\frac 12 \minEnt{X}$, and $\twonmext$ is a left-strong non-malleable extractor.

We will briefly discuss the idea behind that construction. Let $X'$ be a tampering of $X$, and $Y'$ be a tampering of $Y$. We need to argue that if $X\neq X' \; \lor \; Y\neq Y'$ then $\twonmext^*(X,Y)$ remains uniform even given $\twonmext^*(X',Y')$.

If $X\neq X'\; \lor \; Y\neq Y'$ then left-strong non-malleable extractor $\twonmext(X,Y)$ is uniform even given $\twonmext(X',Y'), X$. The final idea crucially relies on the fact that $\sext$ extracts only half of the entropy of $X$: we can reveal $\twonmext(X',Y')$ and then $\sext(X',\twonmext(X',Y'))$ becomes a leakage from $X$ (i.e. it is just a deterministic function of $X$ with a small output). We get that\\
$\avgCondMinEnt{X}{\twonmext(X',Y'),\sext(X',\twonmext(X',Y'))} \ \approx \frac 12 \minEnt{X}$ \\(size of $\twonmext(X',Y')$ is tiny so it's asymptotically irrelevant). Moreover by the left-strong property of $\twonmext$ we get that $X$ and $\twonmext(X,Y)$ remain independent given $\twonmext(X',Y'),\sext(X',\twonmext(X',Y'))$, this means that $\sext(X,\twonmext(X,Y))$ is uniform given $\twonmext(X',Y')$ and\\ $\sext(X',\twonmext(X',Y'))$ which gives the result.


If we make use of $\twonmext$ from the previous section we can obtain a two-source unbalanced non-malleable extractor with rate $\frac{1}{2}$.

\begin{lemma}[Theorem 5 of \cite{AKOOS21}]\label{lma:RateHalfCompiler}
    If $\extractorParam{\twonmext}{(n_1, k_1)}{(n_2, k_2)}{d}{\eps_1}$ is a strong two-source unbalanced non-malleable extractor, with $n_2 = o(n_1)$ and \\
    $\extractorParam{\ext}{(n_1, k_1)}{(d, d)}{\ell}{\eps_2}$ is a strong seeded extractor, then there exists a two source non-malleable extractor $\extractorParam{\twonmext^*}{(n_1, k_1)}{(n_2, k_2)}{\ell}{ \eps_1 + \eps_2}$. Furthermore, if $k_1, \ell < \frac{n_1}{2}$, then $\twonmext^*$ has a rate of $\frac{1}{2}$.
\end{lemma}

\begin{theorem}
    There exists an extractor $\extractorParam{\twonmext^*}{(n_1, k_1)}{(n_2, k_2)}{\ell}{\eps_1 + \eps_2}$ such that:
    \begin{enumerate}
        \item $k_1 \geq \max\{ 12 \log(n_2 - k_2) + 15, \log^3(n_1) \log(1/\eps_2) \}$
        \item $n_2 \geq \max\{ 30 \log(n_2) + 10\log(n_1) + 20, \log^3(n_1) \log(1/\eps_2) \}$
        \item $k_2 \geq (\frac{4}{5} + \gamma)n_2 + 3\log(n_2) + \log(n_1) + 4$
        \item $\eps_1 \leq 3 \cdot 2^{- \frac{9\gamma}{10} n_2} + 40 \cdot \treError$ where $\treError = 2^{-\Omega(d^c)}$ with $c < \frac{1}{2}$
        \item $\ell < \frac{k_1}{2}$
    \end{enumerate}
    Furthermore, if $n_2 = o(n_1)$, $k_1, \ell < \frac{n_1}{2}$, then $\twonmext^*$ has a rate of $\frac{1}{2}$.
\end{theorem}

\begin{proof}
    By Theorem \ref{thm:betterTwoNMExt} there exists an extractor $\extractorParam{\twonmext}{(n_1, k_1)}{(n_2, k_2)}{m}{\eps_1}$ such that:
    \begin{enumerate}
        \item $k_1 \geq 12 \log(n_2 - k_2) + 15$
        \item $n_2 \geq 30 \log(n_2) + 10\log(n_1) + 20$
        \item $k_2 \geq (\frac{4}{5} + \gamma)n_2 + 3\log(n_2) + \log(n_1) + 4$
        \item $\eps_1 \leq 3 \cdot 2^{- \frac{9\gamma}{10} n_2} + 40 \cdot \treError$ where $\treError = 2^{-\Omega(d^c)}$ with $c < \frac{1}{2}$
        \item $m = \Omega( \min\{ n_2, k_1 \} )$
    \end{enumerate}

    Using Lemma \ref{lma:tre}, $\extractorParam{\treExt}{(n_1, k_1)}{(m, m)}{\Omega(k_1)}{\eps_2}$ is a strong seeded extractor with $m = O(\log^2(n_1)\log(1/\eps_2))$. Thus by Lemma \ref{lma:RateHalfCompiler} there exists a two source non-malleable extractor $\extractorParam{\twonmext^*}{(n_1, k_1)}{(n_2, k_2)}{\Omega(k_1)}{\eps_1 + \eps_2}$.

    Furthermore, with $n_2 = o(n_1)$ and $k_1, \ell < \frac{n_1}{2}$, we get that $\twonmext^*$ has a rate of at most $\frac{n_1}{2(n_1 + n_2)} < \frac{1}{2}$.
\end{proof}

\section{Privacy Amplification against Memory Tampering Active Adversaries.}\label{App:PA}

Imagine Alice and Bob sharing some random but not uniform string $W$, they would like to "upgrade" their random string $W$ to uniformly random string. However Eve is fully controlling a channel between Alice and Bob and can arbitrarily tamper with the messages sent. The Privacy Amplification (PA) protocol guarantees that either Alice and Bob will end up with the same uniform string (unknown to Eve), or at least one of them will abort\footnote{If one of the parties, say Alice, aborts but Bob generates random string $R_B$ then we require $R_B$ to be uniform and unknown to Eve.}.

In \cite{AORSS20} the authors consider a stronger version of PA which they call a \emph{privacy amplification resilient against memory-tampering active adversaries}. In their model, Alice and Bob have access to a shared string $W$ and their local sources of (not necessarily uniform) randomness $A$ and $B$ respectively. At the beginning of the protocol Eve can select one party, say Alice, and corrupt her memory $F(W,A)=(\tilde W, \tilde A)$ (or $F(W,B)=(\tilde W, \tilde B)$ if Eve decides to corrupt Bob). If Eve did not corrupt the memory of any of the parties then the standard PA guarantees follow. On the other hand if Eve decides to corrupt one of the parties then either Alice and Bob agree on a uniformly random string (unknown to Eve) or the non-corrupted party will detect the tampering.

The following two definitions are taken verbatim from \cite{AORSS20}.
\begin{definition}[Protocol against memory-tampering active adversaries]
    An \emph{$(r,\ell_1,k_1,\ell_2,\linebreak[1] k_2,m)$-protocol against memory-tampering active adversaries} is a protocol between Alice and Bob, with a man-in-the-middle Eve, that proceeds in $r$ rounds.
    Initially, we assume that Alice and Bob have access to random variables $(W,A)$ and $(W,B)$, respectively, where $W$ is an $(\ell_1,k_1)$-source (the \emph{secret}), and $A$, $B$ are $(\ell_2,k_2)$-sources (the \emph{randomness tapes}) independent of each other and of $W$. The protocol proceeds as follows:
    \begin{description}
        \item In the first stage, Eve submits an arbitrary function $F:\{0,1\}^{\ell_1}\times\{0,1\}^{\ell_2}\to \{0,1\}^{\ell_1}\times\{0,1\}^{\ell_2}$ and chooses one of Alice and Bob to be corrupted, so that either $(W,A)$ is replaced by $F(W,A)$ (if Alice is chosen), or $(W,B)$ is replaced by $F(W,B)$ (if Bob is chosen).
    
   \item  In the second stage, Alice and Bob exchange messages $(C_1,C_2,\dots,C_r)$ over a non-authenticated channel, with Alice sending the odd-numbered messages and Bob the even-numbered messages, and Eve is allowed to replace each message $C_i$ by $C'_i$ based on $(C_1,C'_1,\dots,C_{i-1},C'_{i-1},C_i)$ and independent random coins, so that the recipient of the $i$-th message observes $C'_i$.
    Messages $C_i$ sent by Alice are deterministic functions of $(W,A)$ and $(C'_2,C'_4,\dots,C'_{i-1})$, and messages $C_i$ sent by Bob are deterministic functions of $(W,B)$ and $(C'_1,C'_3,\dots,C'_{i-1})$.
    
 \item    In the third stage, Alice outputs $S_A\in\{0,1\}^m\cup\{\bot\}$ as a deterministic function of $(W,A)$ and $(C'_2,C'_4,\dots)$, and Bob outputs $S_B\in\{0,1\}^m\cup\{\bot\}$ as a deterministic function of $(W,B)$ and $(C'_2,C'_4,\dots)$.

    \end{description}
    \end{definition}

\begin{definition}[Privacy amplification protocol against memory-tampering active adversaries]
    An \emph{$(r,\ell_1,k_1,\ell_2,k_2,m,\eps,\delta)$-privacy amplification protocol against memory-tampering active adversaries} is an $(r,\ell_1,k_1,\ell_2,k_2,m)$-protocol against memory-tampering active adversaries with the following additional properties:
    \begin{itemize}
        \item \textbf{ If Eve is passive:} In this case, $F$ is the identity function and Eve only wiretaps.
        Then, $S_A=S_B\neq \bot$ with $S_A$ satisfying
        \begin{equation}\label{eq:sa}
            S_A,C\approx_\eps U_m,C,
        \end{equation}
        where $C=(C_1,C'_1,C_2,C'_2,\dots,C_r,C'_r)$ denotes Eve's view.
        
        \item \textbf{ If Eve is active:} Then, with probability at least $1-\delta$ either $S_A=\bot$ or $S_B=\bot$ (i.e., one of Alice and Bob detects tampering), or $S_A=S_B\neq\bot$ with $S_A$ satisfying~\eqref{eq:sa}.
    \end{itemize}
\end{definition}

One building block of our extension is MAC:
\begin{definition}\label{def: mac}
 A family of functions $ \mac:\{0,1\}^{\gamma}\times \{0,1\}^{\tau}\rightarrow\{0,1\}^{\delta},\mathtt{Verify}: \{0,1\}^{\gamma} \times \{0,1\}^{\delta}\times \{0,1\}^{\tau} \rightarrow \{0,1\} $ is said to be a $ \mu -$secure one time message authentication code if 
	\begin{enumerate}
		\item For $ k_{a}\in_R\{0,1\}^{\tau},\ \forall \; m\in\{0,1\}^{\gamma} $, $ \Pr[\mathtt{Verify}(m,\mac_{k_{a}}(m),k_{a})=1]=1 $,\\[1mm]
		where for any $(m,t)$, $\mathtt{Verify}(m,t,k_{a}):=$
		$\begin{cases}
		1\text{ if }\ \mac(m,k_{a})=t\\
		0\text{ otherwise}
		\end{cases}$
		\item For any $ m\neq m',t,t' $, $ \Pr\limits_{k_{a}}[\mac(m,k_{a})=t|\mac(m',k_{a})=t']\leq \mu $, where $ k_{a}\in_R\{0,1\}^{\tau}$.
	\end{enumerate}
 \end{definition}

\begin{lemma}\label{lemma-mac}\cite{JKS93,DKKRS12}
	For any $ \gamma,\eps >0 $ there is an efficient $ \eps- $secure one time $\mac$ with $ \delta\leq (\log(\gamma)+\log(\dfrac{1}{\eps})) $, $ \tau\leq 2\delta $, where $ \tau,\gamma,\delta $ are key, message, tag length respectively.
\end{lemma}

\begin{figure*}[h]
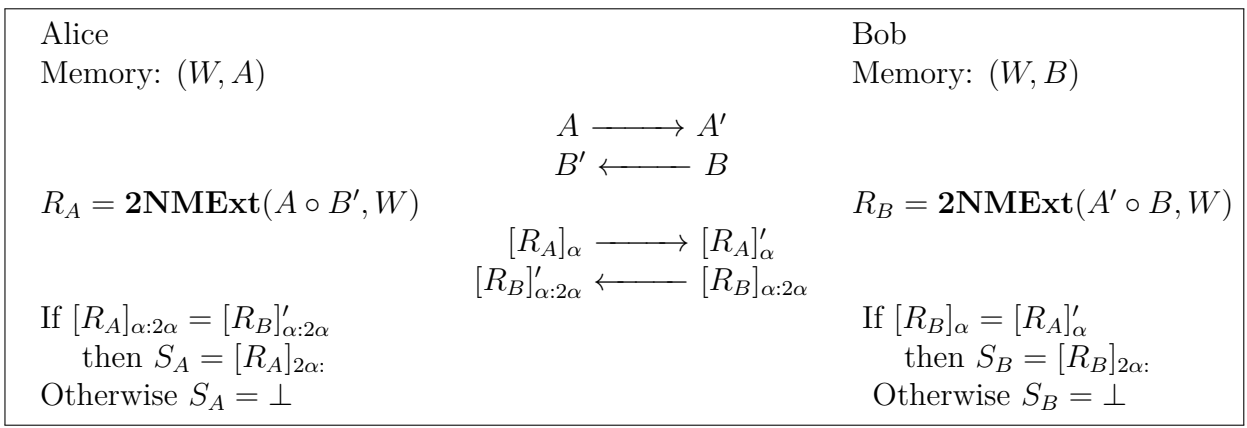

\centering
\fbox{
\begin{minipage}{15.8cm}
\begin{tabular}{lcl}
Alice & ~ &  Bob \\
Memory: $(W, A)$ & ~ &  
Memory: $(W, B)$ \vspace{0.2cm}\\

 ~ & $A$ $\xrightarrow{\quad \quad \hspace{0.5em}  }$ $A'$ ~\\
 ~ & $B'$ $\xleftarrow{\quad \quad \hspace{0.5em} }$ $B$ ~\\
$R_A=\twonmext(A\concat B',W)$  ~ &  ~ &  $R_B=\twonmext(A'\concat B,W)$\\
 ~ & $[R_A]_\alpha$ $\xrightarrow{\quad \quad \hspace{0.5em}  }$ $[R_A]'_\alpha$ ~\\
 ~ & $[R_B]'_{\alpha : 2\alpha}$ $\xleftarrow{\quad \quad \hspace{0.5em}  }$ $[R_B]_{ \alpha : 2\alpha}$ ~\\
If $[R_A]_{\alpha : 2\alpha}=[R_B]'_{\alpha : 2\alpha}$ & ~ & ~If $[R_B]_{\alpha}=[R_A]'_{\alpha}$ \\
$\quad \text{  then  } S_A= [R_A]_{ 2\alpha:}$ & ~ & ~$\quad \text{  then  } S_B= [R_B]_{2\alpha:}$ \\
Otherwise $S_A= \bot$ & ~& ~ Otherwise $S_B= \bot$ \\
\end{tabular}

\end{minipage}

}
\caption{Verbatim from \cite{AORSS20}. Privacy amplification protocol against memory-tampering active adversaries. In the above, for an $n$-bit string $x$ we define $[x]_i=(x_1,x_2,\dots,x_i)$, $[x]_{i:j}=(x_{i+1},\dots,x_j)$, and $[x]_{j:}=(x_{j+1},\dots,x_n)$.}
\label{fig:ITPA}
\end{figure*}

In the \cite{AORSS20} protocol Alice and Bob exchange the random strings $A$ and $B$ and then locally compute $R=\twonmext(A\concat B,W)$. They then split $R$ into $3$ parts, Alice sends the first part to Bob to prove she has gotten the right output, Bob then sends the second part to Alice to do the same. If this phase was successful then last part of $R$ is the shared uniform string. Figure \ref{fig:ITPA} illustrates the protocol.

Since one of the sources of randomness might be faulty, even if the original $A, B$ were uniform, one requires a left-strong non-malleable extractor $\twonmext$ to remain secure for the first source with entropy below $0.5$, the construction of such an extractor prior to this work was unknown\footnote{Authors of \cite{AORSS20} proceed to construct a computational non-malleable extractors with parameters that would allow for this protocol to go through.}.

\begin{figure*}[h]
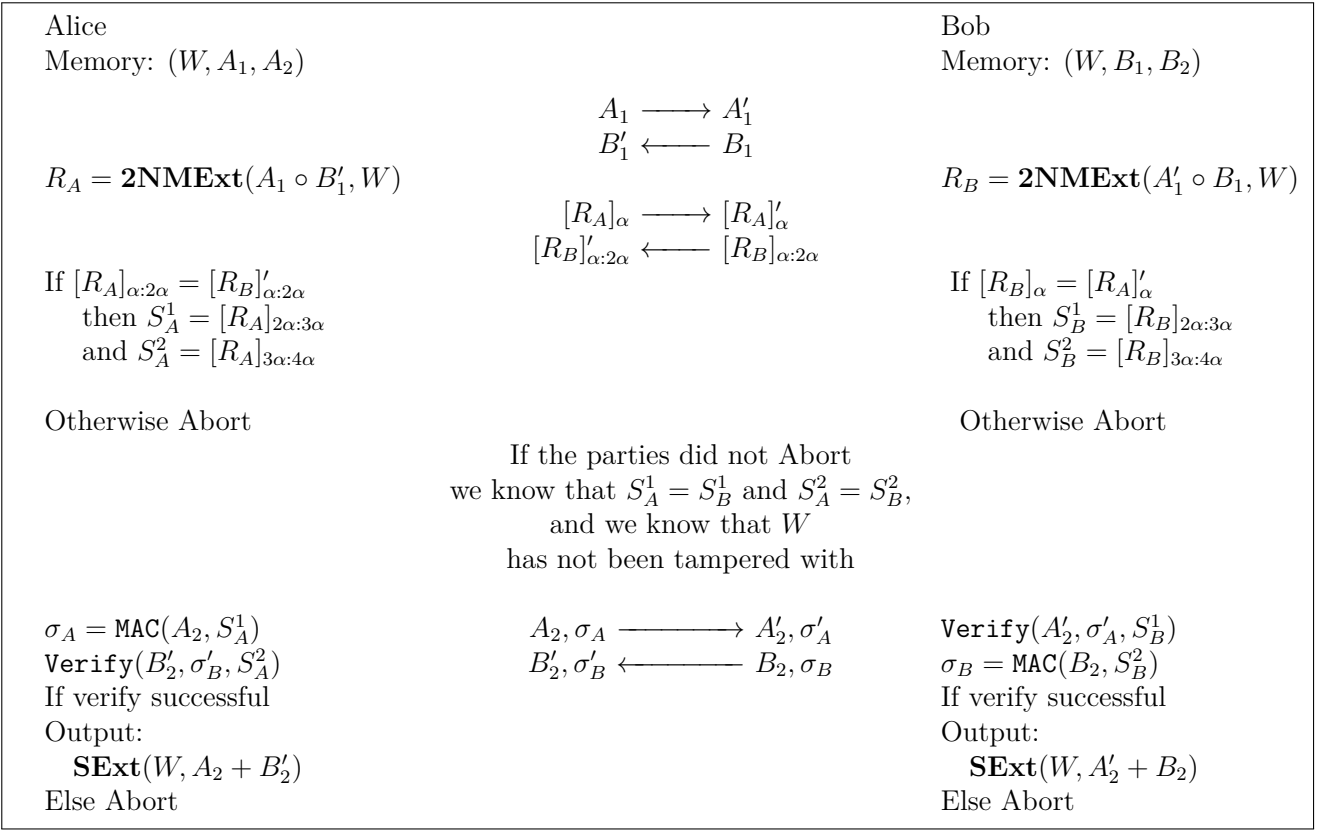

\centering
\hspace*{-3em}
\fbox{
\scalebox{.9}{
\begin{minipage}{18.5cm}
\begin{tabular}{lcl}
Alice & ~ &  Bob \\
Memory: $(W, A_1, A_2)$ & ~ &  
Memory: $(W, B_1, B_2)$ \vspace{0.2cm}\\

 ~ & $A_1$ $\xrightarrow{\quad \quad  }$ $A'_1$ ~\\
 ~ & $B'_1$ $\xleftarrow{\quad \quad   }$ $B_1$ ~\\
$R_A=\twonmext(A_1\concat B'_1, W)$  ~ &  ~ &  $R_B=\twonmext(A'_1\concat B_1, W)$\\
 ~ & $[R_A]_\alpha$ $\xrightarrow{\quad  \quad  }$ $[R_A]'_\alpha$ ~\\
 ~ & $[R_B]'_{\alpha : 2\alpha}$ $\xleftarrow{\quad \quad }$ $[R_B]_{ \alpha : 2\alpha}$ ~\\
If $[R_A]_{\alpha : 2\alpha}=[R_B]'_{\alpha : 2\alpha}$ & ~ & ~If $[R_B]_{\alpha}=[R_A]'_{\alpha}$ \\
$\quad \text{  then  } S^1_A= [R_A]_{ 2\alpha: 3\alpha}$ & ~ & ~$\quad \text{  then  } S^1_B= [R_B]_{2\alpha: 3\alpha}$ \\
$\quad \text{  and  } S^2_A= [R_A]_{ 3\alpha: 4\alpha}$ & ~ & ~$\quad \text{  and  } S^2_B= [R_B]_{3\alpha: 4\alpha}$ \\
~ & & ~ \\
Otherwise Abort & ~& ~ Otherwise Abort \\
~ &If the parties did not Abort  & ~\\
~ &we know that $S^1_A=S^1_B$ and $S^2_A=S^2_B$,  & ~\\
~ &and we know that $W$ & ~\\
~ &has not been tampered with  & ~\\
~ & & ~ \\
$\sigma_A= \mac(A_2,S^1_A)$ & $A_2, \sigma_A$ $\xrightarrow{\quad \quad \quad \quad  }$ $A'_2, \sigma'_A$ & $\mathtt{Verify}(A'_2,\sigma'_A, S^1_B)$\\
 $\mathtt{Verify}(B'_2,\sigma'_B, S^2_A)$ & $B'_2, \sigma'_B$ $\xleftarrow{\quad \quad \quad \quad}$ $B_2, \sigma_B$ & $\sigma_B=\mac(B_2,S^2_B)$\\
 If verify successful & & If verify successful\\
 Output: & & Output: \\
 \quad $\sext(W,A_2+B'_2)$ & & \quad $\sext(W,A'_2+B_2)$\\
 
 Else Abort & & Else Abort\\
\end{tabular}

\end{minipage}

}}
\caption{Extension of the original PA protocol. $R$ is split into $4$ parts instead of $3$. Here $\mac$ is a standard information theoretic message authentication code (MAC). And $\sext$ is any seeded extractor. When party Aborts it stops responding and the final output is $\bot$. }
\label{fig:ITPA2}
\end{figure*}

The above protocol obtains very short output compared to entropy of $W$, whereas ideally we would like to obtain something close to entropy of $W$.
If Alice and Bob have access to uniform randomness, one can extend this protocol to output almost as many bits as $W$'s entropy (see Figure \ref{fig:ITPA2}). After the execution of the \cite{AORSS20} protocol we have the additional guarantee (see proof of Theorem 6, point (b)) that if $S_A\neq \bot $ and $S_B\neq \bot $ then we know that $S_A = S_B$ and are close to uniform and moreover Eve did not tamper with $W$ of either of the parties (this is only achieved with standard notion of non-malleability, not the one from~\cite{GSZ21}). If Alice and Bob have access to some extra uniform bits (if $A$ and $B$ were uniform to start with then we could cut them in half $A=A_1\concat A_2$ and $B=B_1\concat B_2$, use the first half to run the original protocol by \cite{AORSS20} and save the other half for later) then we can continue the protocol (in the spirit of~\cite{DW09}): 
Alice will send $A_2,\sigma_A$ to Bob, where $\sigma_A$ is a Message Autentication Code of $A_2$ with first half of $S_A$ as a key. Bob will do the same: send $B_2, \sigma_B$ to Alice using other half of $S_B$ as a MAC key. There is a one final problem, we know that one of $A_2$ or $B_2$ is uniform but we don't know which (Eve could have left $W$ unchanged but could have tampered with random coins $A$ and $B$), moreover one of them might depend on $W$. Notice that $A_2$ and $B_2$ will remain independent, and one of them is independent of $W$ and uniform. Therefore $A_2+B_2$ is uniform and independent of $W$. Now all we have to do is plug in $W$ and $A_2+B_2$ into seeded extractor $\sext(W,A_2+B_2)$ and we can extract almost whole entropy out of $W$ (and the output remains hidden from the view of Eve).


Let us analyse the protocol described in Figure \ref{fig:ITPA2} (we copy the figure below). Let $\twonmext$ be a $[(\ell_1, k_1-2\ell_2-2\gamma-1), (2\cdot\ell_2, \ell_2-\gamma-1)  \mapsto 4\alpha \sim \epsilon]$ strong non-malleable extractor for some parameter $\gamma>0$. Let shared secret $W\in \{0,1\}^{\ell_1}$ have min-entropy $k_1$, let $A_1,A_2,B_1,B2 \in \{0,1\}^{\ell_2}$ be uniform random variables. 
If Eve is passive the security is straight forward thus we will only consider the case of active Eve. We will follow the original proof \cite{AORSS20} very closely.
Let us focus on the case where Alice is the one with corrupted memory $F(W,(A_1,A_2))= \tilde W, (\tilde A_1, \tilde A_2)$. Since randomness $(\tilde A_1,\tilde A_2)$ is controlled by the adversary we can simply reveal $(\tilde a_1, \tilde a_2)=(\tilde A_1,\tilde A_2)$ it along with original randomness $(a_1,a_2)=(A_1,A_2)$, this makes $\tilde W$ only a function of $W$, let's denote it as $\tilde W= f(W)$, moreover let us denote $B'_1=g(B_1)$. We define  $\cL=\{w:f(w)=w\}$ and $\cR=\{b_1:g(b)=b\}$.

In the proof of Theorem 6 in \cite{AORSS20} in point (2.b) authors prove that if\\ $\Pr(W\notin \cL \lor B_1\notin \cR \lor a_1=\tilde a_1 ) > 2^{-\gamma}$ then $\Pr(S_B\neq \bot\;|\; W\notin \cL \lor B_1 \notin \cR)<\epsilon + 2^{-\alpha}$, thus Bob will abort.

The only case left to analyse is the point (2.a) where $W\in \cL \land B_1\in \cR \land a_1=\tilde a_1$. We assume that  
$\Pr(W\in \cL \land B_1\in \cR \land a_1=\tilde a_1) > 2^{-\gamma}$ (else this case happens with negligible probability). Authors argue that $W$ has enough entropy and thus $R_A$ is $\epsilon$ close to uniform. 
If $[R_A]'_\alpha=[R_A]_\alpha$ and $[R_B]'_{\alpha:2\alpha}=[R_B]_{\alpha:2\alpha}$, then $S^1_A\concat S^2_A=S^1_B\concat S^2_B \neq \bot$ and $S^1_A\concat S^2_A$ is $\epsilon$ close to uniform given Eve's view.
Now we know that $S^1_A\concat S^2_A=S^1_B\concat S^2_B \neq \bot$ and $\tilde W= W$ so we can follow with the analysis of the extension:
First of all the $\avgCondMinEnt{W}{A_1,A_2,\tilde A_1, \tilde A_2, W\in \cL}> k_1-2\ell_2-\gamma$ (where $|A_i|=\ell$, and $\gamma$ penalty comes from probability of the event $W\in \cL$). Now notice that by the security of MAC either $\Pr((A_2\neq A'_2 \lor B_2\neq B'_2) \land \text{ neither Alice or Bob Aborts}) <2\cdot 2^{-\Omega(\alpha)}$.

Further observe that even if Eve controls $A_2$, and $ A_2$ has no entropy and it might depend on $W$, still $B_2$ is uniform and independent of $(A_2)$. Thus $A_2+B_2$ is uniform\footnote{Technically speaking Eve can abort protocol by tampering with $A_2$ or $B_2$, Alice and Bob will simply abort. However $A_2$ and $B_2$ are no longer fully uniform conditioned on the event that Eve let them through. This is not a problem, by Lemma \ref{lma:extReduction}, this only doubles extraction epsilons.} and independent of $W$. Now we have uniform independent seed, all we have to do is extract:

Let $\sext:\{0,1\}^{\ell_1}\times\{0,1\}^{\ell_2}\rightarrow \{0,1\}^{ 0.999\cdot(k_1-2\ell_2-\gamma)}$ is a strong seeded extractor\footnote{Constant $0.999$ is just a placeholder for any constant less then $1$. By \cite{GUV09} we know that such explicit extractor exists.} with the error $2^{-\Omega(\ell_2)}$. 
Since $W$ has enough entropy  $\sext(W,A_2+B_2)$ is $2^{-\Omega(\ell_2)}$ close to uniform given the view of Eve. The analysis for Eve corrupting Bob is symmetrical. Thus we obtain the following:


\begin{theorem}
Let $\twonmext$ be a $[(\ell_1, k_1-2\ell_2-2\gamma-1), (2\cdot\ell_2, \ell_2-\gamma-1)  \mapsto 4\alpha \sim \epsilon]$ strong non-malleable extractor.
Then, there exists an $(r=6,\ell_1,k_1,2\cdot\ell_2,2\cdot \ell_2,0.999\cdot(k_1-2\ell-\gamma) ,2^{-\Omega(\ell_2)} ,\delta=\eps+2^{-\alpha}+2\cdot 2^{-\gamma}+2^{-\Omega(\alpha)})$-privacy amplification protocol against memory-tampering active adversaries.
\end{theorem}
And thus when we plug in our extractor and some example parameters we get:
\begin{corollary}
 For shared secret $W$ with $|W|=n$ and  $\minEnt{W}>0.803 \cdot n$ and $|A_i|=|B_i|=0.001n$ we get privacy amplification protocol that outputs $0.8 \cdot n$ uniform bits, and has a security  $2^{-\Omega(\sqrt{n})}$.
\end{corollary}

\chapter{Stronger 3SUM-Indexing Lower Bounds}

Finally, in this chapter (based on the paper \cite{CL23}), we study one-way functions against pre-processing adversaries.
One-way functions have been studied in this setting first
by Hellman in \cite{Hel80}. A pre-processing adversary has two phases. During the \emph{online phase}, the adversary is first given oracle access to a function 
$F : [N] \to [N]$ and is given unbounded time, and an unbounded number of oracle accesses to $F$, and 
has to output $S$ memory cells of $w$ bits. Subsequently during the \emph{offline phase}, the adversary is then given an input $y$ on which it must output an $x$ for which $F(x) = y$, in $T$ time steps (including number of accesses into the memory cells, and oracle accesses to $F$). Hellman showed that there 
exists a pre-processing adversary with $S$ space and time $T$
with $S^2T = O(N^2)$ when $F$ is chosen uniformly at random.
Fiat and Naor then followed up in \cite{FN00}, showing that 
adversaries exist when $S^3T = O(N^3)$ for any function (not just randomly chosen ones). Lastly, De, Trevisan, and Tulsiani in \cite{DTT10} give an adversary that inverts 
on at least $\eps$ fraction of inputs, with $S$ and $T$ both
being at most $\max\{ \eps^{\frac{5}{4}} N^\frac{3}{4}, \sqrt{\eps N} \}$. We can think of these as the first 
step in creating one-way functions against pre-processing
adversaries --- by using $F$ directly as the one-way function.

\paragraph{(Backdoors, and Immunization)}
Following this, Golovnev et al. in \cite{GGHPV20} observe that there is a ``trivial'' adversary that breaks $F$ by essentially storing
a huge inversion table (and thus having $S = N$). At face value, while this might look
infeasible for most people, it might be possible for big organisations (such as the NSA) to afford that one-time cost, such that
all subsequent inversions are cheap. Worse still, adversaries may take the initiative 
by influencing the design of protocols to use such compromised primitives. An often cited case is when the cryptographic PRG known as Dual\_EC\_DRBG was standardised by NIST, which was widely suspected to contain a backdoor by the NSA \cite{AFMV19,FJM18,DGGJR15}. 
Golovnev et al. in the same work \cite{GGHPV20} also propose an interesting follow-up to address this by creating one-way functions based on $F$ that potentially circumvent this 
shortcoming. They refer to this as \emph{immunization}. They do this by relying on the hardness of the data structure variant of the $3$SUM problem, called $3$SUM-Indexing. 
More concretely, the proposed one-way function $F'$
on input $x \in \bit^n$, cuts the input $x$ into 
two parts $x_1, x_2 = x$, and then outputs $F(x_1) \oplus F(x_2)$. Golovnev et al. further show that inverting $F'$ is essentially equivalent to solving $3$SUM-Indexing \footnote{In \cite{GGHPV20} they give a construction using $k$SUM, but all other complexity related results are for the case of $k = 3$.}. Thus, the hardness of $3$SUM-Indexing is directly linked to the security of this construction.

\paragraph{(The $3$SUM Problem)}
We will first give a brief exposition on the rich history
of the $3$SUM problem in algorithms and complexity.
In the $3$SUM Problem, we are given a set $\cS$ of $n$ group elements from an abelian group $(\cG,+)$ and the goal is to determine whether there is a triple $a,b,c \in \cS$ such that $a+b = c$. The $3$SUM Problem was originally introduced by Gajentaan and Overmars~\cite{3SUM} as a means of establishing hardness of geometric problems. Concretely, it was conjectured that $3$SUM requires $\Omega(n^2)$ time when the underlying group is the set of reals and we use the Real-RAM computational model. By reductions, this conjecture implies similar lower bounds for a wealth of geometric problems, see e.g.~\cite{Geom3SUM1,Geom3SUM2}.

While originally being restricted mostly to geometric problems, the seminal work by \patrasku~\cite{TowardsPoly} showed that a suitable integer version of $3$SUM (e.g. $\cG$ is the integers modulo $n^3$), may be used to prove hardness of numerous fundamental algorithmic problems (see e.g.~\cite{Higher3SUM,Jumbled,WeightSubgraph,TowardsPoly}) in the more realistic word-RAM model. These lower bounds are based on the so-called $3$SUM Conjecture, asserting that no $n^{2-\delta}$ time $3$SUM algorithm exists for any constant $\delta>0$. To date, the fastest $3$SUM algorithm runs in time $O(n^2 (\lg \lg n)^{O(1)}/\lg^2 n)$~\cite{Fast3SUM}, which is far from refuting the conjecture. The $3$SUM Conjecture is now one of the pillars in fine-grained complexity and much effort has gone into understanding its implications for algorithm lower bounds. In fact, recently 
the fine-grained hardness of $k$SUM was used 
by LaVigne, Lincoln, and Williams in \cite{LLW19} to construct fine-grained one-way functions.

Highly related to algorithm lower bounds is lower bounds for data structures, which is the setting that we are interested in. 
While more progress has been made on proving unconditional lower bounds for data structures compared to algorithms, current state-of-the-art lower bounds are still only polylogarithmic~\cite{DynamicLB,ButterDynamic,Static}. This lack of progress motivates fine-grained conditional lower bounds also for data structures. The first approach in this direction, is via the Online Matrix-Vector Problem by Henzinger et al.~\cite{OMV}. Their framework yields polynomial conditional lower bounds for dynamic data structures via reductions from multiplication of a boolean matrix and a boolean vector, with addition replaced by OR and multiplication replaced by AND. However, their framework is inherently tied to \emph{dynamic} data structure problems, where a data set is to be maintained under update operations. As a means to addressing \emph{static} data structure problems, Goldstein, Kopelowitz, Lewenstein, and Porat in \cite{GKLP17} introduced the $3$SUM-Indexing Problem.

\paragraph{($3$SUM-Indexing)}
The $3$SUM-Indexing problem was first defined by Demaine and Vadhan in an unpublished manuscript \cite{DV01} and then by Goldstein, Kopelowitz, Lewenstein, and Porat in \cite{GKLP17} and is as follows:
\begin{definition}[$3$SUM-Indexing]\label{defn:3sumIndexing}
    Let $(\cG, +)$ be a finite abelian group. Preprocess two sets of group elements $\cA_1, \cA_2 \subseteq \cG$ each of size $n$ into a data structure of $S$ memory cells of $w$ bits so that given any query group element $z$, deciding whether there exists $a_1 \in \cA_1$ and $a_2 \in \cA_2$ such that $a_1 + a_2 = z$ is done by accessing at most $T$ memory cells.
\end{definition}

A number of hardness conjectures were provided together with the definition of the $3$SUM-Indexing Problem. Combined with reductions, these conjectures allow establishment of conditional lower bounds for static data structures. To be consistent with the terminology used for unconditional data structure lower bounds, which are typically proved in the cell probe model~\cite{CellProbe}, we refer to accessing a memory cell as \emph{probing} the cell. The following conjectures were made regarding the hardness of $3$SUM-Indexing:

\begin{conjecture}[\cite{GKLP17}]
\label{cnj:a}
Any data structure for $3$SUM-Indexing with space $S$ and $T=O(1)$ probes must have $S = \tilde{\Omega}(n^2)$.
\end{conjecture}

\begin{conjecture}[\cite{DV01}]
\label{cnj:b}
Any data structure for $3$SUM-Indexing with space $S$ and $T$ probes must have $ST = \tilde{\Omega}(n^2)$.
\end{conjecture}

\begin{conjecture}[\cite{GKLP17}]
\label{cnj:c}
Any data structure for $3$SUM-Indexing with space $S$ and $T=O(n^{1-\delta})$ probes must have $S = \tilde{\Omega}(n^2)$.
\end{conjecture}
Clearly the last conjecture is the strongest, and in general, we have the following implications:
\[
\textrm{Conjecture~\ref{cnj:c}} \Rightarrow \textrm{Conjecture~\ref{cnj:b}}  \Rightarrow \textrm{Conjecture~\ref{cnj:a}} 
\]
These conjectures have been successfully used to prove fine-grained hardness of several natural static data structure problems ranging from Set Disjointness, Set Intersection, Histogram Indexing to Forbidden Pattern Document Retrieval~\cite{GKLP17}. Furthermore, the security of 
the aforementioned construction of one-way function 
$F'$ relies on the hardness of $3$SUM-Indexing.

Very surprisingly, Golovnev et al.~\cite{GGHPV20} showed that the strongest of these conjectures, Conjecture~\ref{cnj:c}, is false. Concretely, they gave a data structure for $3$SUM-Indexing with $T = \tilde{O}(n^{3\delta})$ and $S = \tilde{O}(n^{2-\delta})$ for any constant $\delta>0$. This refutes Conjecture~\ref{cnj:c}, but not the remaining two conjectures. Their data structure is based on an elegant use of Fiat and Naor's~\cite{FuncInv} general time-space tradeoff for function inversion.

The refutation of Conjecture~\ref{cnj:c} only makes it more urgent that we replace these conjectured lower bounds by unconditional ones. However, depressingly little is still known in terms of unconditional hardness of $3$SUM-Indexing. First,~\cite{DV01} proved Conjecture~\ref{cnj:a} in the special case of $T=1$. Secondly, in the recent work by Golovnev et al.~\cite{GGHPV20}, the following was proved for \emph{non-adaptive} data structures:
\begin{theorem}[\cite{GGHPV20}]
\label{thm:previous}
    Any \emph{non-adaptive} cell probe data structure answering $3$SUM-Indexing queries for input sets of size $n$ from an abelian group $G$ of size $O(n^2)$ using $S$ words of $w$ bits must have query time $T = \Omega( \log n / \log(Sw/n) )$.
\end{theorem}
A non-adaptive data structure is one in which the cells to probe are chosen beforehand as a function \emph{only} of the query element $z$. That is, the data structure is not allowed to choose which memory cells to probe based on the contents of previously probed cells. Proving lower bounds for non-adaptive data structures is often easier than allowing adaptivity, see e.g.~\cite{AdaptOrDie,NonAdaptPred,NonAdaptSunflower}, and Golovnev et al. remark:
"\emph{It is crucial for our proof that the input is chosen at random after the subset of data structure cells, yielding a lower bound only for non-adaptive algorithms}."~\cite{GGHPV20}. Golovnev et al. explicitly raised it as an interesting open problem (Open Question 3 in~\cite{GGHPV20}) whether a similar lower bound can be proved also for adaptive data structures.

\subsection{Our Results}
Our main contribution is a lower bound for $3$SUM-Indexing that holds also for adaptive data structures:
\begin{theorem}
\label{thm:main}
    Any cell probe data structure answering $3$SUM-Indexing queries for input sets of size $n$ for abelian groups $([m], + \mod m)$ with $m = O(n^2)$ and $(\bit^{2\log(n)+O(1)}, \oplus)$ using $S$ words of $w = \Omega(\lg n)$ bits must have query time $T = \Omega( \log n / \log(Sw/n) )$.
\end{theorem}
Our lower bound matches the previous bound from~\cite{GGHPV20}, this time however allowing adaptivity. Moreover, it (essentially) matches the strongest known lower bounds for static data structures (the strongest lower bounds peak at $T = \Omega(\log n/\log(Sw/n))$~\cite{Static}), thus ruling out further progress without a major breakthrough (also in circuit complexity~\cite{ViolaBarrier,DvirBarrier}).

Our proof is based on a novel reduction from \patrasku's Reachability Oracles in the Butterfly graph problem~\cite{P11}. This problem, while rather abstract, has been shown to capture the hardness of a wealth of static data structure problems such as 2D Range Counting, 2D Rectangle Stabbing, 2D Skyline Counting and Range Mode Queries, see e.g.~\cite{ButterDistance,ButterSkyline,ButterMode} as well as for dynamic data structure problems, including Range Selection and Median~\cite{ButterDynamic} and recently also all dynamic problems that the Marked Ancestor Problem reduces to~\cite{FurtherUnifying,MarkedAncestor}, which includes 2D Range Emptiness, Partial Sums and Worst-Case Union-Find. Our work adds $3$SUM-Indexing and all problems it reduces to, to the list.

\paragraph{Even Smaller Universes.}
The reduction from Reachability Oracles in the Butterfly Graph problem gives lower bounds for abelian groups of size $\Omega(n^2)$, leaving open the possibility of more efficient data structures for smaller groups. Indeed, $\Omega(n^2)$ cardinality of the groups seems like a natural requirement for hardness, as there are $n^2$ pairs of elements $a_1 \in \cA_1$ and $a_2 \in \cA_2$ and thus for smaller groups, one might start to exploit structures in the sumset $\cA_1 + \cA_2$ to obtain more efficient data structures. We therefore investigate whether the lower bound in Theorem~\ref{thm:main} can be generalized to smaller groups. Quite surprisingly, we show that:
\begin{theorem}
\label{thm:smalluni}
    Any cell probe data structure answering $3$SUM-Indexing queries for input sets of size $n$ for abelian groups $([m], + \mod m)$, with $m = O(n^{1+\delta})$ and $(\bit^{(1+\delta)\log(n)+O(1)}, \oplus)$ for a constant $\delta>0$, using $S$ words of $w = \Omega(\lg n)$ bits must have query time $T = \Omega( \log n / \log(Sw/n) )$.
\end{theorem}
Thus we get logarithmic lower bounds for linear space data structures, even when the group has size only $n^{1+\delta}$.

To prove Theorem~\ref{thm:smalluni}, we revisit \patrasku's Lopsided Set Disjointness (LSD) communication game, which he also used to prove his lower bound for Reachability Oracles in the Butterfly graph problem. We give a careful reduction from LSD to $3$SUM-Indexing on small universes, thereby establishing Theorem~\ref{thm:smalluni}.

\paragraph{Non-Adaptive Data Structures.}
As another contribution, we revisit the non-adaptive setting considered by Golovnev et al.~\cite{GGHPV20}. Here we present a significantly shorter proof of their lower bound and also improve it from $T=\Omega(\lg n/\lg(Sw/n))$ to $T=\Omega(\lg |G|/\lg(Sw/n))$. Concretely, we prove the following theorem:
\begin{theorem}
\label{thm:nonadaptive}
    Any non-adaptive cell probe data structure answering $3$SUM-Indexing queries for input sets of size $n$ for an abelian group $\mathcal{G}$ of size $\omega(n^2)$, using $S$ words of $w = \Omega(\lg n)$ bits must have query time $T = \Omega(\min\{ \log |\cG| / \log(Sw/n), n/w\})$.
\end{theorem}
We remark that the proof of Golovnev et al.~\cite{GGHPV20} cannot be extended to a $\lg |\cG|$ (technically, they require $|\cG|/n$ queries to survive a cell sampling, whereas we only require $n$ queries to survive). 

Our improvement has a subtle, but interesting consequence. Concretely, if the size of the group grows to sub-exponential in $n$, say $|\cG| = 2^{\sqrt{n}}$, then the lower bound becomes $T=\Omega(\min\{ \sqrt{n} / \log(\frac{Sw}{n}), n/w\})$. Since it is most natural to assume the cell size is large enough to store a group element, i.e. $w = \Omega(\lg |\cG|) = \Omega(\sqrt{n})$, the lower bound is still at least $T = \Omega(\sqrt{n}/\lg S)$. While such large groups are perhaps unrealistic, one can also interpret the result as saying that if we are non-adaptive and attempt to design a data structure that does not exploit the size of the underlying group, then we are doomed to have a slow query time.

\paragraph{Non-Adaptive $2$-Bit-Probe Data Structures.}
Finally, we consider non-adaptive data structures restricted to $T=2$ probes in the \emph{bit probe model}, meaning that each memory cell has $w=1$ bits. The lower bound from Theorem~\ref{thm:previous} by~\cite{GGHPV20} in this case is $S = \tilde{\Omega}(n^{3/2})$ (see the paper~\cite{GGHPV20} for the general formulation $S = \tilde{\Omega}(n^{1+1/T})$) and our lower bound from Theorem~\ref{thm:nonadaptive} is $S = \tilde{\Omega}(n (|\cG|/n)^{1/T}) = \tilde{\Omega}(\sqrt{n |\cG|})$. We significantly strengthen this result by proving an $S = \Omega(|\cG|)$ lower bound for an abelian group $(\cG,+)$, completely ruling out any non-trivial data structure with $2$ non-adaptive bit probes (with $|\cG|$ space, we can trivially store a bit vector representing the sumset $\cA_1 + \cA_2$ and have $T=1$ while being non-adaptive):
\begin{theorem}
\label{thm:bitprobe}
    Any non-adaptive data structure for $3$SUM-Indexing such that $T = 2$ and $w = 1$ requires $S = \Omega(|\cG|)$ for an abelian group $(\cG,+)$.
\end{theorem}
Our proof takes an interesting new approach to data structure lower bounds and we find that the proof itself is a valuable contribution to data structure lower bounds. The basic idea is to view the memory cells of the data structure as a graph with one node per cell. The queries then become edges corresponding to the $T=2$ memory cells probed. If the number of memory cells is $o(|\cG|)$, then the graph has a super-linear number of edges. This implies that its girth is at most logarithmic and hence we can find a short cycle in the graph. A cycle is a set of $m$ queries being answered by $m$ memory cells. The standard cell sampling lower bounds (often used in data structure lower bounds) cannot derive a contradiction from this, as the $m$ memory bits intuitively are sufficient to encode the $m$ query answers. However, our novel contribution is to examine the different types of possible query algorithms (i.e. which function of the two bits probed does it compute) and argue that in all cases, such a short cycle is impossible. Directly examining the types of query algorithms has not been done before in data structure lower bounds and we find this a valuable contribution that we hope may prove useful in future work.

\section{Preliminaries}
In this section we will formally define 
data structure problems, and the $3$SUM Indexing problem.

\subsection{Data Structures}
A data structure problem $\cP$ is a subset of $\cS \times \cQ$. A cell probe data structure solving $P$ is a pair of algorithms $(\cA_1, \cA_2)$ such that $\cA_1$ is a computationally
unbounded algorithm that takes as input $D \in \cS$
and outputs $S$ memory cells, where each cell consists $w$ bits. Then, $\cA_2$ is a computationally 
unbounded algorithm that takes $q \in \cQ$ and uses at most 
$T$ accesses to the data structure output by $\cA_1$ in deciding if $(D, q) \in \cP$.

\subsection{3SUM-Indexing}
The $3$SUM-Indexing problem was first defined by Demaine and Vadhan in an unpublished manuscript \cite{DV01} and then by Goldstein, Kopelowitz, Lewenstein, and Porat in \cite{GKLP17}. 
\begin{definition}[$3$SUM-Indexing]
    Let $(\cG, +)$ be an finite abelian group. Pre-process two sets of group elements $A_1, A_2 \subseteq \cG$ each of size $n$ into a data structure $D$ of $S$ memory cells of $w$ bits so that given any query group element $z$, deciding whether there exists $a_1 \in A_1$ and $a_2 \in A_2$ such that $a_1 + a_2 = z$ is done by accessing at most $T$ memory cells.
\end{definition}

Golovnev, Guo, Horel, Park, and Vaikuntanathan in~\cite{GGHPV20} then defined a similar variant of the problem which additionally asserted that the input sets $A_1$ and $A_2$ should be the same.
We first show that the two definitions are essentially equivalent up to a small change in the group. 
i.e. Any solution that solves the~\cite{GGHPV20} variant of $3$SUM-Indexing can also be adapted to solve the variant in Definition~\ref{defn:3sumIndexing}.

\begin{lemma}
    Any pre-processing solution that solves $3$SUM-Indexing as defined in Definition \ref{defn:3sumIndexing} for input sets of size $n$ and universe sizes of $u$ can also be used to solve $3$SUM-Indexing as defined in \cite{GGHPV20}. 
\end{lemma}

\begin{proof}
    For any abelian group $(\cG', +')$, we define $(\cG, +)$ to be the group obtained as the direct product of $(\bit, +)$ and $(\cG', +')$. Then, given any two sets $\cA_1$ and $\cA_2$, we construct set $\cA = \{0\} \times A_1 \cup \{1\} \times \cA_2$. For any query value $z$, we then form the query as $(1, z)$. In our transformation, the group size and input array length is doubled.

    Assuming that there exists a pair $a_1 \in \cA_1$ and $a_2 \in \cA_2$ such that $a_1 + a_2 = z$, then $(0, a_1), (1, a_2)$ is in $A$ and so $(0, a_1) + (1, a_2) = (1, z)$. On the other hand, assuming that no such pairing exists, then no such pairing $(0, a_1), (1, a_2) \in \cA$ sums to $(1, z)$ as well.
\end{proof}

\section{Reduction from Reachability Oracles in the Butterfly Graph}
In this section, we give a reduction from the problem of \emph{Reachability Oracles in the Butterfly Graph} to $3$SUM-Indexing with the cyclic group and the XOR group, proving Theorem~\ref{thm:main}. In both cases, the size of the group is at most quadratic with respect to the input set sizes.

\begin{definition}[Butterfly Graphs]
    A Butterfly graph of degree $B$ and depth $d$ is a directed graph with $d + 1$ layers, each comprising of $B^d$ nodes. For each layer, the $i^{th}$ node can be associated with a $d$-digit number in base $B$ which we will refer to as its \emph{label} $\numberRep{i}$ where $\numberRep{i}[0]$ denotes the least significant digit. Then there is an edge from node $i$ on the $k^{th}$ layer to node $j$ on the $(k+1)^{th}$ layer if and only if $\numberRep{i}[h] = \numberRep{j}[h]$ for all $h \neq k$. That is to say, that there is an edge if and only if $i$ and $j$ may differ only on the $k^{th}$ digit of their labels. We will denote such an edge by $\butterflyEdge{k}{i}{j}$.
    
    Nodes in the layer $0$ of the graph are called \emph{source nodes}, whereas nodes in layer $d$ of the graph are called \emph{sink nodes}.
\end{definition}

\begin{definition}[Reachability Oracles in the Butterfly Graph]
    The problem of Reachability Oracles in the Butterfly Graph is that one has to pre-process into a data structure a subset of the edges $E$ of the butterfly graph of degree $B$ and depth $d$. Queries come in the form of $(s, t)$ and the goal is decide if there exists a path from source node $s$ to sink node $t$ using the subset of edges $E$.
\end{definition}

\patrasku\ proved the following lower bound for the problem in the cell probe model:
\begin{lemma}[Section 5 of \cite{P11}]\label{lemma:butterflyLowerBound}
    Any cell probe data structure answering reachability queries in subgraphs of the butterfly graph with degree $B$ and depth $d$, using $S$ words of $w$ bits must have query time $t = \Omega(d)$, assuming that $B = \Omega(w^2)$ and $\log(B) = \Omega(\log(Sd/N))$ where $N = d B^d$.
\end{lemma}

A few remarks about reachability in the Butterfly graph are in order. Firstly, note that for any source-sink pair $(s, t)$, there exists a unique path from source $s$ to sink $t$ in the Butterfly graph. Namely, the path uses exactly edges of the form $\butterflyEdge{k}{i}{j}$ such that for $k \in [d]$, $\butterflyEdge{k}{i}{j}$ is the edge from node $i$ on layer $k$ to node $j$ on layer $k + 1$ such that:
\begin{enumerate}
    \item $\numberRep{s}[h] = \numberRep{i}[h]$ for all $h \geq k$. That is to say that the $d - k$ most significant digits of the labels of nodes $s$ and $i$ are the same. 
    \item $\numberRep{t}[h] = \numberRep{j}[h]$ for all $h \leq k$. That is to say that the $k + 1$ least significant digits of the labels of nodes $t$ and $j$ are the same. 
\end{enumerate}
Conversely, we can also say that the edge $\butterflyEdge{k}{i}{j}$ connects all pairs of nodes $s, t$ such that the label for $s$ shares the most significant $d - k$ digits with $i$ and the label for $t$ shares the least significant $k + 1$ digits with $t$. 

Intuitively, this is because the traversing from node $i$ in the $k^{th}$ layer to node $j$ in the $(k+1)^{th}$ layer can be seen as ``setting'' the $k^{th}$ digit of the label for node $i$ into the $k^{th}$ digit of the label for node $j$ while leaving the rest of the digits unaltered.

The general idea of the reduction to $3$SUM-Indexing is to test whether all the required edges are present when querying for $s$ and $t$. This should be done by asking one $3$SUM-Indexing query. We will design it such that a sum $z = a_1 + a_2$ exists for our query $z$ if and only if there is at least one edge missing on the path from $s$ to $t$. 

\paragraph{Constructing $\cA_1$.}
Our basic idea is to take every edge $\butterflyEdge{k}{i}{j}$ in the Butterfly graph and encode it into a group element $g$ in $\cA_1$. We construct $g$ such that its digits can be broken up into 5 blocks so that conceptually the:
\begin{enumerate}
    \item first block encodes the layer the edge is from;
    \item second block encodes the presence of edge $\butterflyEdge{k}{i}{j}$ in $E$;
    \item third block encodes the $d - k$ most significant bits of $i$ followed by $k$ zeroes;
    \item fourth block holds $d - k - 1$ zeroes followed by the $k + 1$ least significant digits of $j$;
    \item fifth block holds $2$ zeroes.
\end{enumerate}
In short, for every edge $\butterflyEdge{k}{i}{j}$, we add group element to $\cA_1$ whose digits are in the following form:
\begin{equation*}
    (k, \indicate{\butterflyEdge{k}{i}{j} \in E}, \numberRep{i}[d-1], \ldots, \numberRep{i}[k], \underbrace{0, \ldots, 0}_\text{d - 1}, \numberRep{j}[k], \ldots, \numberRep{j}[0], 0, 0)
\end{equation*}
where $\indicate{\butterflyEdge{k}{i}{j} \in E}$ is $1$ if $e_k(i, j) \in E$ and $0$ otherwise. Note that the Butterfly graph has $dB^d$ nodes with degree $B$, hence a total of $n=dB^{d+1}$ edges. Since $\cA_1$ has one element for each such edge, we have $|\cA_1|=n$.

\paragraph{Constructing $\cA_2$.}
Next, we construct the set $\cA_2$ of group elements such that for every $k$, it ``helps'' any group element $g$ in $\cA_1$, originating from an edge $\butterflyEdge{k}{i}{j}$, to sum to any value where the third block shares the $d - k$ most significant bits with $i$ and the fourth block shares the $k + 1$ least significant digits of $j$. This can be done by adding into set $\cA_2$ every group element such that the:
\begin{enumerate}
    \item first block holds some value $-k$;
    \item second block is zero;
    \item third block is $d - k$ zeroes followed by any possible $k$ digit value;
    \item fourth block holds any possible $d - k - 1$ digit value followed by $k + 1$ zeroes;
    \item fifth block holds any possible digit value from $[0, B - 1]$.
\end{enumerate}

Thus for $k \in [0, d - 1]$, we add any number of the following form into $\cA_2$:
\begin{equation*}
    (-k, 0, \underbrace{0, \ldots, 0}_{d - k} , \underbrace{\star, \ldots, \star}_{d - 1}, \underbrace{0, \ldots, 0}_{k+1}, \star, \star)
\end{equation*}
where $\star$ denotes wildcard. Note that the least significant digits is not strictly necessary but is included to enforce that the size of the sets $\cA_1$ and $\cA_2$ are the same. Observe that $|\cA_2| = dB^{d+1} = n = |\cA_1|$.

\paragraph{Different Groups.}
For the reduction to $3$SUM-indexing in the cyclic group, we will consider the set of integers in $[(d B^{d+1})^2]$. To that end, the encoding works by understanding the $2(d + 2)$ digits as specifying a mixed-radix number, where the most significant digit is in base $4d$, the second most significant digit is in base $3$ and the remaining digits are in base $B$. In which case, we can take $-k$ to be $4d-k$.

On the other hand, for the XOR group, assuming that $d$ and $B$ are powers of $2$, we can then also naturally transform each digit into their binary representation with the exception of the most significant digit whose bit representation should be based on the number's complement and the second most significant digit may be in base $2$.



\paragraph{Translating a Query.}
What remains is to explain how we answer a reachability query $(s,t)$. We will first consider the reduction for the group $([(d B^{d+1})^2], + \mod (d B^{d+1})^2)$ and subsequently argue that the same reduction basically holds for the XOR group assuming that $d$ and $B$ are powers of $2$. We claim that there exists $a_1 \in A_1$ and $a_2 \in A_2$ whose sum is
    \begin{equation*}
        z_{(s,t)} = (0, 0, \numberRep{s}[d-1], \ldots, \numberRep{s}[0], \numberRep{t}[d-1], \ldots, \numberRep{t}[0], 0, 0)
    \end{equation*}
    if and only if there does not exist a path from $s$ to $t$ in the Butterfly graph. 
    
    To see this, we first argue that for a pair $a_1 + a_2$ that could potentially sum to $z_{(s,t)}$, we need not worry about carries amongst the digits of the numbers. To see this, we start by observing that $a_1 + a_2$ must have its most significant digit equal to $0$. We claim this is only possible if $a_1$'s most significant digit is $k$ and $a_2$'s is $4d-k=-k$. To see this, observe that the second most significant digit of $a_1$ is at most $1$ and the second most significant of $a_2$ is always $0$. Since the second most significant digit is in base $3$, this means that we cannot get a carry from these digits. Now that we have established this, we observe that for all remaining digits of any valid pair $a_1$ and $a_2$ (pairs where the most significant digit in the sum is $0$), there is at most one of the elements that has a non-zero digit, hence we will not see any carries.
    
    Now assume there does not exists a path from some source node $s$ to some sink node $t$. This must mean that there exists a $k \in [0, d - 1]$ and an edge $e_k(i, j)$ not in $E$ where:
    \begin{align*}
        \numberRep{i} &= ( \numberRep{s}[d-1], \ldots, \numberRep{s}[k], \numberRep{t}[k - 1], \ldots, \numberRep{t}[0] ) \\
        \numberRep{j} &= ( \numberRep{s}[d-1], \ldots, \numberRep{s}[k+1], \numberRep{t}[k], \ldots, \numberRep{t}[0] )
    \end{align*}
    By construction, this implies that the following group element exists in the set $A_1$:
    \begin{equation*}
        (k, 0, \numberRep{s}[d-1], \ldots, \numberRep{s}[k], \underbrace{0, \ldots, 0}_\text{d - 1}, \numberRep{t}[k], \ldots, \numberRep{t}[0], 0, 0)
    \end{equation*}
    Furthermore, the following group element always exists in $A_2$:
    \begin{equation*}
        (-k, 0, 0, \ldots, 0, \numberRep{t}[k - 1], \ldots, \numberRep{t}[0], \numberRep{s}[d-1], \ldots, \numberRep{s}[k+1], 0, \ldots, 0, 0)
    \end{equation*}
    This means that the value $(0, 0, \numberRep{s}[d-1], \ldots, \numberRep{s}[0], \numberRep{t}[d-1], \ldots, \numberRep{t}[0], 0)$ is obtainable as a sum $a_1 + a_2$. If on the other hand there is a path between $s$ and $t$, then all elements in $A_1$ of the form
    \begin{equation*}
        (k, \star, \numberRep{s}[d-1], \ldots, \numberRep{s}[k], \underbrace{0, \ldots, 0}_\text{d - 1}, \numberRep{t}[k], \ldots, \numberRep{t}[0], 0)
    \end{equation*}
    must have $\star = 1$ and thus it is not possible to write $z_{(s,t)}$ as $a_1 + a_2$.
\paragraph{The XOR Group.}
    For a reduction to the XOR group setting, we consider each element coordinate-wise using their binary representations with the exception that in the first coordinate the value is represented using the number's complement representation. Using the previous remark we also assert that for any pair $a_1 \in \cA_1$, $a_2 \in \cA_2$, the only common digit that is both non-zero is the most significant digit and thus the addition being done digit-wise. For that reason, the sum behaves exactly the same way over the XOR group as it does over the cyclic group that we have defined. Thus the size of the universe and input sets $\cA_1, \cA_2$ remain unchanged and the reduction holds in the XOR group as well.

\paragraph{Analysis.}
    Now by setting $B = \frac{Sw^2}{n}$, note that $B = \Omega(w^2)$ and:
    \begin{align*}
        \log(Sd/N) \leq \log\left(\frac{S B\log(n)}{n}\right) \leq \log(S B w/n) \leq \log(B^2) = O(\log B)
    \end{align*}

    Furthermore, it holds that
    \begin{align*}
        \log(Sw/n) = \frac{1}{2} \log( (Sw/n)^2) \geq \frac{1}{2} \log(Sw^2/n) \geq \frac{1}{2} \log B
    \end{align*}

    Using Lemma \ref{lemma:butterflyLowerBound}, it then follows that for any cell-probe solution for $3$SUM-Indexing for the cyclic group $([m], + \mod m)$ where $m = O(n^2)$ and XOR group $(\bit^{2\log(n) + O(1)}, \oplus)$ any static data structure that uses $S \geq n$ cells of $w \geq \log(n)$ bits has query time $T = \Omega(d) = \Omega(\lg_B n) = \Omega( \log n / \log(Sw/n))$.


\section{Reduction from Lopsided Set Disjointness}
In this section, we prove Theorem~\ref{thm:smalluni}, establishing hardness of $3$SUM-Indexing also for abelian groups of size $\Delta = n^{1+\delta}$. For the proof, we focus on the integers modulo $\Delta$, but remark that the proof readily adapts to the XOR group as well.

For the proof, we use \patrasku's Blocked Lopsided Set Disjointness (Blocked LSD) problem. In this problem, there are two players, Alice and Bob. Bob receives as input a set $\cX$, which is an arbitrary subset of a universe $[N] \times [B]$. Alice receives a set $\cY \subset [N] \times [B]$ with the restriction that $\cY$ contains exactly one element $(i,b_i)$ for every $i=0,\dots,N-1$. The goal for Alice and Bob is to determine whether $\cX \cap \cY = \emptyset$ while minimizing communication. The following is known regarding the communication complexity of Blocked LSD:
\begin{lemma}[Theorem 4 of \cite{P11}]\label{lemma:lsdLowerBound}
    Fix $\delta > 0$. Any communication protocol for Blocked LSD requires either Alice sending at least $\delta N \log B$ bits, or Bob sending at least $NB^{1-O(\delta)}$ bits.
\end{lemma}
The basic idea in the reduction, is to have Bob interpret his set $\cX$ as two input sets $\cA_1,\cA_2$ of $n=NB$ group elements to $3$SUM-Indexing (we may have $|\cA_1|$ and $|\cA_2|$ smaller than $NB$, but we can always pad with dummy elements, so we assume $n=NB$). Given a data structure $D$ for $3$SUM-Indexing, Bob then builds $D$ on this input. Alice on the other hand interprets her set $\cY$ (which has cardinality $N$) as a set of $N/\ell$ queries to $3$SUM-Indexing, where $\ell$ is a parameter to be determined. The key property of the reduction, is that the answers to all $N/\ell$ queries of Alice on $D$, determines whether $\cX \cap \cY = \emptyset$.

\paragraph{Communication Protocol.}
Assume for now that we can give such a reduction. Alice and Bob then obtains a communication protocol for Blocked LSD as follows: Let $T$ be the query time of $D$. For $i=1,\dots,T$, Alice simulates the $i$'th step of the query algorithm for each of her $N/\ell$ queries, \emph{in parallel}. This is done by asking Bob for the set of at most $N/\ell$ cells that they probe in the $i$'th step. This costs $O(\lg \binom{S}{N/\ell}) = O((N/\ell) \lg(S\ell/N))$ bits of communication by specifying the required cells as a subset of the $S$ memory cells of $D$. Bob replies with the contents of the cells, costing $((N/\ell)w)$ bits. This is done for $T$ rounds, resulting in a communication protocol where Alice sends $O((N/\ell) T \lg(S\ell/N))$ bits and Bob sends $O((N/\ell)Tw)$ bits. If we fix $B = w^4$ and $\delta$ as a small enough constant, then Lemma~\ref{lemma:lsdLowerBound} says that either Alice sends $\Omega(N \lg w)$ bits or Bob sends $\Omega(N \sqrt{B}) = \Omega(N w^2)$ bits. In our protocol, Bob's communication is $O((N/\ell)T w) = O(N Tw)$ bits. We assume $w = \Omega(\lg n)$, thus we conclude that either $NTw = \Omega(N w^2) \Rightarrow T = \Omega(\lg n)$, or Alice's communication must be $\Omega(N \lg B) = \Omega(N \lg w)$ bits. In the first case, we are done with the proof, hence we examine the latter case. Alice's communication is $O((N/\ell) T \lg(S \ell/N))$ bits, which implies $T = \Omega(\ell \lg w/\lg(S \ell /N))$. Thus to derive our lower bound, we have to argue that it suffices for Alice to answer $N/\ell$ queries for a large enough $\ell$.

\paragraph{Asking Few Queries.}
We will show that it suffices for Alice to ask $N/\ell$ queries with $\ell = \eps \lg n/\lg w$. Here $\eps > 0$ is a small constant depending on $\delta$ in the group size $\Delta = n^{1+\delta}$. Thus we get a lower bound of $T = \Omega(\lg n/\lg((S \lg n)/(N \lg w)))$. Since $N = n/B = n/w^4$, this simplifies to $T = \Omega(\lg n/\lg(Sw/n))$ as claimed in Theorem~\ref{thm:smalluni}.

Thus what remains is to show how Alice and Bob computes the input and queries. For this, they conceptually partition the universe $[N] \times [B]$ into groups $\{i \ell,\dots,(i+1)\ell-1\} \times [B]$ for $i=0,\dots,N/\ell$. Alice will ask precisely one query for each such group. Denote by $\cY_i$ the subset of $\cY$ that falls in the $i$'th group and denote by $\cX_i$ the subset of $\cX$ that falls in the $i$'th group. Clearly $\cX \cap \cY = \emptyset$ if and only if $\cX_i \cap \cY_i = \emptyset$ for all $i$. Thus Alice will use her $i$'th query to determine whether $\cX_i \cap \cY_i = \emptyset$. 

\paragraph{Constructing $\cA_1$ and $\cA_2$.}
To support this, Bob first constructs the set $\cA_1$ based on his elements $\cX$. He examines each group $\cX_i$, and for every $(j,b) \in \cX_i$, he adds the integer $i(2B+1)^{\ell+1} + (b+1)(2B+1)^{j-i\ell}$ to $\cA_1$. Next, he constructs the set $\cA_2$. For this, he considers all vectors $Z=(b_0,\dots,b_{\ell - 1})$ for which the numbers are all between $0$ and $B$ and precisely one of them is $0$. He adds the integer $\sum_{j \in [\ell]} b_j (2B+1)^j$ to $\cA_2$. This completes Bob's construction of the input sets $\cA_1$ and $\cA_2$. We have $|A_1| \leq NB = n$ and $|\cA_2| \leq (2B+1)^\ell$.

\paragraph{Asking the Queries.}
We next describe how Alice translates her set $Y$ into queries. For each $Y_i$, she needs to construct one query $z_i$ whose answer determines whether $X_i \cap Y_i = \emptyset$. Recall that $Y_i$ is of the form $\{(i\ell,b_0),(i\ell+1,b_1),\dots,((i+1)\ell-1,b_{\ell-1})\}$. She starts by subtracting off $i \ell$ from the first index in each pair, obtaining the set $\{(0,b_0),(1,b_1),\dots,(\ell-1,b_{\ell-1})\}$. Alice now asks the query $z_i = i(2B+1)^{\ell+1} + \sum_{j=0}^{\ell-1} (b_j+1) (2B+1)^j$. 

\paragraph{Correctness.}
We claim that $z_i$ is part of a $3$SUM if and only if $\cX_i \cap \cY_i \neq \emptyset$. To see this, observe first that to write $z_i$ as $a_1 + a_2$, it must be the case that $a_1$ was constructed from $\cX_i$ as otherwise we cannot obtain the $i(2B+1)^{\ell+1}$ parts of $z_i$. Next, observe that if we write the integers in base $2B+1$, then $\cA_2$ contains precisely every integer of the form where there is a single digit $j \in [\ell]$ that is zero and all remaining are non-zero. Also, the numbers obtained from $(j,b) \in \cX_i$ are of the form $i(2B+1)^{\ell  + 1} + (b+1)(2B+1)^{j-i\ell}$ and thus have exactly one non-zero digit among the first $\ell$. Since $z_i$ has exclusive non-zero digits in the first $\ell$, it follows that $z_i=i(2B+1)^{\ell+1} + \sum_{j=0}^{\ell-1} (b_j+1) (2B+1)^j$ can be written as $a_1 + a_2$ if and only if $a_1$ was obtained from a $(j,b) \in X_i$ for which $b$ is equal to $b_{j-i\ell}$. This is the case if and only if $\cX_i$ and $\cY_i$ intersect in $(j,b)$.

\paragraph{Analysis.}
We now determine $\ell$. Recall that $B = w^4$ and observe that all possible integers are bounded by $N (2B+1)^{\ell + 2}  \leq n (2B+1)^{\ell + 2}$. If we insist on a group of size $\Delta = n^{1+\delta}$, this means we can set $\ell = \delta \lg n/\lg(2B+1) - 2 \geq \eps \lg n/\lg w$ for a sufficiently small constant $\eps>0$. This also implies that $|A_2| \leq n^\delta \leq n$ and thus completes the proof of Theorem~\ref{thm:smalluni}.
\section{Lower Bound for Non-Adaptive Data Structures}
In this section, we prove an $\Omega(\min\{\lg|\cG|/\lg(Sw/n),n/w\})$ lower bound for non-adaptive $3$SUM-Indexing data structures when $|\cG| = \omega(n^2)$. Similarly to the previous approach by Golovnev et al.~\cite{GGHPV20}, we use a cell sampling approach. 

Consider a data structure using $S$ memory cells of $w$ bits and answering queries non-adaptively in $T$ probes. Consider all subsets of $\Delta = n/(2w)$ memory cells. There are $\binom{S}{\Delta}$ such subsets. We say that a query $z$ is answered by a set of cells $\cC$, if all the (non-adaptively chosen) cells it probes are contained in $\cC$. Any query $z$ is answered by at least $\binom{S-T}{\Delta-T}$ sets of $\Delta$ cells, namely all those containing the $T$ cells probed on $z$. It follows by averaging over the $|\cG|$ queries that there is a set of cells $\cC^*$ answering at least
\[
|G|\binom{S-T}{\Delta-T}/\binom{S}{\Delta} = |\cG| \frac{\Delta(\Delta-1)\cdots(\Delta-T+1)}{S(S-1)\cdots(S-T+1)} \geq |G| \left(\frac{\Delta - T + 1}{S}\right)^T
\]
queries.

If $T \geq \Delta/2$, we are already done as we have proven $T = \Omega(n/w)$. Otherwise, $T \leq \Delta/2$ and thus the above is at least $|\cG|(\Delta/(2S))^T = |\cG|(n/(4Sw))^T$. If we assume for contradiction that $T = o(\lg|\cG|/\lg(Sw/n))$, this is at least $|\cG|^{1-o(1)} > n$. Let $\cQ$ be the group elements corresponding to an arbitrary subset of $n$ of those queries. We argue that we can construct a distribution over inputs $\cA_1,\cA_2$ such that the queries $\cQ$ cannot be answered from few cells, contradicting that we have answered them from $\cC^*$. More precisely, we show:

\begin{lemma}\label{lemma:independence}
    Let $(\cG, +)$ be an abelian group with $\omega(n^2)$ elements. Given any subset $\cQ \subseteq \cG$ of at most $n$ elements, there exists an input distribution $D$ of $\cA_1, \cA_2$ such that, all the events of the form $q \in (\cA_1 + \cA_2)$ (defined as $\{a_1 + a_2 : a_1 \in \cA_1, a_2 \in \cA_2\}$) for all $q$ in $\cQ$ is fully independent. That is, for any subset $\cS = \{ s_1, s_2 \ldots, s_r\}$ of $\cQ$ of $r$ elements, and any sequence of $r$ events $E_1, E_2, \ldots, E_r$ either of the form $s_i \in (A_1 + A_2)$ or the form $s_i \notin (A_1 + A_2)$, it holds that $\Pr[\bigwedge_{i = 1}^r E_i] = \prod_{i = 1}^r \Pr[E_i]$ . Furthermore, for any $q \in \cQ$, it is the case that $\Pr_{(A_1,A_2) \sim D}[q \in (A_1 + A_2)] = \frac{1}{2}$.
\end{lemma}
The proof is deferred to the end of the section.

We now use Lemma~\ref{lemma:independence} to derive a contradiction to the assumption that $T=o(\lg|\cG|/\lg(Sw/n))$. Concretely, we invoke the lemma with the $Q$ defined above. This implies that the answers to the queries in $\cQ$ has entropy $n$ bits. However, they are being answered from a fixed set of $n/2w$ cells. These cells together have $n/2$ bits. Since their addresses are fixed, their contents must uniquely determine the $n$ query answers, yielding the contradiction and hence $T = \Omega(\min\{\lg|\cG|/\lg(Sw/n), n/w\})$. This completes the proof of Theorem~\ref{thm:nonadaptive}. What remains is to prove Lemma~\ref{lemma:independence}:

\paragraph{Proof of Lemma \ref{lemma:independence}.}
We prove the lemma by first showing that given $\cQ \subseteq \cG$ of $n$ elements, for any $\cP \subseteq \cQ$ there exists an input pair $\cA_1^\cP$ and $\cA_2^\cP$ such that $\cP \subseteq (\cA_1^\cP + \cA_2^\cP)$ and $(\cQ \setminus \cP) \cap (\cA_1^\cP + \cA_2^\cP) = \emptyset$. That is to say that for every possible subset $\cP$ of $\cQ$, there exists a pair of sets $(\cA_1^\cP, \cA_2^\cP)$ such that $(\cA_1^\cP + \cA_2^\cP)$ contains all the pair sums of $\cP$ and none of the pair sums outside of $\cP$ and in $\cQ$. Then $D$ is the distribution that is uniform over all possible pairs of sets $(\cA_1^\cP, \cA_2^\cP)$ with $\cP$ ranging over all subsets of $\cQ$. Another way to view $D$ is the distribution that first randomly samples $\cP \subseteq \cQ$ before deterministically outputing pairs of sets $(\cA_1^P, \cA_2^P)$.

Given any $\cP$, we build the sets $\cA_1^\cP$ and $\cA_2^\cP$ iteratively, where they are both initially empty. Let $p_1, p_2, \ldots$ enumerate the elements of $\cP$. At each iteration, let $p_i$ be the first value not in $(\cA_1^\cP + \cA_2^\cP)$. There are $\lvert \cG \rvert$ ordered pairs of elements $(a_1, a_2)$ such that $a_1 + a_2 = p_i$. To see this, note that letting $a_1 = p_i + (-t)$ and $a_2 = t$ for any $t \in \cG$ yields us a distinct pair of elements for which the sum holds. We want to show that we can add $n$ pairs of elements (thus enumerating all of the elements in $\cP$ and beyond) without ever having any pair sum to an element in $(\cQ \setminus \cP)$. For each element $q \in (\cQ \setminus \cP)$, and each element in $a \in \cA_1$, there is exactly one element $b \in \cG$ such that $a + b = q$ (likewise for each element $a \in \cA_2$). Therefore, for any given $q$, there are $\lvert \cA_1 \rvert$ elements $b \in \cG$ that if added into set $\cA_2$, would imply that $q \in (\cA_1 + \cA_2)$ (likewise for set $\cA_2$). Since $\lvert \cQ \setminus \cP \rvert \leq n$, and at every iteration $\lvert \cA_1 \rvert = \lvert \cA_2 \rvert \leq n$, we have that there are at most $2n^2$ elements that cannot be added into either set $\cA_1$ or set $\cA_2$ (otherwise sets $(\cQ \setminus \cP)$ and $(\cA_1 + \cA_2)$ are no longer disjoint).

Therefore there must still exist a pair $(a_1, a_2)$ such that $a_1 + a_2 = p_i$ and $(\{a_1\} \cup \cA_1^\cP + \{a_2\} \cup \cA_2^\cP) \cap (\cQ \setminus \cP) = \emptyset$, assuming that $\groupSize = \omega(n^2)$. In the case that every element in $\cP$ is enumerated before we have added $n$ pairs, we can still pad with more arbitrary pairs of elements from $\cG$ whilst avoiding creating any element in $(\cQ \setminus \cP)$ for the same reason as laid out above.

It remains to show that our distribution $D$ indeed witnesses full independence and that each individual event occurs with probability $\frac{1}{2}$. Let $\cS$ be an arbitrary subset of $\cQ$ of size $r \leq n$. Further, let $E_i$ be either the event that $s_i \in (\cA_1 + \cA_2)$ or $s_i \notin (\cA_1 + \cA_2)$, and let $\cS' \subseteq \cS$ contain the elements $s_i$ such that $E_i$ is the event that $s_i \in (\cA_1 + \cA_2)$ (so $\cS \setminus \cS'$ is precisely the set of elements $s_i$ for which there is the event $s_i \notin (\cA_1 + \cA_2)$). In the support of $D$, there are exactly $2^n$ pairs of sets $(\cA_1^P, \cA_2^P)$, each realising a distinct subset $\cP \subseteq \cQ$ of elements such that $\cP \subseteq (\cA_1 + \cA_2)$ and $(\cQ \setminus \cP) \cap (\cA_1 + \cA_2) = \emptyset$. Thus, given any set $\cS' \subseteq \cS \subseteq \cQ$, there are $2^{n - r}$ pairs of sets $(\cA_1^\cP, \cA_2^\cP)$ each with for set $\cP$ such that $\cS' \subseteq \cP$ and $(\cS \setminus \cS') \subseteq (\cQ \setminus \cP)$. Thus we argue that
\begin{equation*}
    \Pr\left[ \bigwedge_{i = 1}^r E_i \right] = \frac{2^{n - r}}{ 2^{n} } = 2^{-r}.
\end{equation*}

Note that for individual events, we can take the subset $\cS$ to contain only a single element $q$ from $\cQ$ and the above argument would imply that $\Pr[ q \in (\cA_1, \cA_2) ] = \frac{1}{2}$ and that $\Pr[ q \notin (\cA_1, \cA_2) ] = \frac{1}{2}$. Thus the conclusion readily follows from the fact that
\begin{equation*}
    \prod_{i = 1}^r \Pr\left[ E_i \right] = 2^{-r}.
\end{equation*}
\section{Bit Probe Lower Bound for $3$SUM-Indexing}
In this section we give the bit probe lower bound for $3$SUM-Indexing stated in Theorem~\ref{thm:bitprobe}.

The proof idea is based on an incompressibility argument. We will inspect the way the queries are structured and construct a specific input distribution that the data structure algorithm end up using too few bits for and therefore derive a contradiction. For this, we will again use Lemma~\ref{lemma:independence} from the previous section. The key difference between this proof and the proof in the previous section, lies in how we find a set of queries answered by too few cells. Moreover, in this proof, we will derive a contradiction even with $m$ queries being answered by $m$ cells, and thus intuitively the cells actually have enough information, but yet cannot answer the queries. We start by introducing some graph theory that we need: 

\begin{lemma}\label{lemma:girthBound}[Theorem 1 of \cite{AHL02}]
    Let $(V, E)$ be a graph with $n$ nodes, average degree $d > 2$ and girth $r$. Then $n \geq 2(d - 2)^{r / 2 - 2}$.   
\end{lemma}

From Lemma \ref{lemma:girthBound} we conclude that for a graph with $o(\groupSize)$ nodes and $\groupSize$ edges, it is the case that the graph has a girth of $O(\log (n))$. To see this, note that the average degree $d$ of such a graph is $\omega(1)$ and thus it follows that for some constant $c > 1$:
\begin{eqnarray*}
    o(\groupSize) \geq 2(d - 2)^{r / 2 - 2} &\Rightarrow& \\
    o(\groupSize) \gg 2(c)^{r / 2} &\Rightarrow &\\
    r \in O(\log_c(n))\\
\end{eqnarray*}

Given any non-adaptive pre-processing algorithm with $T = 2$, $S = o(\groupSize)$, and $w = 1$, define $\cV$ to be the set of $S$ nodes each representing a memory cell and let $E$ be the set of edges such that an edge $e_g = (u, v)$ is in the edge set if and only there exists some group element $g \in \cG$ such that the querying algorithm on input $g$ accesses both memory cell $u$ and $v$. Furthermore, associate with each edge $e_g$ a function $f_g : \bit^2 \to \bit$ that defines the output behaviour of the querying algorithm upon reading the bits at node $u$ and $v$. We broadly categorise the possible functions $f_g$ into $4$ types:
\begin{enumerate}
    \item \textbf{Copy type functions}. The type of functions $f_g$ that depend only one of its two inputs. There are $4$ of such functions.
    \item \textbf{Constant type functions}. The type of functions $f_g$ that are completely independent of its two inputs. There are $2$ such functions. 
    \item \textbf{AND type functions}. The type of functions $f_g$ whose truth table is such that exactly $3$ of the $4$ possible inputs leads to the same output where the last input differs. There are $8$ such functions.
    \item \textbf{XOR type functions}. The type of functions $f_g$ that are either the XOR of its $2$ inputs or the negation of the XOR of its $2$ inputs. There are $2$ such functions. 
\end{enumerate}

Note that none of the edges can be the constant type, since this means that the querying algorithm's answer is independent of the input set. Also, by an averaging argument there is at least one type of function that $\Omega(\groupSize)$ edges are associated with. Furthermore, Lemma \ref{lemma:independence} asserts that there can be at most $2$ edges that are parallel to each other, otherwise we can construct an input distribution $D$ such that the data structure manages to use $2$ bits to encode the outcome of a random variable that has Shannon entropy at least $3$, which is a contradiction. Thus there are $\Omega(\groupSize)$ many edges that are not parallel to each other and are all of the same type. We analyse the different types separately. We start with the simplest COPY type:

\textbf{(COPY type)} Assuming that there are $\Omega(\groupSize)$ edges that are associated with the copy type function, there must exist at least one node $u$ such that $\omega(1)$ edges $e_g$ are such that the associated function $f_g$ depend only on the bit at this node. Letting $\cQ$ contain two such group elements, this yields a contradiction using Lemma \ref{lemma:independence} to construct a distribution over $\cA_1$ and $\cA_2$ such that the entropy of the two query answers in $\cQ$ is $2$ bits.

For the remaining types, we look for a short cycle. Using Lemma \ref{lemma:girthBound}, we get that there is a cycle of $O(\log n)$ length using only edges associated with functions of the same type. Denote by $\cY$ the set of group elements $g$ such that $e_g$ is in the cycle and $\blocks{y}{t}$ enumerates the elements of $\cY$ based on a traversal of the cycle. That is, the edge corresponding to $y_i$ shares endpoints with edges corresponding to $y_{i-1}$ and $y_{i+1}$, where $y_{t+1} = y_1$ and $y_0 = y_t$. We use Lemma~\ref{lemma:independence} with $\cQ=\cY$ to get a distribution $D$ over $(\cA_1,\cA_2)$ such that the answers to queries in $\cY$ are independent and they are all uniform random. We now handle the two remaining types separately.

\textbf{(AND type)}
Let $b$ be the output of $f_{y_1}$ that is only obtainable by exactly $1$ of the $4$ possible inputs. Consider the distribution $D$ conditioned on the event that $\indicate{y_1 \in (A_1 + A_2)} = b$. Since only $1$ of the $4$ inputs to $f_{y_1}$ is consistent with this output, this fixes the two input bits to $f_{y_1}$. Therefore, there are $t - 2$ bits left to encode $t - 1$ independent and fully random outputs (namely, whether $y_2, \ldots, y_t$ are in $A_1 + A_2$), which yields us the desired contradiction. 

\textbf{(XOR type)}
Let $(\cA_1, \cA_2) \sim D$ be drawn from the input distribution constructed using Lemma \ref{lemma:independence} with $\cQ = \cY$. Let the endpoints of $y_i$ be $u_i, v_i$, such that $v_t = u_1$. Note for all $i$, it is necessarily the case that $\indicate{y_i \in (\cA_1 + \cA_2)} = u_i \oplus v_i$ or $\indicate{y_i \in (\cA_1 + \cA_2)} = 1 \oplus u_i \oplus v_i$. Then $\bigoplus_2^{t} \indicate{y_i \in (A_1 + A_2)}$ is either $u_1 \oplus v_1$ or $1 \oplus u_1 \oplus v_1$ which means that $\indicate{y_1 \in (\cA_1 + \cA_2)}$ is either $\bigoplus_2^{t} \indicate{y_i \in (\cA_1 + \cA_2)}$ or its negation. Then Lemma \ref{lemma:independence} yields the desired contradiction.

\chapter{Conclusion and Future Work}
\label{ch:concl}

\section{Conclusion}
In this thesis, we have worked on giving ``negative'' (or impossibility) results in the form of randomness requirements
for extensions of secret sharing, and ``positive'' (or feasibility) results in the form of either giving privacy amplification with better guarantees, or confirming a certain level of security guarantees for one-way functions against pre-processing adversaries from random oracles.

\section{Future Work}

\subsection{Randomness Requirements}
\paragraph{(Does plain secret sharing need extractable randomness?)} The main question posed by Dodis in \cite{BD07} remains: Do we need extractable randomness to share even a single bit to two parties?
Given that secret sharing is a fundamental cryptographic primitive used as a building block in many other primitives, being able to 
find an answer to this will also shed light on the complexity (in terms of randomness) of other primitives as well.

\subsection{Non-malleable Extractors}
\paragraph{(Even better entropy rates)} After our work, Li in \cite{L23} improved the entropy requirement from $> 0.8n$ from one source to $> \frac{2}{3}n$.
This falls slightly short of the $\frac{1}{2}n$ threshold witnessed by the Raz extractor.
Are we able to close the gap any further to the halfway point? 
We also note that we get a $t$-non-malleable extractor instead. 
As mentioned by \cite{AORSS20}, ``good enough'' $t$-non-malleable extractors will yield a breakthrough 
in two-source extractors as well. While our extractor falls short of accomplishing that, a question might be 
whether our ideas can be improved upon to yield such an extractor.

\paragraph{(Better Multi-Tampering for Two-Source Extractors)}
In Theorem 7 of \cite{AORSS20}, it was mentioned that two-source
non-malleable extractors that have entropy-rate $m(1-\gamma) - 3\log(m)$ entropy in both sources, and can handle $\frac{C_\delta}{\gamma^5})$ tamperings will lead to two source extractors with for entropy rates $\delta n$. Both extractors here in the statement extract a single bit.

Our extractor in Chapter 4, at a one-sided rate of $(1-\frac{1}{2t + 3})$,
allows for $t$ tampering functions, instead of $t^5$ tamperings. On
the other hand, our very same extractor is lopsided (in that 
the other entropy requirement is $\approx n^\alpha$, for any $\alpha > 0$
in the second source). Can we somehow perform a trade-off to 
increase the entropy requirement in the second source to also 
increase the number of tampering functions that we can allow for (from $t$ to $O(t^5)$)?

\subsection{Collision Resistant Extractors}
\paragraph{(Improving the seed length for collision resistant extractors)} One other thing to note is that our result 
of constructing collision resistant extractors from seeded extractors incurs a blowup in the seed by an \emph{additive} term of
\\ $\log(1/\eps)\log(1/\delta)\log^2(\log(1/\delta))$ for a collision probability of $\delta$, and distance $\eps$ from $m$ uniform bits.

Using Trevisan's extractor \cite{RRV02}, we note that for $\delta \leq \eps$, 
this additive term is dominated by the original seed length requirement of the extractor anyway, and thus we get collision resistance ``for free''. 
However, Trevisan's extractor excels at extracting $k$ bits from a source with
min-entropy $k$. If one were content with extracting $\Omega(k)$ bits instead,
they can use the GUV extractor from \cite{GUV09}. In this case,
the seed requirement is actually quadratically smaller --- $d = O(\log(n) + \log(1/\eps))$, and thus our term dominates instead (by a factor of $\log(1/\delta)\log^2(\log(1/\delta))$). It seems pretty clear from our construction that 
we will not be able to shave off this additional factor. But it does beg the question
(since we were able to match, up to constant factors, the parameters in the $m = k$ regime) whether it is also possible to create an extractor with the same parameters as
the GUV extractor, but while retaining collision resillience?

\subsection{Pre-processing Adversaries}
\paragraph{(Improving lower bounds for $3$SUM-Indexing)} While we have some results based on the hardness of $3$SUM, this is far less than desirable in a few ways:
\begin{enumerate}
    \item The lower bounds are worst-case hardness, rather than average case. For this to be useful in a cryptographic setting, 
    average case results would be required. Can we get lower bounds for a random queries?
    \item The initial construction introduced by Guo et al. in \cite{GGHPV20} used $k$SUM. While all confirmed hardness results
    used $3$SUM. Can we derive potentially stronger results using $k$SUM?
    \item Our lower bounds only work in a low-space setting. Are we able to obtain lower bounds for the other extreme: High space but 
    low oracle access. One of our results takes a step in this direction but still in a highly restricted setting: non-adaptive, and for $T = 2, w = 1$. Can we at least increase $w$ to $O(\log (n))$? What about $T = O(1)$? 
\end{enumerate}

We want to additionally note that slight improvements to the results
will lead to explicit circuit lower bounds (c.f. works by Dvir, Golovnev, and Weinstein \cite{DGW19}, and Viola \cite{V19}). Thus, we view our
results as nearly tight without new techniques.

\bookmarksetup{startatroot}
\printbibliography[heading=bibintoc]



\end{document}